\documentclass[apj,twocolumn, twocolappendix]{openjournal}
\usepackage{amsmath}
\usepackage{booktabs}
\usepackage{multirow}
\usepackage{color}
\usepackage{soul}
\usepackage{booktabs} % for \addlinespace

\usepackage[dvipsnames]{xcolor} %added later for color

\usepackage[breaklinks,colorlinks,citecolor=blue,urlcolor=blue]{hyperref}

\newcommand{\red}[1]{\textcolor{red}{#1}}
\newcommand{\green}[1]{\textcolor{green}{#1}}
\newcommand{\blue}[1]{\textcolor{blue}{#1}}

\defcitealias{Shen2018}{S18}

\renewcommand{\arraystretch}{1.2}  
\usepackage{listings}
\usepackage{color}
\definecolor{dkgreen}{rgb}{0,0.6,0}
\definecolor{gray}{rgb}{0.5,0.5,0.5}
\definecolor{mauve}{rgb}{0.58,0,0.82}
\definecolor{golden}{rgb}{0.86,0.65,0.01}
\usepackage{orcidlink}

\begin{document}

%\preprint{}

\title{A population of neutron star candidates in wide orbits from Gaia astrometry}

\author{\vspace{-1.0cm}Kareem El-Badry\,\orcidlink{0000-0002-6871-1752}$^{1,2}$}
\author{Hans-Walter Rix\,\orcidlink{0000-0003-4996-9069}$^{2}$}
\author{David W. Latham\,\orcidlink{0000-0001-9911-7388}$^{3}$}
\author{Sahar Shahaf\,\orcidlink{0000-0001-9298-8068}$^{4}$}
\author{Tsevi Mazeh\,\orcidlink{0000-0002-3569-3391}$^{5}$}
\author{Allyson Bieryla\,\orcidlink{0000-0001-6637-5401}$^{3}$}
\author{Lars A. Buchhave\,\orcidlink{0000-0003-1605-5666}$^{6}$}
\author{René Andrae\,\orcidlink{0000-0001-8006-6365}$^{2}$}
\author{Natsuko Yamaguchi\,\orcidlink{0000-0001-6970-1014}$^{1}$}
\author{Howard Isaacson\,\orcidlink{0000-0002-0531-1073}$^{7,8}$}
\author{Andrew W. Howard\,\orcidlink{0000-0001-8638-0320}$^{1}$}
\author{Alessandro Savino\,\orcidlink{0000-0002-1445-4877}$^{7}$}
\author{Ilya V. Ilyin\,\orcidlink{0000-0002-0551-046X}$^{9}$}

\affiliation{$^1$Department of Astronomy, California Institute of Technology, 1200 E. California Blvd., Pasadena, CA 91125, USA}
\affiliation{$^2$Max-Planck Institute for Astronomy, K\"onigstuhl 17, D-69117 Heidelberg, Germany}
\affiliation{$^3$Center for Astrophysics ${\rm \mid}$ Harvard \& Smithsonian,  60 Garden Street, Cambridge, MA 02138, USA}
\affiliation{$^4$Department of Particle Physics and Astrophysics, Weizmann Institute of Science, Rehovot 7610001, Israel}
\affiliation{$^5$School of Physics and Astronomy, Tel Aviv University, Tel Aviv, 6997801, Israel}
\affiliation{$^6$DTU Space, National Space Institute, Technical University of Denmark, Elektrovej 328, DK-2800 Kgs. Lyngby, Denmark}
\affiliation{$^{7}$Department of Astronomy, University of California Berkeley, Berkeley, CA 94720, USA}
\affiliation{$^{8}$Centre for Astrophysics, University of Southern Queensland, Toowoomba, QLD, Australia}
\affiliation{$^{9}$Leibniz-Institut für Astrophysik Potsdam (AIP), An der Sternwarte 16, D-14482 Potsdam, Germany}
\email{Corresponding author: kelbadry@caltech.edu}

\begin{abstract}
We report discovery and spectroscopic follow-up of 21 astrometric binaries containing solar-type stars and dark companions with masses near $1.4\,M_{\odot}$. The simplest interpretation is that the companions are dormant neutron stars (NSs), though ultramassive white dwarfs (WDs) and tight WD+WD binaries cannot be fully excluded.  We selected targets from {\it Gaia} DR3 astrometric binary solutions in which the luminous star is on the main sequence and the dynamically-implied mass of the unseen companion is (a) more than $1.25\,M_{\odot}$ and (b) too high to be any non-degenerate star or close binary. We obtained multi-epoch radial velocities (RVs) over a period of 700 days, spanning a majority of the orbits' dynamic range in RV. The RVs broadly validate the astrometric solutions and significantly tighten constraints on companion masses. Several systems have companion masses that are unambiguously above the Chandrasekhar limit, while the rest have masses between 1.25 and 1.4\,$M_{\odot}$. The orbits are significantly more eccentric at fixed period than those of typical WD + MS binaries, perhaps due to natal kicks. Metal-poor stars are overrepresented in the sample: three out of 21 objects (14\%) have $\rm [Fe/H] \sim -1.5$ and are on halo orbits, compared to $\sim 0.5\%$ of the parent {\it Gaia} binary sample. The metal-poor stars are all strongly enhanced in lithium. The formation history of these objects is puzzling: it is unclear both how the binaries escaped a merger or dramatic orbital shrinkage when the NS progenitors were red supergiants, and how they remained bound when the NSs formed. {\it Gaia} has now discovered 3 black holes (BHs) in astrometric binaries with masses above $9\,M_{\odot}$, and 21 NSs with masses near $1.4\,M_{\odot}$. The lack of intermediate-mass objects in this sample is striking and significant, supporting the existence of a BH/NS mass bimodality over four orders of magnitude in orbital period.

\end{abstract}

\maketitle

\section{Introduction}
\label{sec:intro}
Most stars with initial masses $\gtrsim 8\,M_{\odot}$ leave behind neutron stars (NSs) when they die. Several thousand NSs are known in the Milky Way, a large majority of which are radio pulsars.   Most ($> 99\%$; \citealt{Lorimer2008}) young pulsars are isolated. Yet, a large majority of the massive stars from which NSs form are in binaries, triples, and higher-order multiples \citep[e.g.][]{Sana2012, Moe2017}. The apparent mismatch in multiplicity properties of NSs and their progenitors hints that most massive binaries are destroyed during or prior to the formation of a NS. This destruction can come as a result of binary interaction leading to a stellar merger, or due to the binary becoming unbound during a supernova (SN), which can impart a kick of order $250\,\rm km\,s^{-1}$ on the newborn NS \citep[e.g.][]{Hobbs2005, Faucher2006}.

All known companions to young pulsars are massive OB stars \citep{Kaspi1996,Bassa2011, Shannon2014, Lyne2015, Andersen2023, vanderWateren2023}. The same is true for all detached  NS + main sequence (MS) binaries detected in X-rays \citep[e.g.][]{Reig2011}. This likely reflects the fact that binaries containing a massive star are more likely to survive mass transfer and a SN when the companion is also a massive star.

However, NSs with low-mass stellar companions do exist, and in fact make up the majority of all known binary NSs. The known systems are all currently accreting from a binary companion (``low-mass X-ray binaries''; LMXBs) or recycled, meaning that past accretion from a companion spun up the pulsar and buried its magnetic field, slowing subsequent spin-down \citep[e.g.][]{Bhattacharya1991}. LMXBs and recycled pulsars are over-represented in observed samples because they can have long lifetimes (up to and exceeding the age of the Universe) compared to normal young pulsars, which are only detectable for $\sim 10^7$ yrs. The companions to most recycled pulsars are white dwarfs (WDs), low-mass stars, and brown dwarfs that have transferred mass to the NS via stable Roche lobe overflow. Models predict that the initial masses of typical companions were at most $(1-2)\,M_{\odot}$ \citep{Podsiadlowski2002}.

Although they have not yet been unambiguously detected, there is little doubt that non-interacting binaries containing a low-mass MS star and a NS  exist: such systems are the progenitors of LMXBs and millisecond pulsars.
A few candidates for such objects have been identified, including (a) young pulsars with roughly solar-mass companions and eccentric orbits \citep[PSR B1820-11 and PSR J1954+2529][]{Phinney1991, Parent2022}, and (b) MS stars with unseen companions that may be NSs \citep[e.g.][]{Mazeh2022, Yuan2022, Zheng2022, Yi2022, Escorza2023,  Lin2023, Zhao2023}. The nature of these candidates as NS + MS binaries is quite uncertain. Among the objects in group (a), there is no doubt of the presence of a NS, but the companions -- which have not been detected electromagnetically -- may be massive WDs. Among those in group (b), the NS has not been detected, and the minimum dynamically implied mass is well below the Chandrasekhar limit. Many of the candidates in group (b) are likely to host massive WDs rather than NSs. 

 %The period distribution of LMXB progenitors is uncertain, but there are at a handful of known symbiotic X-ray binaries containing a low-mass giant and a wind-accreting neutron star \citep{Hinkle2006, Hinkle2019, Yungelson2019}. These objects are almost certainly formed from wide (separations $a\gtrsim\,1$\,au) NS+MS binaries, which are potentially detectable with astrometry. 

By precisely monitoring the astrometric light-centroid ``wobble'' of nearly two billion stars, the {\it Gaia} mission opens a new window on the Galactic binary population \citep[see][for a recent review]{ElBadry2024_gaia_binaries}. The mission's 3rd data release (``DR3'') in June 2022 included orbital solutions for about $1.7\times 10^5$ astrometric binaries, including $3\times 10^4$ joint astrometric + radial velocity (RV) solutions \citep{GaiaCollaboration2023}. Although DR3 employed stringent quality cuts \citep{Halbwachs2023} and represents only a small fraction of the binary sample that will be accessible in future data releases, the DR3 binary sample was already more than an order of magnitude larger than all samples of binary orbits in the previous literature. {\it Gaia's} particular sensitivity to long-period orbits ($P_{\rm orb}\sim (1-3)$ years in DR3)  -- and the fact that astrometric data provides constraints on binary inclinations that are not accessible with RVs alone -- has already enabled the discovery of  unexpected binary populations, including three stellar-mass black holes (BHs) in au-scale orbits \citep{El-Badry2023_bh1, El-Badry2023_bh2, Panuzzo2024} and a population of WD + MS binaries with similar orbital separations \citep{Shahaf2023, Shahaf2023b, Yamaguchi2023}. The population of wide NS + MS binaries studied in this paper is closely related to these two populations. 

Here we present results from a follow-up program of {\it Gaia} astrometric binaries suspected to contain NSs. One of our candidates, Gaia NS1 (J1432-1021), was already studied in detail by \citet{El-Badry2024}. This object has the highest inferred dark companion mass of any of the objects in our sample, and it is the only object for which the minimum companion mass from RVs alone is well above the Chandrasekhar mass, independent of astrometric constraints on the inclination.  We suspect that most of the other binaries in the sample also host NSs, but because the unseen companions have masses near the maximum WD mass, other possibilities cannot be ruled out definitively. 

The remainder of this paper is organized as follows. Section~\ref{sec:sample} describes selection of our initial sample from {\it Gaia} DR3, and Section~\ref{sec:spec_follow_up} summarizes our spectroscopic follow-up and measurement of metallicities.  We infer parameters of the luminous stars by fitting their spectral energy distributions in Section~\ref{sec:seds}. In Section~\ref{sec:thiele_innes}, we carry out joint fits of the astrometry and our follow-up RVs. In Section~\ref{sec:disc}, we discuss the nature of the dark companions, the binaries' possible formation histories, their Galactic orbits and possible abundance anomalies, and the BH/NS mass distribution. We summarize our findings in Section~\ref{sec:conclusions}. We discuss spurious astrometric solutions in Appendix~\ref{sec:spurious} and limits on possible optical contamination from WD companions in Appendix~\ref{sec:appendix_color_excess}. Tables of RVs and orbital parameters are provided in Appendices~\ref{appendix:all_fits} and ~\ref{appendix:all_rvs}. 

\section{Sample selection}
\label{sec:sample}
We selected targets from {\it Gaia} DR3 following the general approach outlined by \citet{Shahaf2019}. In brief, the astrometric ``triage'' algorithm seeks to identify sources whose astrometric orbits are so large -- given their orbital period -- that they cannot be explained by any luminous star companion, or by a companion that is a close binary containing two luminous stars. \citet{Shahaf2023} applied this algorithm to astrometric binaries published in {\it Gaia} DR3, producing a catalog of 177 candidates in which the astrometric solution and assumed luminous star mass implies the secondary must be a WD, NS, or BH. These classifications are, however, contingent on the validity of the {\it Gaia} astrometric solutions, which are in some cases spurious.  

Calculating the mass of a star's unseen companion from its astrometric solution requires an estimate of the mass of the star. In constructing their candidate sample, \citet{Shahaf2023} used the \texttt{IsocLum} mass estimates calculated by \citet{GaiaCollaboration2023}. These estimates, which are available through the {\it Gaia} archive in the \texttt{gaiadr3.binary\_masses} catalog, were inferred by comparison of the extinction-corrected colors and absolute magnitudes to a grid of PARSEC isochrones, with a prior that the metallicity is close to solar. These masses can thus be overestimated if the metallicities are subsolar, or underestimated if they are supersolar. Their validity is also contingent on the validity of the extinction estimates and on the assumption that a single star contributes to the observed photometry. We improve the mass estimates in Section~\ref{sec:seds}.  

\citet{Shahaf2023} noted that their compact object binary candidate sample appeared to fall within two populations in the mass-eccentricty plane: one with a mean mass close to $0.6\,M_{\odot}$ and eccentricities below 0.2, and another with an apparent mean mass of $\sim 1.3\,M_{\odot}$ and a broad eccentricity distribution. It would be natural, they noted, to identify these two populations with WDs and NSs. Our follow-up has shown that the division is probably not so simple: some systems in the high-mass, high-eccentricity sample unambiguously host white dwarfs, as revealed by strong UV excess \citep[e.g.][]{Ganguly2023}. Other objects have high eccentricities but companion masses below $1\,M_{\odot}$, which are implausibly low for a NS.  Despite these caveats, our follow-up has shown that most of the low-eccentricity, lower-mass compact object candidates identified by \citet{Shahaf2023} are WDs, and at least some of the higher-eccentricity, higher-mass companions are NSs. We return to this discussion in Section~\ref{sec:period_ecc}.

We initiated spectroscopic follow-up for a majority of the NS candidates identified by \citet{Shahaf2023} in June 2022. We also carried out RV follow-up observations of some astrometrically-selected NS candidates not included in the \citet{Shahaf2023} sample because their orbital periods slightly exceed 1000 days. Several of our candidates were also listed in other samples of compact object candidates from {\it Gaia} DR3, including the sample curated by \citet{Andrews2022}. The observations presented here were obtained before June 2024, but our program is ongoing.

\subsection{Rejection of spurious astrometric solutions}
For some candidates, our RV follow-up soon showed the {\it Gaia} astrometric solution to be spurious or to have significantly underestimated uncertainties. Examples are shown in Appendix~\ref{sec:spurious}. About a quarter of candidates with good astrometric quality flags turned out to be spurious.\footnote{Our follow-up suggests that the fraction of spurious solutions among {\it all} astrometric binaries is lower than this. Spurious solutions are overrepresented in regions of parameter space where genuine binaries are rare.} While the fraction of all astrometric solutions that are spurious is small, RV follow-up over a significant fraction of an orbit is critical for vetting astrometric solutions of unusual objects: our follow-up has demonstrated that incorrect solutions do exist, even among solutions with favorable \texttt{goodness\_of\_fit} and other {\it Gaia} quality flags.

\subsection{Completeness of the sample}
\label{sec:completeness}
Among the NS candidates identified by \citet{Shahaf2023} and \citet{Andrews2022}, our sample includes all systems that (a) are brighter than $G=15$, (b) have best-fit companion masses $M_2 > 1.25\,M_{\odot}$ from joint fitting of astrometry and RVs, (c) were not found to have spurious solutions or significantly underestimated astrometric uncertainties through RV follow-up, and (d) were observed over at least half an orbit. Properties of these sources are summarized in Table~\ref{tab:sample}. The \citet{Shahaf2023} and \citet{Andrews2022} samples mainly contain sources near the main sequence with inferred luminous star masses $M_\star \lesssim 1.3\,M_{\odot}$.\footnote{One candidate in the \citet{Andrews2022} sample contains an sdB star \citep{Geier2023}. Follow-up of this source will be presented in a separate publication.} While NS companions to evolved stars and more massive MS stars are likely to exist, these in most cases cannot be distinguished from luminous stars or tight luminous-star binaries based on astrometry alone.

Table~\ref{tab:all_cands} in Appendix~\ref{sec:spurious} provides a summary of our RV follow-up and our current assessment of the viability of all candidates from the \citet{Shahaf2023} and \citet{Andrews2022} samples. We defer a full description of our follow-up program -- including RVs of suspected WDs and all objects that turned out to have spurious astrometric solutions -- to future work. We suspect that our sample contains most of the NS-hosting binaries with $M_2 \gtrsim 1.25\,M_{\odot}$ and astrometric solutions published in DR3.  A handful of likely good candidates are not included because our RV follow-up has not yet covered enough of the orbits to confirm the astrometric solution with high confidence, and another handful were excluded because they are too faint ($G>15$) to be amenable for RV follow-up with the instruments at our disposal. Our sample does not contain any low-mass NSs with $M< 1.25\,M_{\odot}$, a mass limit below which some NSs likely do exist \citep{Ferdman2014, Martinez2015}. As we discuss throughout the paper, it becomes increasingly challenging to distinguish between NSs and massive WDs at lower masses.

\subsection{Summary of the sample}
\label{sec:sample_summary}

\begin{figure*}
    \centering
    \includegraphics[width=\textwidth]{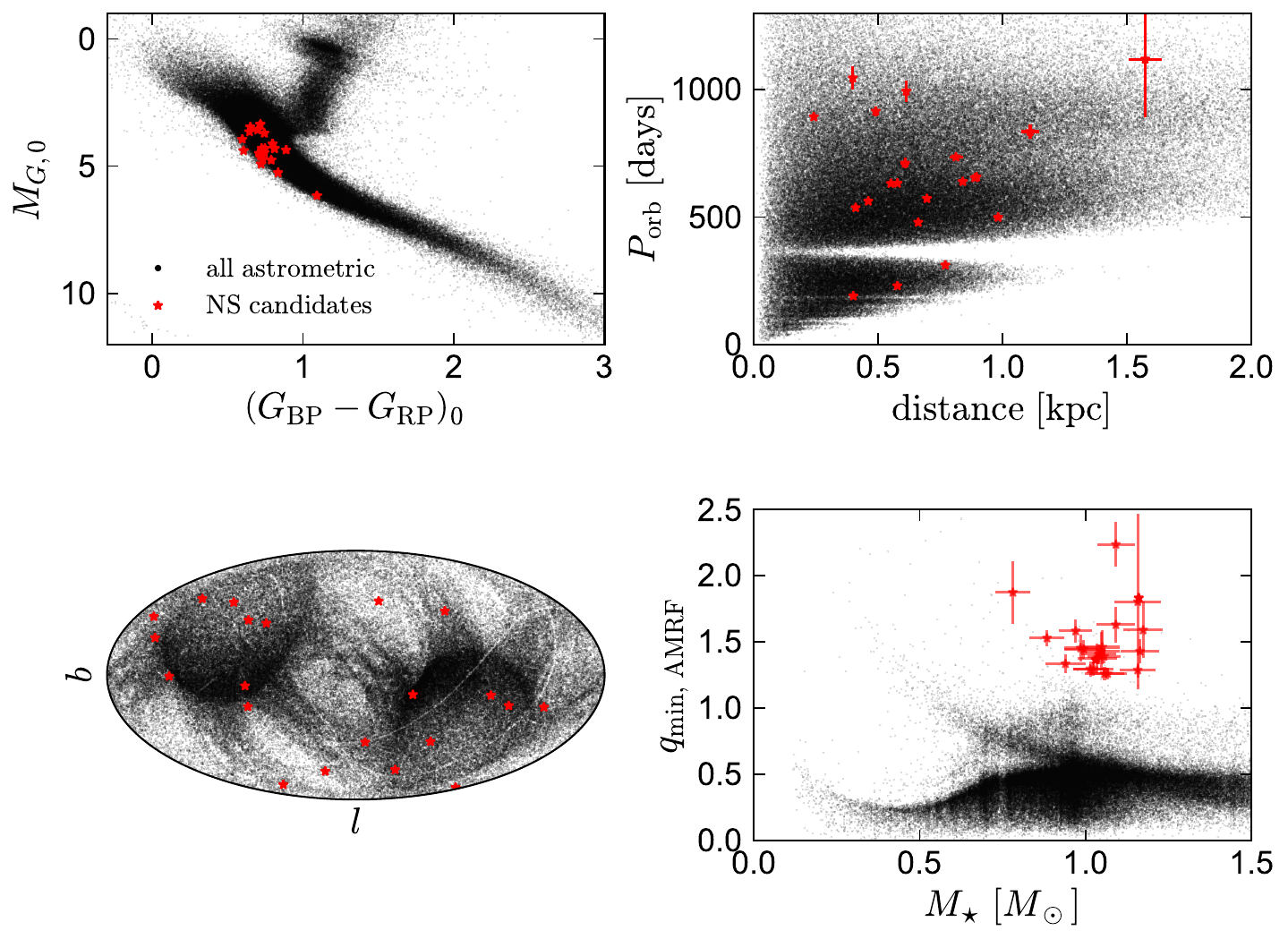}
    \caption{Black points show all binaries with astrometric orbital solutions published in {\it Gaia} DR3. Red points show the 21 objects presented in this work, which have astrometrically-inferred companion masses $M_2 > 1.25\,M_{\odot}$. {\it Upper left}: dereddened color-magnitude diagram. Most candidates are solar-type stars near the main-sequence. {\it Upper right}: orbital period and distance. Candidates have orbital periods of $\sim (100-1000)$ d and are within $\sim1$\,kpc of the Sun. {\it Lower left}: Galactic coordinates, with the Galactic center in the middle of the plot. Imprints of the {\it Gaia} scanning law are visible; most NS candidates are at high latitude. {\it Bottom right}: luminous star mass and minimum mass ratio, $M_2/M_\star$. The NS candidates have some of the highest estimated mass ratios in the astrometric binary sample.  }
    \label{fig:cmd}
\end{figure*}

\begin{table*}
\begin{tabular}{lllllllll}
Name & $P_{\rm orb}$ & $M_{\star}$ & $M_2$ & eccentricity & $\varpi$ & $G$ & $N_{\rm RVs}$ & {\it Gaia} DR3 ID \\
 & [days] & [$M_{\odot}$] & [$M_{\odot}$] &  & [mas] & [mag] &  \\
\hline
J0553-1349 & $189.10 \pm 0.05$ & $0.98 \pm 0.06$ & $1.33 \pm 0.05$  & $0.3879 \pm 0.0007$ & $2.505 \pm 0.015$ & 13.00 & 20 & 2995961897685517312 \\
J2057-4742 & $230.15 \pm 0.07$ & $1.048 \pm 0.031$ & $1.31 \pm 0.04$  & $0.3095 \pm 0.0026$ & $1.745 \pm 0.019$ & 13.58 & 11 & 6481502062263141504 \\
J1553-6846 & $310.17 \pm 0.11$ & $1.04 \pm 0.05$ & $1.323 \pm 0.032$  & $0.5314 \pm 0.0021$ & $1.344 \pm 0.012$ & 14.19 & 16 & 5820382041374661888 \\
J2102+3703 & $481.04 \pm 0.26$ & $1.03 \pm 0.03$ & $1.473 \pm 0.034$  & $0.448 \pm 0.009$ & $1.521 \pm 0.013$ & 13.70 & 10 & 1871419337958702720 \\
J0742-4749 & $497.6 \pm 0.4$ & $0.90 \pm 0.05$ & $1.28 \pm 0.04$  & $0.168 \pm 0.004$ & $1.035 \pm 0.014$ & 14.60 & 8 & 5530442371304582912 \\
J0152-2049 & $536.14 \pm 0.18$ & $0.782 \pm 0.03$ & $1.291 \pm 0.024$  & $0.6615 \pm 0.0010$ & $2.453 \pm 0.017$ & 12.05 & 15 & 5136025521527939072 \\
J0003-5604 & $561.83 \pm 0.29$ & $0.802 \pm 0.03$ & $1.34 \pm 0.04$  & $0.795 \pm 0.005$ & $2.183 \pm 0.016$ & 14.48 & 12 & 4922744974687373440 \\
J1733+5808 & $570.94 \pm 0.31$ & $1.16 \pm 0.05$ & $1.362 \pm 0.030$  & $0.3093 \pm 0.0010$ & $1.452 \pm 0.010$ & 13.65 & 13 & 1434445448240677376 \\
J1150-2203 & $631.81 \pm 0.22$ & $1.18 \pm 0.06$ & $1.39 \pm 0.04$  & $0.552 \pm 0.004$ & $1.738 \pm 0.016$ & 12.66 & 20 & 3494029910469026432 \\
J1449+6919 & $632.65 \pm 0.21$ & $0.91 \pm 0.05$ & $1.258 \pm 0.032$  & $0.2668 \pm 0.0010$ & $1.812 \pm 0.010$ & 13.20 & 19 & 1694708646628402048 \\
J0217-7541 & $636.1 \pm 0.7$ & $0.996 \pm 0.033$ & $1.396 \pm 0.033$  & $0.3228 \pm 0.0033$ & $1.193 \pm 0.012$ & 14.01 & 10 & 4637171465304969216 \\
J0639-3655 & $654.6 \pm 0.6$ & $1.32 \pm 0.06$ & $1.70 \pm 0.07$  & $0.721 \pm 0.013$ & $1.130 \pm 0.011$ & 13.36 & 10 & 5580526947012630912 \\
J1739+4502 & $657.4 \pm 0.6$ & $0.781 \pm 0.03$ & $1.38 \pm 0.04$  & $0.6777 \pm 0.0018$ & $1.126 \pm 0.013$ & 13.52 & 18 & 1350295047363872512 \\
J0036-0932 & $719.8 \pm 0.9$ & $0.94 \pm 0.04$ & $1.362 \pm 0.034$  & $0.3993 \pm 0.0021$ & $1.661 \pm 0.019$ & 13.02 & 16 & 2426116249713980416 \\
J1432-1021 & $730.9 \pm 0.5$ & $0.79 \pm 0.03$ & $1.898 \pm 0.030$  & $0.1203 \pm 0.0022$ & $1.367 \pm 0.011$ & 13.34 & 34 & 6328149636482597888 \\
J1048+6547 & $827 \pm 5$ & $0.99 \pm 0.05$ & $1.52 \pm 0.07$  & $0.357 \pm 0.009$ & $0.916 \pm 0.016$ & 14.52 & 9 & 1058875159778407808 \\
J2145+2837 & $889.5 \pm 0.7$ & $0.95 \pm 0.05$ & $1.396 \pm 0.035$  & $0.5840 \pm 0.0035$ & $4.137 \pm 0.016$ & 12.19 & 11 & 1801110822095134848 \\
J2244-2236 & $938.3 \pm 0.5$ & $1.002 \pm 0.03$ & $1.443 \pm 0.023$  & $0.5666 \pm 0.0011$ & $2.079 \pm 0.019$ & 13.35 & 13 & 2397135910639986304 \\
J0824+5254 & $1026.7 \pm 3.3$ & $1.102 \pm 0.03$ & $1.604 \pm 0.034$  & $0.686 \pm 0.012$ & $1.643 \pm 0.015$ & 13.59 & 13 & 1028887114002082432 \\
J0230+5950 & $1029 \pm 5$ & $1.114 \pm 0.03$ & $1.401 \pm 0.034$  & $0.753 \pm 0.011$ & $2.523 \pm 0.015$ & 13.09 & 15 & 465093354131112960 \\
J0634+6256 & $1046.0 \pm 2.1$ & $1.18 \pm 0.06$ & $1.48 \pm 0.09$  & $0.564 \pm 0.011$ & $0.689 \pm 0.019$ & 14.62 & 10 & 1007185297091149824 \\
\end{tabular}
\caption{Basic properties of the sample. $M_{\star}$ is the inferred mass of the luminous star from fitting the SED and spectroscopic metallicity. $P_{\rm orb}$ and $M_2$ are the orbital period and mass of the unseen companion. These quantities, as well as the eccentricity and parallax, $\varpi$, are from joint fits of the {\it Gaia} astrometry and our RV follow-up. $G$ is the apparent magnitude measured by {\it Gaia}, and $N_{\rm RVs}$ is the number of follow-up RVs we have measured. The source J1432-1021 was already studied by \citet{El-Badry2024} under the name Gaia NS1. }
\label{tab:sample}
\end{table*}

Basic properties of our NS candidates are listed in Tables~\ref{tab:sample} and~\ref{tab:allfits}.
Figure~\ref{fig:cmd} compares our candidates to the full sample of astrometric binaries (solution types \texttt{Orbital} and \texttt{AstroSpectroSB1}) published in DR3. Three candidates have \texttt{AstroSpectroSB1} solutions: J0152-2049, J2145+2837, and J1150-2203; the rest have \texttt{Orbital}  solutions. The upper left panel of Figure~\ref{fig:cmd} shows the sources on the extinction-corrected color-magnitude diagram. We estimate the extinction for sources in the north ($\delta > -30$\,deg) using the 3D dust map from \citet{Green2019}; we use the map from \citet{Lallement2022} for sources farther south. All our targets are solar-type stars on the main sequence, with absolute magnitudes and colors suggesting luminous star masses of $(0.7-1.3)\,M_{\odot}$. 
Several candidates are near the blue edge of the main sequence. A potential concern is that this could be due to blue excess from a hot WD companion. However, our analysis of the sources' full spectral energy distributions --  in particular, the lack of UV excess -- speaks against this possibility (Appendix~\ref{sec:appendix_color_excess}).

The upper right panel of Figure~\ref{fig:cmd} shows orbital periods and distances. Most of the binaries in our sample are within 1 kpc of the Sun and have periods between 100 and 1000 days. There are no binaries with orbital periods close to 1 year owing to the degeneracy between such orbits and parallactic motion. The period distribution peaks near 600 days, likely because short-period orbits are smaller and can only be resolved at close distances, while significantly longer orbits would not have been well-sampled during the $\sim$1000-day observing window for DR3 solutions. The period distribution of the NS candidate sample is fairly similar to that of all astrometric binaries. 

The lower left panel of Figure~\ref{fig:cmd} shows the sources' distribution on the sky in Galactic coordinates, with the Galactic center at the center. Both our candidates and the full astrometric binary sample are distributed all across the sky. Some evidence of the Galactic disk is evident in the distribution of all binaries, but the distribution is heavily affected by the {\it Gaia} scanning law, and most of the NS candidates are at high latitude. 

Finally, the lower right panel of Figure~\ref{fig:cmd} shows the masses of the luminous stars in our sample and the minimum mass ratio inferred from their astrometric mass ratio function \citep[AMRF;][]{Shahaf2019}. The luminous star masses plotted here are taken from the \texttt{gaiadr3.binary\_masses} table following \citet{Shahaf2023}; more accurate masses for our candidates are measured in Section~\ref{sec:seds}. The astrometric mass ratio functions are also calculated based on {\it Gaia} data alone following \citet{Shahaf2023}, without accounting for our follow-up RVs. By virtue of our selection, the objects in our sample have among the largest minimum mass ratios of binaries with solutions published in DR3.

\begin{figure}
    \centering
    \includegraphics[width=\columnwidth]{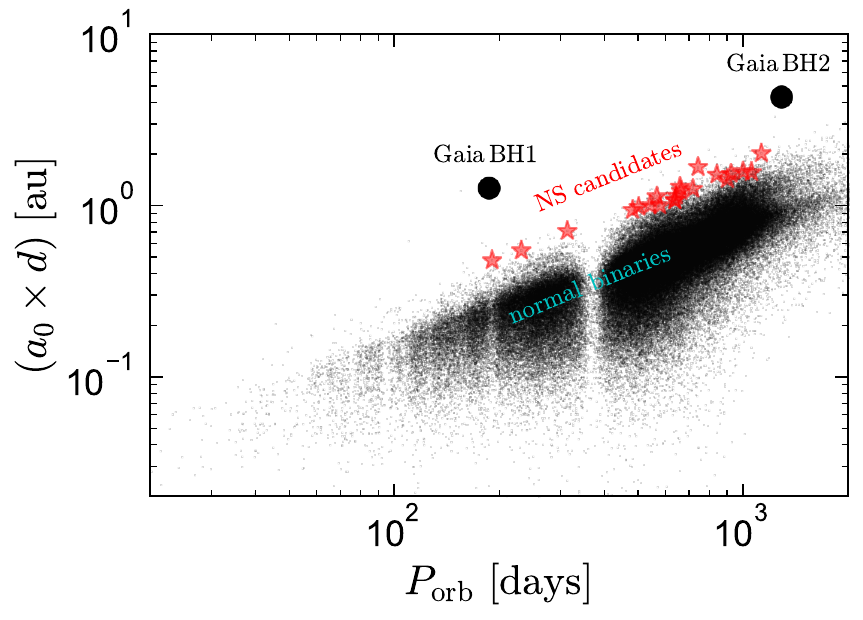}
    \caption{Comparison of NS candidates presented here (red stars) to the rest of the {\it Gaia} DR3 astrometric binary sample. At fixed period, dark and massive companions produce larger photocenter orbits than normal stellar companions. For typical solar-type primaries, NS companions produce photocenter orbits that are smaller than those of BH binaries, but larger than those of luminous binaries and triples. }
    \label{fig:keplers_law}
\end{figure}

That the unseen companions to objects in our sample are massive can be appreciated intuitively from Figure~\ref{fig:keplers_law}, which compares their periods and the physical size of their photocenter orbits to other binaries with astrometric solutions in DR3, including Gaia BH1 and BH2. According to Kepler's 3rd law, more massive and darker companions are found above and to the left of the population of luminous binaries and triples in this parameter space \citep[see][]{El-Badry2023_bh1}. NS companions are expected to be found between the populations of luminous binaries and BH companions, and this is indeed where our candidates are clustered. Close inspection of Figure~\ref{fig:keplers_law} will reveal that a few additional binaries fall above the candidates studied here, suggestive of higher masses. These are primarily sources where our RV follow-up showed the astrometric solution to be spurious, and red giants, for which a more massive MS companion could not be ruled out.

\section{Spectroscopic follow-up}
\label{sec:spec_follow_up}
We measured multi-epoch RVs for all targets, primarily using high-resolution spectrographs on two 2m-class telescopes. Our targets are bright, with most having apparent magnitudes $G=13-14$. We additionally obtained single-epoch higher-SNR spectra for most targets with 8-10m class telescopes. We describe all the spectroscopic observations below.

\subsection{FEROS}
\label{sec:feros}
We obtained 129 spectra with the Fiberfed Extended Range Optical Spectrograph \citep[FEROS;][]{Kaufer1999} on the 2.2m ESO/MPG telescope at La Silla Observatory (programs P109.A-9001, P110.A-9014, P111.A-9003, P112.A-6010, and P113.26XB). Some observations used $2\times 2$ binning to reduce readout noise at the expense of spectral resolution; the remainder used $1\times 1$ binning. The resulting spectra have resolution $R\approx 40,000$ ($2\times 2$ binning) and $R\approx 50,000$ ($1\times 1$ binning). Exposure times ranged from 1200 to 3600 seconds, depending on source brightness.  We reduced the data using the CERES pipeline \citep{Brahm2017}, which performs bias-subtraction, flat fielding, wavelength calibration, and optimal extraction. The pipeline measures and corrects for small shifts in the wavelength solution during the course a night via simultaneous observations of a ThAr lamp with a second fiber.

\subsection{TRES}
\label{sec:tres}
We obtained 155 spectra using the Tillinghast Reflector Echelle Spectrograph \citep[TRES;][]{Furesz2008} mounted on the 1.5 m Tillinghast Reflector telescope at the Fred Lawrence Whipple Observatory (FLWO) on Mount Hopkins, Arizona. TRES is a fibrefed echelle spectrograph with a wavelength range of 390–910 nm and a resolving power of $R\sim 44,000$. The spectra were extracted as described in \citet{Buchhave2010}.

\subsection{MIKE}
\label{sec:MIKE}

We obtained 7 spectra with the Magellan Inamori Kyocera Echelle (MIKE) spectrograph on the Magellan Clay telescope at Las Campanas Observatory \citep{Bernstein2003SPIE}. We used the 0.5'' slit and exposure times ranging from 600 to 1200 seconds, yielding spectral resolution $R \sim 40,000$ on the blue side and $R \sim 55,000$ on the red side. The typical SNR at 5800\,\AA\, was $\sim 30$ and the total wavelength coverage was $\sim 3330 - 9680$ \AA.
We reduced the spectra with the MIKE Pipeline within \texttt{CarPy} \citep{Kelson2000ApJ, Kelson2003PASP} and subsequently flux-calibrated them using observations of a standard star. We merged the orders into a single spectrum, weighting by inverse variance in the overlap regions. 

\subsection{HIRES}
\label{sec:HIRES}

We obtained 6 spectra using the High Resolution Echelle Spectrometer \citep[HIRES;][]{Vogt1994} on the 10m Keck I telescope on Maunakea. The data were obtained and reduced using the standard California Planet Survey setup \citep[CPS;][]{Howard2010}, including use of the C2 decker (0.86 arcseconds $\times$ 14 arcseconds), which yields spectra with $R\approx 55,000$ and wavelength coverage over most of 3700--8000\,\AA. We used 600 second exposures, yielding a typical SNR of  40 per pixel at 6000\,\AA. The CPS reduction includes sky-subtraction using the long C2 decker. We merged spectra from individual orders using the same procedure as with the MIKE data. 

\subsection{PEPSI}
\label{sec:PEPSI}

We obtained 10 spectra using the Potsdam Echelle Polarimetric and Spectroscopic Instrument (PEPSI; \citealt{Strassmeier2015}) spectrograph on the Large Binocular Telescope in binocular mode. We used the 300\,$\mu$m fiber, which has a diameter of 2.3 arcsec on sky, and the CD2 and CD5 cross-dispersers on the blue and red side, respectively. Exposure times ranged from 300 to 1200s.  The spectra were reduced and orders were merged as described in \citet{Strassmeier2018}; they cover the wavelength ranges of 4222-4792\,\AA\,\,and 6236-7433\,\AA\,\,with spectral resolution $R\approx 50,000$.

\subsection{ESI}
\label{sec:ESI}
We obtained 1 spectrum with the Echellette Spectrograph and Imager \citep[ESI;][]{Sheinis2002} on the 10m Keck II telescope on Maunakea. We used a 300 second exposure with the 0.3 arcsec slit, yielding a resolution $R\approx 12000$ and $\rm SNR \approx 50$, with useful wavelength coverage of 3900-10000\,\AA. We reduced the data using the MAuna Kea Echelle Extraction (MAKEE) pipeline, which performs bias-subtraction, flat fielding, wavelength calibration, and sky subtraction, and we refined the wavelength solution using telluric absorption lines.

\subsection{Other spectroscopy}
\label{sec:other}
Early in our follow-up program, we measured RVs for several candidates using spectra from lower-resolution spectrographs, including Keck/DEIMOS \citep{Faber2003},  Magellan/MagE \citep{Marshall2008}, and Palomar P200/DBSP \citep{Oke1982}. These RVs were used to rule out some candidates with spurious solutions but were not included in our final analysis due to their larger uncertainties ($\gtrsim 3\,\rm km\,s^{-1}$). A few objects in our sample also have archival spectra from the LAMOST survey \citep{Cui2012}. However, these typically have RV uncertainties of at least a few $\rm km\,s^{-1}$ -- about 100 times larger than our observations -- so we opted not to include them in our analysis. 

\subsection{RVs and offsets}
\label{sec:offsets}

We measure RVs for each echelle order by cross-correlating a synthetic spectral template with the normalized spectrum. We report the median across orders as the measured RV for each epoch and calculate the uncertainty as the standard deviation across orders divided by the square root of the number of orders.  We use a Kurucz spectral template from the \texttt{BOSZ} library \citep{Bohlin2017} with surface gravity $\log(g/{\rm cm\,s^{-2}}) = 4.0$ and effective temperature and metallicity matched to each target (Section~\ref{sec:metallicity}). We use non-rotating templates for all targets except J0230+5950, which is rotating with $v \sin i \approx 15\,\rm km\,s^{-1}$.

For the FEROS spectra, we use 15 orders covering wavelengths between 450 and 670 nm when calculating RVs. For the TRES spectra, we use 31 orders covering 420-670 nm. The median uncertainty of the FEROS RVs is $\approx 0.06$ km s$^{-1}$, while the median uncertainty of the TRES RVs is $\approx 0.05$ km s$^{-1}$.

As described by \citet{El-Badry2024}, we fit for a single global RV offset between TRES and FEROS. We found an offset of $0.16\,\rm km\,s^{-1}$, in the sense that ${\rm RV}_{{\rm FEROS}}={\rm RV}_{{\rm TRES}}-0.16\,{\rm km\,s^{-1}}$. This offset has already been applied to our reported RVs.  Because we generally only obtained one spectrum per target for MIKE, HIRES, PEPSI, and ESI, we did not fit for offsets for these instruments, but instead adopted a conservative 0.5\,$\rm km\,s^{-1}$ minimum uncertainty for all the RVs measured with these instruments. All the RVs are listed in Appendix~\ref{appendix:all_rvs} and provided in machine-readable form as supplementary material.

\subsection{Metallicities}
\label{sec:metallicity}
The radius and temperature of a MS star of a given mass depend on its metallicity, so metallicity estimates are important for reliable mass estimates of the luminous stars, and in turn, mass estimates of their companions. We collate metallicity measurements from several sources:

\subsubsection{BACCHUS}
For the MIKE, HIRES, and FEROS spectra, we measured metallicities and atmospheric parameters using the Brussels Automatic Code
for Characterizing High accUracy Spectra \citep[BACCHUS;][]{Masseron2016ascl, Hayes2022ApJS}. The code performs 1D LTE spectral synthesis to determine stellar parameters from Fe excitation/ionization balance; i.e., the requirement that lines with different excitation potentials all imply the same Fe abundance. The reported metallicity [Fe/H] is the mean Fe abundance calculated over lines in the VALD atomic linelist with a wavelength coverage of 4200 to 6700\,\AA. Here we assume that the detailed abundance pattern traces solar values; detailed abundances of these stars will be investigated in future work. The errors reported by BACCHUS represent the scatter in the implied abundances between the different lines and methods of abundance calculations but do not take into account other systematic uncertainties.

\subsubsection{SPC}
We fit the TRES spectra using the Stellar Parameter Classification (SPC) tool \citep{Buchhave2012}, which cross-correlates the normalized TRES spectra with a grid of synthetic spectra in the wavelength range of 5050 to 5360\,\AA, centered on the  Mg I b triplet. SPC infers the metallicity [M/H], effective temperature, $T_{\rm eff}$, and surface gravity $\log g$ by fitting the peaks of the cross-correlation functions with a three-dimensional polynomial in stellar parameters. Given systematic uncertainties in the synthetic stellar spectra, error floors on the derived [M/H] and $T_{\rm eff}$ values are $\sim$ 0.08 dex and $\sim$ 50 K, respectively \citep[][]{Buchhave2012, Furlan2018}.

\subsubsection{Gaia XP metallicities}
\label{sec:mets}
All of the stars in our sample have metallicity estimates (along with $T_{\rm eff}$ and $\log g$) calculated as described by  \citet{Andrae2023ApJS} using {\it Gaia} XP low-resolution spectra. For bright stars within the temperature range of our sample, the expected precision of these metallicities is $\sim 0.1$ dex. However, the metallicities published by \citet{Andrae2023ApJS} are expected to be less reliable for astrometric binaries -- even those with dark companions -- because they are calculated based on the \texttt{gaiadr3.gaia\_source} parallax, which comes from a 5-parameter astrometric solution that neglects orbital motion. These parallaxes can be significantly biased for astrometric binaries with large photocenter wobbles. We therefore recalculate metallicities for our targets using the parallaxes in the \texttt{gaiadr3.nss\_two\_body\_orbit} catalog, otherwise using the same empirically trained machine-learning models  as \citet{Andrae2023ApJS}.

\begin{table*}
    \makebox[\textwidth][c]{
        \begin{tabular}{c|ccc|ccccc}
            \hline
              & \multicolumn{3}{c}{Metallicity} & \multicolumn{5}{c}{SED fit} \\
            \cmidrule(lr){2-4}\cmidrule(lr){5-9}
            Name & [Fe/H], High-res & Instrument &[M/H], Gaia XP & $E(B-V)$  & $R \,[R_{\odot}]$ & $T_{\rm eff}$\,[kK]  & $M\,[M_{\odot}]$ & $[\rm Fe/H]_{\rm init}$ \\ \hline
J0553-1349 &  $0.12 \pm 0.07$ & MIKE & $0.04 \pm 0.03$ & $0.06 \pm 0.02$ & $0.988 \pm 0.015$  & $5.59 \pm 0.07$  & $0.98 \pm 0.06$ & $0.14 \pm 0.04$ \\ 
J2057-4742 &  $0.13 \pm 0.08$ & MIKE & $0.22 \pm 0.06$ & $0.01 \pm 0.01$ & $0.941 \pm 0.010$  & $5.834 \pm 0.027$  & $1.05 \pm 0.03$ & $0.096 \pm 0.030$ \\ 
J1553-6846 &  $0.13 \pm 0.10$ & MIKE & $0.16 \pm 0.06$ & $0.05 \pm 0.02$ & $0.971 \pm 0.013$  & $5.88 \pm 0.05$  & $1.04 \pm 0.05$ & $0.141 \pm 0.034$ \\ 
J2102+3703 &  $-0.37 \pm 0.08$ & HIRES & $-0.30 \pm 0.05$ & $0.06 \pm 0.01$ & $0.947 \pm 0.010$  & $6.27 \pm 0.06$  & $1.035 \pm 0.03$ & $-0.31 \pm 0.04$ \\ 
J0742-4749 &  $-0.13 \pm 0.12$ & MIKE & $-0.11 \pm 0.07$ & $0.16 \pm 0.02$ & $1.259 \pm 0.027$  & $5.68 \pm 0.07$  & $0.90 \pm 0.05$ & $0.10 \pm 0.05$ \\ 
J0152-2049 &  $-1.28 \pm 0.08$ & TRES & $-1.31 \pm 0.17$ & $0.04 \pm 0.00$ & $1.169 \pm 0.013$  & $6.47 \pm 0.04$  & $0.780 \pm 0.03$ & $-1.243 \pm 0.015$ \\ 
J0003-5604 &  $-0.09 \pm 0.08$ & FEROS & $-0.03 \pm 0.05$ & $0.00 \pm 0.00$ & $0.752 \pm 0.006$  & $4.885 \pm 0.013$  & $0.794 \pm 0.03$ & $-0.121 \pm 0.024$ \\ 
J1733+5808 &  $0.17 \pm 0.10$ & TRES & $0.15 \pm 0.09$ & $0.05 \pm 0.02$ & $1.081 \pm 0.019$  & $6.01 \pm 0.08$  & $1.17 \pm 0.05$ & $0.23 \pm 0.04$ \\ 
J1150-2203 &  $0.42 \pm 0.07$ & MIKE & $0.39 \pm 0.05$ & $0.05 \pm 0.01$ & $1.495 \pm 0.019$  & $5.834 \pm 0.019$  & $1.18 \pm 0.06$ & $0.297 \pm 0.004$ \\ 
J1449+6919 &  $-0.65 \pm 0.09$ & HIRES & $-0.47 \pm 0.06$ & $0.05 \pm 0.02$ & $1.060 \pm 0.024$  & $6.03 \pm 0.14$  & $0.91 \pm 0.05$ & $-0.33 \pm 0.05$ \\ 
J0217-7541 &  $0.14 \pm 0.10$ & FEROS & $0.15 \pm 0.07$ & $0.02 \pm 0.01$ & $1.238 \pm 0.018$  & $5.58 \pm 0.05$  & $0.99 \pm 0.03$ & $0.272 \pm 0.025$ \\ 
J0639-3655 &  $0.04 \pm 0.10$ & MIKE & $0.09 \pm 0.04$ & $0.09 \pm 0.03$ & $1.43 \pm 0.04$  & $6.48 \pm 0.21$  & $1.32 \pm 0.06$ & $0.17 \pm 0.05$ \\ 
J1739+4502 &  $-1.82 \pm 0.10$ & MIKE & $-1.46 \pm 0.23$ & $0.06 \pm 0.00$ & $1.424 \pm 0.016$  & $6.289 \pm 0.032$  & $0.78 \pm 0.03$ & $-1.58 \pm 0.04$ \\ 
J0036-0932 &  $-0.42 \pm 0.10$ & HIRES & $-0.25 \pm 0.05$ & $0.03 \pm 0.02$ & $1.32 \pm 0.04$  & $5.84 \pm 0.09$  & $0.94 \pm 0.04$ & $-0.10 \pm 0.05$ \\ 
J1432-1021 &  $-1.29 \pm 0.11$ & MIKE & $-1.48 \pm 0.26$ & $0.10 \pm 0.00$ & $1.491 \pm 0.029$  & $6.049 \pm 0.023$  & $0.79 \pm 0.03$ & $-1.239 \pm 0.014$ \\ 
J1048+6547 &  $-0.14 \pm 0.11$ & HIRES & $-0.19 \pm 0.08$ & $0.02 \pm 0.02$ & $1.15 \pm 0.04$  & $5.96 \pm 0.10$  & $1.00 \pm 0.05$ & $-0.03 \pm 0.06$ \\ 
J2145+2837 &  $-0.03 \pm 0.06$ & HIRES & $0.21 \pm 0.04$ & $0.01 \pm 0.01$ & $0.846 \pm 0.005$  & $5.500 \pm 0.023$  & $0.948 \pm 0.03$ & $0.022 \pm 0.023$ \\ 
J2244-2236 &  $0.06 \pm 0.08$ & TRES & $0.03 \pm 0.07$ & $0.00 \pm 0.01$ & $0.881 \pm 0.009$  & $5.892 \pm 0.026$  & $0.990 \pm 0.03$ & $-0.127 \pm 0.034$ \\ 
J0824+5254 &  $0.27 \pm 0.10$ & TRES & $0.13 \pm 0.07$ & $0.04 \pm 0.02$ & $0.997 \pm 0.013$  & $5.89 \pm 0.05$  & $1.104 \pm 0.03$ & $0.19 \pm 0.04$ \\ 
J0230+5950 &  $0.15 \pm 0.08$ & HIRES & $0.11 \pm 0.06$ & $0.21 \pm 0.01$ & $1.001 \pm 0.008$  & $5.87 \pm 0.04$  & $1.114 \pm 0.03$ & $0.230 \pm 0.034$ \\ 
J0634+6256 &  $-0.03 \pm 0.11$ & TRES & $-0.23 \pm 0.11$ & $0.12 \pm 0.02$ & $1.68 \pm 0.07$  & $6.05 \pm 0.12$  & $1.17 \pm 0.06$ & $0.08 \pm 0.06$ \\ 
\hline
        \end{tabular}
    }
    \caption{Parameters of the luminous stars. We list both metallicities measured from high-resolution spectra and those measured from low-resolution {\it Gaia} XP spectra following \citet{Andrae2023ApJS}. The high-resolution metallicities are used as a prior when fitting the broadband SEDs. The metallicities are compared in Figure~\ref{fig:metallicities}, and the SED fits are shown in Figure~\ref{fig:seds}. }
    \label{tab:star_params}%
\end{table*}%

\subsubsection{Metallicity comparisons}
\begin{figure}
    \centering
    \includegraphics[width=\columnwidth]{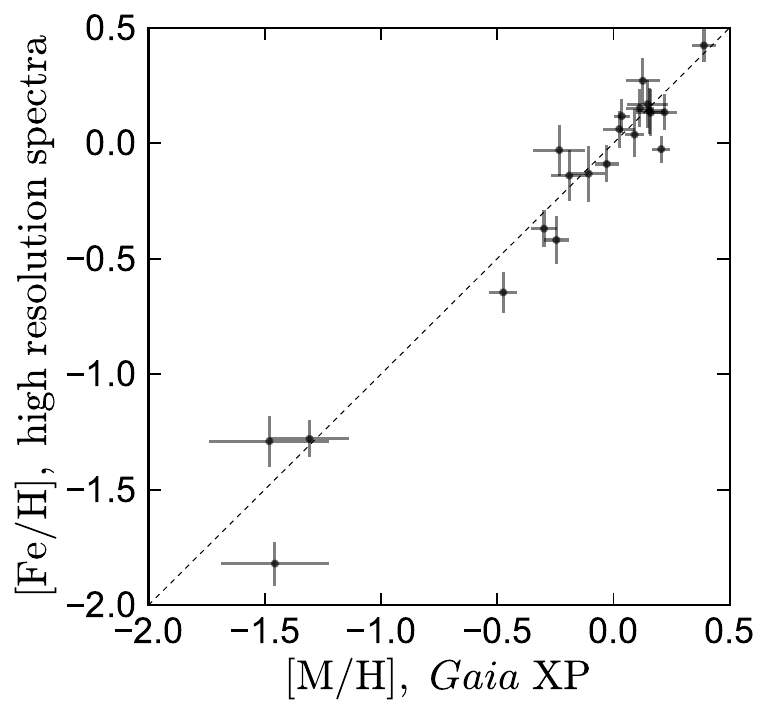}
    \caption{Metallicities measured from high-resolution spectra (y-axis), and from low-resolution {\it Gaia} XP spectra (x-axis). The XP metallicities are inferred with a modified version of the \citet{Andrae2023ApJS} model that uses the parallax from the astrometric binary solution rather than the one from the 5-parameter solution reported in the \texttt{gaia\_source} catalog. The two independent metallicity measurements are in good agreement.}
    \label{fig:metallicities}
\end{figure}

 For each source, we report both the {\it Gaia} XP metallicity and a metallicity measured from a high-resolution spectrum in Table~\ref{tab:star_params}. Several sources were observed with more than one high-resolution spectrograph. In these cases, we adopt the metallicity from the highest-SNR spectrum. The results are reported in Table~\ref{tab:star_params} and shown in Figure~\ref{fig:metallicities}. The agreement between the {\it Gaia} XP and high-resolution metallicities is good, implying that the XP metallicities are robust for this sample and that high-resolution spectra will not be essential for bulk metallicity measurements in future follow-up efforts, though they are required for RV measurements. 

We also investigated the consistency of metallicities measured from high-resolution spectra by different pipelines. For sources with more than one high resolution spectrum, we found a median absolute deviation of 0.08 dex between measurements, which is comparable to the reported uncertainties.

\section{Spectral energy distributions and luminous star masses}
\label{sec:seds}
We inferred the masses and evolutionary states of the luminous stars by fitting their broadband spectral energy distributions (SEDs) with single-star models. In the optical, we use the synthetic photometry in the SDSS $ugriz$ bands constructed from {\it Gaia} BP/RP spectra and empirically calibrated as described by \citet{GaiaCollaboration2023phot}. We supplement this with near-infrared photometry from the  2MASS \citep{Skrutskie2006} and WISE \citep{Wright2010} surveys. To search for possible light contributions from hot WD companions, we also retrieved UV photometry from GALEX \citep{Martin2005}, but we did not include it in our SED fits. 

%Because the UV photometry could potentially be contaminated by a hot WD companion, we do not include it in our SED fits. 

\begin{figure*}
    \centering
    \includegraphics[width=\textwidth]{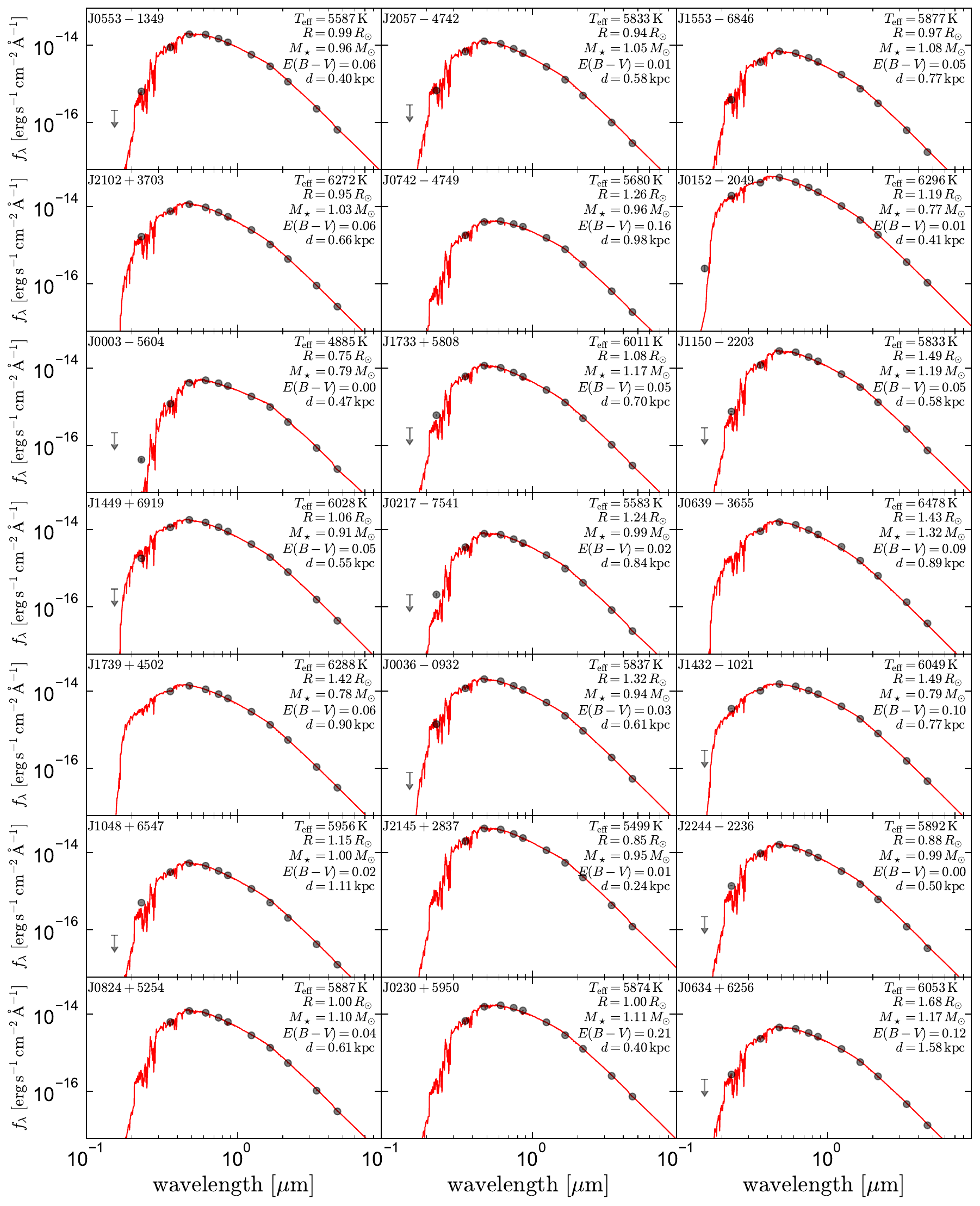}
    \caption{Spectral energy distributions and best-fit single-star models. The parameters of the best-fit model are listed in each panel. UV photometry is plotted where available, but not included in the fit. All targets' SEDs are reasonably well-fit by a single-star model. Parameters from SED fitting are listed in Table~\ref{tab:star_params}. }
    \label{fig:seds}
\end{figure*}

The SEDs of all sources are shown in Figure~\ref{fig:seds}. While optical and near-infrared photometry is available for all sources, several sources are outside the published footprints of GALEX in one or both of its UV bands. In cases where a source was within the footprint of a GALEX observation but was was not detected, we plot the $3\sigma$ upper limit. Sources with no upper limits or detection shown in the NUV or FUV were not observed in that band. 

We fit the SEDs using \texttt{MINESweeper} \citep{Cargile2020}, a code designed for joint modeling of stellar photometry and spectra. We only use the photometric modeling capabilities of the code but place a prior on the present-day metallicity from spectroscopy. The free parameters of the fit are the initial mass and metallicity of each star, its age (as parameterized by the ``equivalent evolutionary point''), the parallax, and the foreground extinction. For each call to the likelihood function, the mass, metallicity, and equivalent evolutionary point are converted to a radius and effective temperature using MIST isochrones \citep{Choi2016}, and are then used to predict a model SED that is compared to the data. We place priors on the parallax from the {\it Gaia} 12-parameter binary solution and set an age upper limit of 13 Gyr. We also use a metallicity prior from the high-resolution spectra (Table~\ref{tab:star_params}), and an extinction prior from 3D dust maps. We again use the \citet{Green2019} map for sources with declination $\delta > -30$ deg, and the \citet{Lallement2022} map for sources farther south. 

The SEDs and best-fit models of all candidates are shown in Figure~\ref{fig:seds}, and best-fit parameters for each target are listed in the upper right corner of each panel. Most of the luminous stars have masses near $1\,M_{\odot}$. This is also true for the full astrometric binary catalog published in {\it Gaia} DR3 and mostly reflects the distance and flux limits of the astrometric binary sample. Several of the MS stars are somewhat evolved (i.e., their radii are up to ~70\% larger than stars of the same mass at the zero-age main sequence).   

We removed one initial target, {\it Gaia} DR3 ID 747174436620510976, from the sample because it was not possible to obtain a good single-star fit to the SED without changing the {\it Gaia} parallax. The star has near solar-metallicity star and an SED that suggests $T_{\rm eff}\approx 5450$\,K and $R\approx 0.60\,R_{\odot}$. However, this radius is too small for any main sequence star of the observed effective temperature. A possible explanation is that the {\it Gaia} parallax is underestimated and the true radius is $\approx 0.9 R_{\odot}$, as expected for a main-sequence star. This would also cast doubt on the other parameters of the astrometric solution, and our follow-up RVs are in modest tension with the predictions of the astrometric solution (Appendix~\ref{sec:spurious}), so we removed the source from the sample.

The best-fit parameters from SED fitting are listed in Table~\ref{tab:star_params}. We set a minimum uncertainty of $0.03\,M_{\odot}$ on $M_\star$ to account for systematic uncertainties in the stellar models. This uncertainty propagates to the uncertainties on $M_2$ calculated in the next section.

Our mass constraints are based on single-star evolutionary models. This is reasonable because the expected lifetimes of the companions' progenitors are much shorter than the lifetimes of the luminous stars. This means than any mass transfer would have occurred before the luminous stars experienced any significant evolution. As discussed by \citet{El-Badry2024}, there is also no plausible evolutionary scenario in which the luminous stars have significantly lower masses than implied by single-star models.  

\section{Joint fitting of astrometry and RVs}
\label{sec:thiele_innes}
Our joint modeling of the RVs and {\it Gaia} astrometry assumes a Keplerian 2-body orbit. The luminous star has mass $M_{\star}$, while the companion has mass $M_2$. The orbit is specified by its period, $P_{\rm orb}$, eccentricity, $e$, periastron time, $T_p$,  and three angles describing the orientation: the inclination, $i$, longitude of the ascending node, $\Omega$, and argument of periastron, $\omega$. The center-of-mass motion of the binary is described by a systemic velocity, $\gamma$, and five astrometric parameters: the right ascension $\alpha$ and declination $\delta$, the parallax $\varpi$, and the proper motions $\mu_{\alpha}^*$ and $\mu_{\delta}$.

The semimajor axis $a$ is set by Kepler's 3rd law: 
\begin{equation}
    \label{eq:kepler}
    a=\left(\frac{P_{\rm orb}^{2}G\left(M_{\star}+M_{2}\right)}{4\pi^{2}}\right)^{1/3},
\end{equation}
while the semimajor axis of the luminous star's orbit is 
\begin{equation}
    \label{eq:a1}
    a_{1}=a\left(\frac{q}{1+q}\right),
\end{equation}
where $q=M_2/M_\star$ is the mass ratio. The RV semi-amplitude of the star's orbit is

\begin{align}
    \label{eq:Kstar}
    K_{\star}=\frac{2\pi a_{1}}{P_{{\rm orb}}\sqrt{1-e^{2}}}\sin i.
\end{align}
Finally, the angular photocenter semimajor axis is given by 
\begin{equation}
    \label{eq:a0}
    \mathring{a}_{0}=\frac{a}{d}\left(\frac{q}{1+q}-\frac{\epsilon}{1+\epsilon}\right).
\end{equation}
Here $d$ is the distance to the binary and $\epsilon$ is the flux ratio in the $G$-band. We define $\epsilon=F_{G,2}/F_{G,\star}$, where $F_{G,2}$ and $F_{G,\star}$ represent the $G$-band flux of the companion and the luminous star, respectively. For a dark companion ($\epsilon=0$), the photocenter simply traces the primary (i.e., the star whose RVs are being measured). A nonzero value of $\epsilon$ increases the primary semimajor axis, $a_1$, that corresponds to a given photocenter semimajor axis. 

The {\it Gaia} astrometric solutions are expressed as constraints on 4 or 6 Thiele-Innes parameters, for \texttt{Orbital} and \texttt{AstroSpectroSB1} solutions, respectively:
\begin{align}
\label{eq:Apred}
    A	&=\mathring{a}_{0}\left(\cos\omega\cos\Omega-\sin\omega\sin\Omega\cos i\right) \\
B	&=\mathring{a}_{0}\left(\cos\omega\sin\Omega+\sin\omega\cos\Omega\cos i\right) \\
F	&=-\mathring{a}_{0}\left(\sin\omega\cos\Omega+\cos\omega\sin\Omega\cos i\right) \\
G	&=-\mathring{a}_{0}\left(\sin\omega\sin\Omega-\cos\omega\cos\Omega\cos i\right) \\
C	&=a_1\sin\omega\sin i \\
\label{eq:Hpred}
H	&=a_1\cos\omega\sin i.
\end{align}
In the convention used here and in the {\it Gaia} archive, the astrometric parameters $A,B,F$, and $G$ have angular units (mas), while the spectroscopic parameters $C$ and $H$ have physical units (au). In the {\it Gaia} data processing, $C$ and $H$ are constrained by the measured RVs, while $A,B,F$, and $G$ are constrained by astrometry \citep[see][]{Pourbaix2022}. Our standard fit has 14 free parameters: $P_{\rm orb}$, $e$, $M_{\star}$, $M_2$, $i$, $\Omega$, $\omega$, $\varpi$, $\alpha$, $\delta$, $\mu_{\alpha}^*$, $\mu_{\delta}$, $T_{p}$, and $\gamma$. In the fiducial modeling, we assume a dark companion ($\epsilon=0$). We also experimented with a 15-parameter fit in which $\epsilon$ is left free (Section~\ref{sec:flux_ratio}). 

For each call to the likelihood function, we construct the predicted vector of {\it Gaia}-constrained parameters, $\boldsymbol \theta_{\rm Gaia}$. For objects with \texttt{Orbital} solutions, this is given by
\begin{equation}
    \label{eq:gaia_vec_orbital}
    \boldsymbol \theta_{\rm Gaia}=\left[\alpha,\delta,\varpi,\mu_{\alpha}^{*},\mu_{\delta},A,B,F,G,e,P_{{\rm orb}},T_{p}\right],
\end{equation}
while for those with \texttt{AstroSpectroSB1} solutions,
\begin{equation}
    \label{eq:gaia_vec}
    \boldsymbol \theta_{\rm Gaia}=\left[\alpha,\delta,\varpi,\mu_{\alpha}^{*},\mu_{\delta},A,B,F,G,C,H,\gamma,e,P_{{\rm orb}},T_{p}\right].
\end{equation}
We then calculate a likelihood term that quantifies the difference between these quantities and the {\it Gaia} constraints: 
\begin{equation}
    \label{eq:lnL_ast}
    \ln L_{{\rm Gaia}}=-\frac{1}{2}\left(\boldsymbol  \theta_{{\rm Gaia}}-\boldsymbol \mu_{{\rm Gaia}}\right)^{\intercal}\boldsymbol\Sigma_{{\rm Gaia}}^{-1}\left(\boldsymbol \theta_{{\rm Gaia}}-\boldsymbol \mu_{{\rm Gaia}}\right).%-\frac{1}{2}\ln\left|\Sigma_{{\rm ast}}\right|-\frac{12}{2}\ln\left(2\pi\right).
\end{equation}
Here $\boldsymbol \mu_{{\rm Gaia}}$ and $\boldsymbol \Sigma_{\rm Gaia}$ represent the vector of best-fit parameters constrained by {\it Gaia} and their covariance matrix, which we construct from the \texttt{corr\_vec} parameter reported in the {\it Gaia} archive. 

We additionally predict the RVs of the luminous star at the array of times $t_i$ at which we obtained spectra. This requires calculating $K_{\star}$ from Equation~\ref{eq:Kstar} and iteratively solving Kepler's equation \citep[e.g.][]{Murray1999}. When then calculate a RV term in the likelihood, 

\begin{equation}
    \label{eq:lnL_RVs}
    \ln L_{{\rm RVs}}=-\frac{1}{2}\sum_{i}\frac{\left({\rm RV_{{\rm pred}}}\left(t_{i}\right)-{\rm RV}_{i}\right)^{2}}{\sigma_{{\rm RV,}i}^{2}},
\end{equation}
where ${\rm RV}_i$ and ${\rm RV}_{{\rm pred}}\left(t_{i}\right)$ are the measured and predicted RVs, with their uncertainties $\sigma_{\rm RV,i}$. 
The full likelihood is then given by
\begin{equation}
    \label{eq:lnL_tot}
    \ln L = \ln L_{{\rm Gaia}} +  \ln L_{{\rm RVs}}.
\end{equation}

To assess the relative importance of the {\it Gaia} data and our follow-up RVs in constraining the orbits, we also carry out a fit in which we omit the $\ln L_{\rm RVs}$ term from Equation~\ref{eq:lnL_tot}. This essentially samples from the {\it Gaia} covariance matrix, but it incorporates our constraints on $M_{\star}$, transforms from the Thiele-Innes coefficients to a more physically interpretable set of parameters, and allows a direct constraint on $M_2$.

We use flat priors on all parameters except $M_{\star}$, for which we use a normal distribution informed by our SED fit (Section~\ref{sec:seds}). We sample from the posterior using \texttt{emcee} \citep[][]{emcee2013} with 64 walkers, taking 3000 steps per walker after a burn-in period of 3000 steps. We choose the initialization point for the sampler by transforming the best-fit {\it Gaia} parameters to Campbell elements using \texttt{nsstools} \citep{Halbwachs2023}. 
 
\subsection{Inclination sign degeneracy}
\label{sec:sign_degen}
Astrometric orbits describe the plane-of-the-sky motion of a binary's photocenter. With astrometry alone, the data are always equally consistent with an orbit  that has inclination $+i$ and one that has inclination $-i$. Mathematically, this is evident from the fact that $A$, $B$, $F$, and $G$ depend only on $\cos i$, an even function. RVs break this degeneracy. However, because orbit-fitting posteriors are usually multi-modal, our MCMC sampling often fails to converge if initialized far from the best-fit solution. For this reason, we separately initialize two sets of MCMC chains near the positive and negative astrometry-only solutions, and retain only the chain that reaches a higher maximum likelihood.

\begin{figure*}
    \centering
    \includegraphics[width=\textwidth]{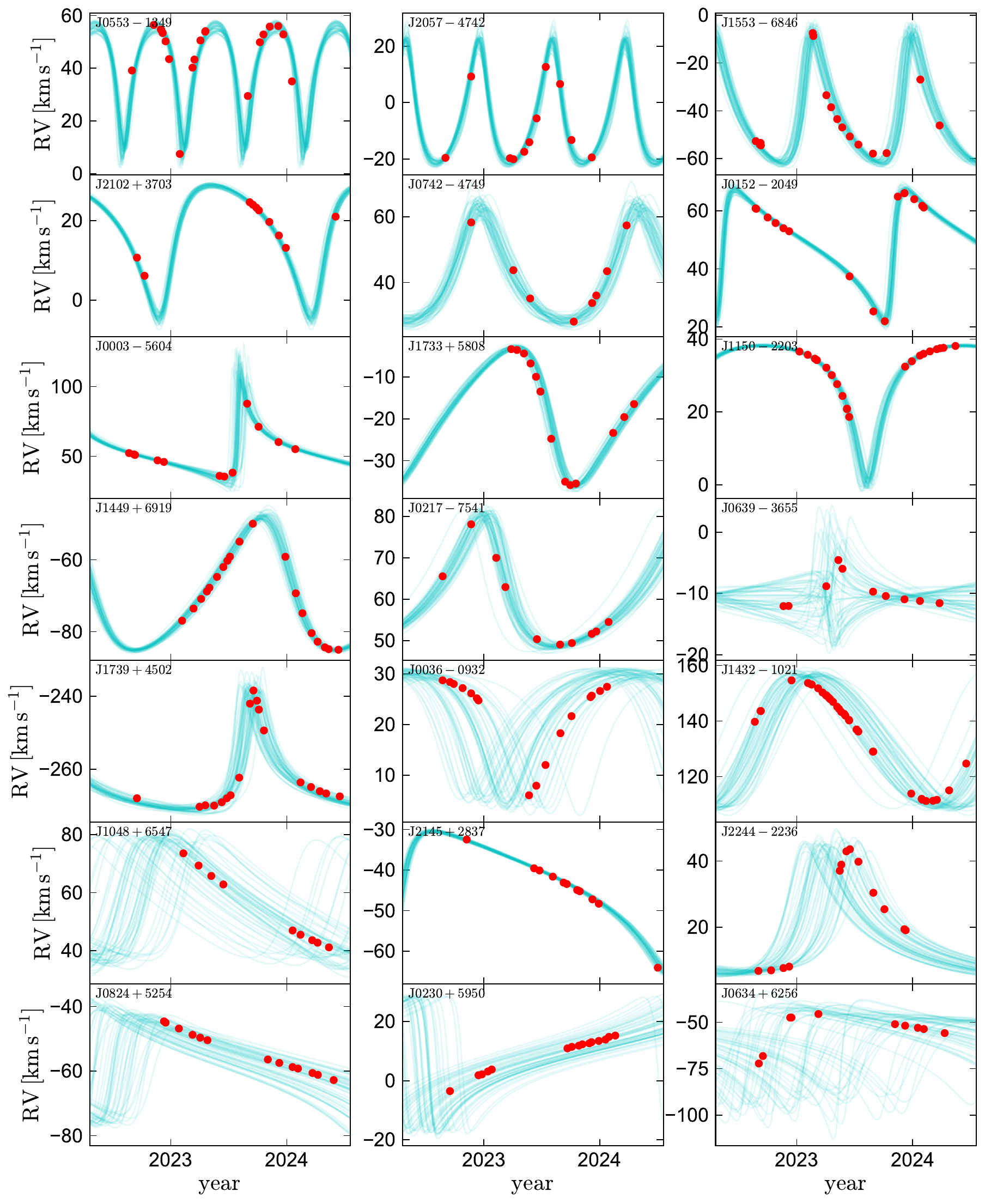}
    \caption{Comparison of our measured RVs (red) to predictions of the {\it Gaia} astrometric solutions (cyan). Individual lines show random draws from the posterior. For \texttt{AstroSpectroSB1} solutions, the systemic RV is set to the value measured by {\it Gaia}; for \texttt{Orbital} solutions, it is set to the best-fit value from our joint astrometry+RV fits (Figure~\ref{fig:black}).  Systems with long periods ($P_{\rm orb} \gtrsim 700$ days) have significant uncertainties in their {\it Gaia}-only periods and phases at the time of our observations. The RVs are broadly consistent with the {\it Gaia} predictions.} 
    \label{fig:cyan}
\end{figure*}

\begin{figure*}
    \centering
    \includegraphics[width=\textwidth]{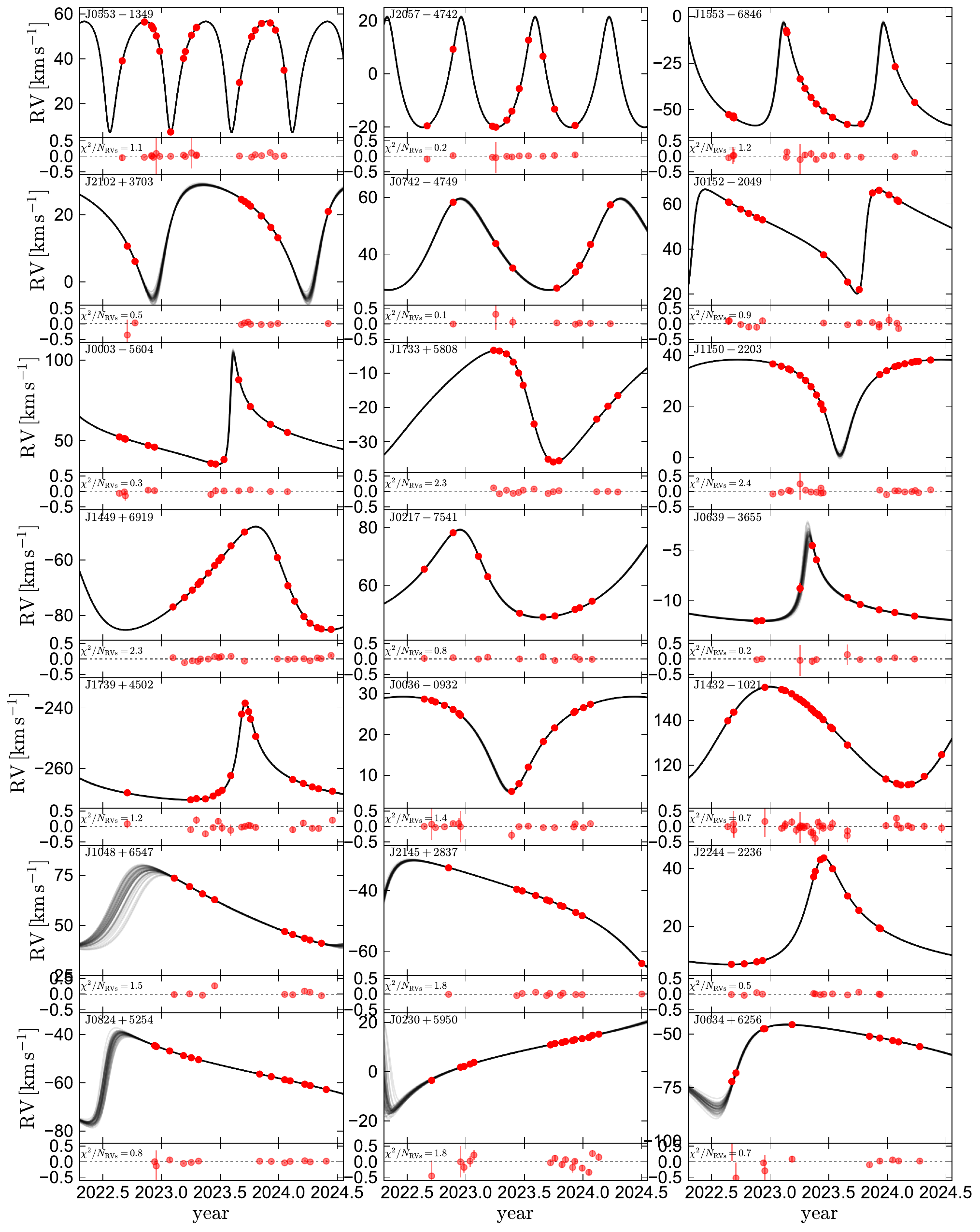}
    \caption{Joint fits of our RVs (red) and the {\it Gaia} constraints. Lower sub-panels show residuals compared to the best-fit solution. Inclusion of RVs in the fit yields much tighter constraints than those from {\it Gaia} alone (Figure~\ref{fig:cyan}). The RV residuals are generally consistent with 0 within their $(1-2)\sigma$ uncertainties. }
    \label{fig:black}
\end{figure*}

\subsection{Results}
\label{sec:results}

The results of all our joint fits are reported in Table~\ref{tab:allfits} in Appendix ~\ref{appendix:all_fits}. Constraints on a few parameters are also listed in Table~\ref{tab:sample}. Figures~\ref{fig:cyan} and~\ref{fig:black} compare our measured RVs to predictions of the {\it Gaia}-only solution (cyan) and the {\it Gaia}+RVs solution (black). We plot predicted RV curves for 50 random samples from the posterior in order to show the uncertainty in these predictions. For the {\it Gaia}-only fits, we fix the center-of-mass RV (which is not constrained by astrometry) to the maximum-likelihood value from the joint fit. For all the targets shown here, the RVs are consistent with predictions of the {\it Gaia}-only solution, meaning that at least some samples predict RV curves that overlap with the observed RVs. 

Not surprisingly, the predicted RV curves from the joint RVs+astrometry fits are much more tightly constrained than those from the {\it Gaia}-only fits. The residuals in Figure~\ref{fig:black} show that for most targets, the RVs are generally consistent with the joint solution within $(1-2)\sigma$. Some systems have $\chi^2/N_{\rm RVs} < 1.0$, suggesting that their RV uncertainties are overestimated. This could occur if a small number of outlier orders dominate the standard deviation in the RV uncertainty calculation. A few candidates have $\chi^2/N_{\rm RVs} = 1.0 - 2.0$, implying modestly underestimated RV uncertainties. 

\begin{figure*}[!ht]
    \centering
    \includegraphics[width=\textwidth]{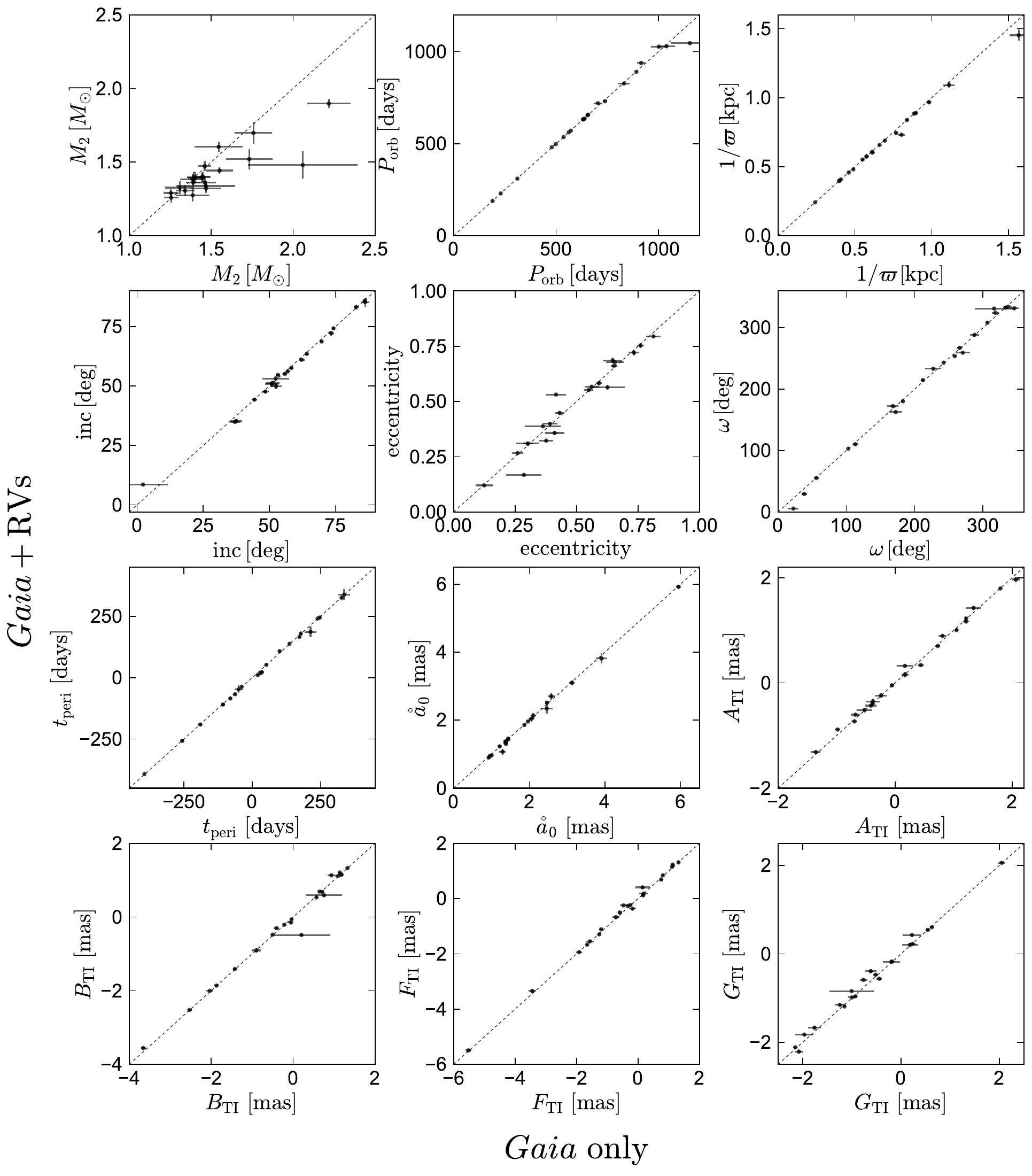}
    \caption{Orbital and astrometric parameters inferred from joint fits of the {\it Gaia} solution and our RVs (vertical axis) and the {\it Gaia} solution alone (horizontal axis). 
    The photocenter semi-major axis, $\mathring{a}_0$, and the Thiele-Innes coefficients, $A_{\rm TI}$, $B_{\rm TI}$, $F_{\rm TI}$, and $G_{\rm TI}$, are derived quantities, while the other parameters are free parameters in our fits. Uncertainties on most parameters are much smaller in the joint fits, but the two sets of constraints are generally consistent at the $(1-2)\sigma$ level. Inclinations are mirrored about 90 deg (i.e., 100 deg is shown as 80 deg) for visualization only.}
    \label{fig:comparisons}
\end{figure*}

\subsubsection{Consistency of the astrometric and RV constraints}
Figure~\ref{fig:comparisons} compares our constraints on several fitting parameters, as well as parameters that are transformations of them, from the {\it Gaia}-only and {\it Gaia}+RVs fits. For most systems, the uncertainties in all parameters are much smaller in the {\it Gaia}+RVs fits than in the {\it Gaia}-only fits. This is true even for parameters such as the inclination that are not constrained directly by RVs, because the {\it Gaia}-only constraints include significant covariances between parameters.

The two sets of parameters are generally consistent within $(1-2)\sigma$, suggesting that the astrometric solutions and their uncertainties are generally reliable. There is little evidence of systematic biases in most of the {\it Gaia}-only parameters, though the inferred $M_2$ values are on average lower in the joint fits than in the {\it Gaia}-only fits. As we discuss in Section~\ref{sec:mass_fns}, this is likely a consequence of how the candidates were selected.

\citet{El-Badry2024} found that the astrometry-only and astrometry+RVs constraints for Gaia NS1 (J1432-1021) were inconsistent at the $3\sigma$ level, suggesting that the astrometric uncertainties for that source were underestimated. Similarly, \citet{Chakrabarti2023} and \citet{Nagarajan2023} found evidence for underestimated uncertainties in the astrometric orbit of Gaia BH1. Figure~\ref{fig:comparisons} implies that most of our candidates have more robust astrometric uncertainties: indeed, Gaia NS1 has the {\it most} discrepant solution among all the objects in the sample.  This could arouse suspicion that the companion's mass is simply overestimated. However, \citet{El-Badry2024} showed that RVs alone require the companion to be the most massive in the sample, even if the astrometric orbit constraints are disregarded. 

Considering the 4 Thiele-Innes parameters $A$, $B$, $F$, and $G$, only 22 out of 84 parameters across 21 objects differ by more than 1$\sigma$, and only 5 differ by more than 2$\sigma$. The {\it Gaia}-only and {\it Gaia}+RVs parallaxes are consistent within $1\sigma$ for 17 candidates, and within $2\sigma$ for 20 candidates.

\subsection{Mass functions}
\label{sec:mass_fns}

\begin{figure*}[!ht]
    \centering   \includegraphics[width=\textwidth]{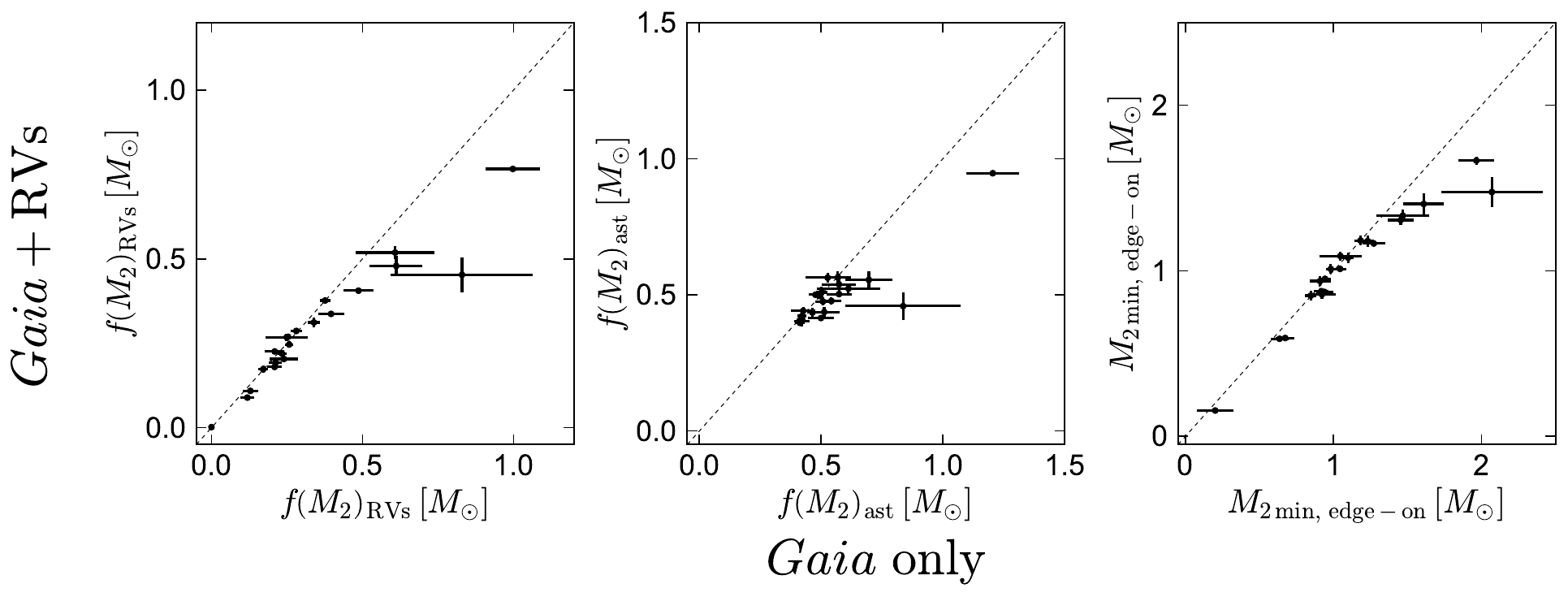}
    \caption{Comparison of mass functions for the unseen companions inferred from the {\it Gaia} solution alone (horizontal axis) and joint fitting of that solution and our RVs (vertical axis). Left and middle panels show the spectroscopic and astrometric mass functions. Right panel shows the minimum companion mass implied by the RV mass function {\it if the orbit were edge-on}. The mass functions inferred from {\it Gaia}+RVs  are in most cases consistent with those from {\it Gaia} alone, but they are systematically lower. Because our candidates were selected on the basis of high astrometric mass functions, and massive companions are intrinsically rare, their companion masses are overestimated on average. That is, systems with underestimated masses did not enter the sample.}
    \label{fig:mass_functions}
\end{figure*}

Figure~\ref{fig:mass_functions} compares several limits on the unseen companion masses between the {\it Gaia}-only and {\it Gaia}+RVs fits. The left panel shows the RV mass function, 
\begin{align}
    f\left(M_{2}\right)&=\frac{P_{{\rm orb}}K_{\star}^{3}}{2\pi G}\left(1-e^{2}\right)^{3/2} \\ 
    &=M_{2}\left(\frac{M_{2}}{M_{2}+M_{\star}}\right)^{2}\sin^{3}i
    \label{eq:fm_spec}
\end{align}
 which represent a lower limit on the companion mass that corresponds to the limiting case where the orbit is edge-on and the luminous star is massless. $K_\star$ is the RV semi-amplitude as measured from the joint fit (Equation~\ref{eq:Kstar}). 

The middle panel shows the astrometric mass function, 
\begin{equation}
    \label{eq:fm_ast_def_}
    f\left(M_{2}\right)_{{\rm ast}}=\left(\frac{\mathring{a}_{0}}{\varpi}\right)^{3}\left(\frac{P_{{\rm orb}}}{{\rm yr}}\right)^{-2}.
\end{equation}
For a dark companion, 
\begin{equation}
    \label{eq:fm_ast_meaning}
    f\left(M_{2}\right)_{{\rm ast}}=M_{2}\left(\frac{M_{2}}{M_{2}+M_{\star}}\right)^{2}.
\end{equation}
That is, $f\left(M_{2}\right)_{{\rm ast}}$ represents a lower limit on the companion mass that incorporates the known inclination from astrometry but treats the luminous star as massless.  Finally, the right panel shows the minimum companion mass implied by the RV mass function and our assumed luminous star mass, for an edge-on orbit. We obtain this by setting $i=90$ deg and solving Equation~\ref{eq:fm_spec} for $M_2$, setting $M_\star$ to the value inferred from the SED (Table~\ref{tab:star_params}). We propagate uncertainties on all parameters by calculating these mass functions directly from the MCMC samples.

Although the {\it Gaia}-only and {\it Gaia}+RVs mass functions are generally consistent, the {\it Gaia}+RVs  constraints on average favor lower mass functions. This means that, on average, the observed RV semi-amplitudes are somewhat lower than predicted by the astrometric solutions. Since our targets are selected from the tails of the companion mass function distribution (e.g. Figure~\ref{fig:cmd}), it is expected that more objects in the sample will scatter toward higher masses than toward lower masses. That is, binaries with noise scattering their best-fit masses toward lower values would not have entered the sample in the first place.  

\subsection{Flux ratio constraints}
\label{sec:flux_ratio}
Our modeling thus far has assumed that the companion is dark. In this case, the semimajor axis of the photocenter orbit, $\mathring{a}_0$, will be identical to the semimajor axis of the star whose RVs are being measured, $\mathring{a}_1$. To explore the possibility of a luminous companion, we tried repeating the fit while adding the flux ratio $\epsilon$ (Equation~\ref{eq:a0}) as a free parameter. 

We first tried fitting with no bounds on $\epsilon$. In this case, we found {\it negative} best-fit values of $\epsilon$ for a majority of the sources, with a median and $\pm 1 \sigma$ range of $\epsilon = -0.01 \pm 0.03$ across the full sample. Most sources have $\epsilon$ constrained with a $1\sigma$ uncertainty of $\sigma_{\epsilon} \lesssim 0.02$. Negative values of $\epsilon$ are unphysical, but mathematically allowed (e.g. Equation~\ref{eq:a0}). A preference for negative $\epsilon$ simply reflects observed RV variability with a lower amplitude than predicted by the best-fit astrometric solution with $\epsilon =0$; i.e., it is a consequence of the same selection effect that causes the mass functions in Figure~\ref{fig:mass_functions} to be overestimated. 

Next, we imposed a constraint that $\epsilon$ must be positive. In this case, the posterior constraints on $\epsilon$ pile up against 0 for most sources. The median upper limit on $\epsilon$ is $\epsilon < 0.012$ (1$\sigma$), or $\epsilon < 0.032$ ($2\sigma$). That is, the combination of RVs and astrometry strongly disfavor luminous companions, because luminous companions would imply a larger $a_1$ for fixed $\mathring{a}_0$, and this would imply a larger RV variability amplitude than is observed. 

\begin{figure*}[!ht]
    \centering
    \includegraphics[width=0.8\textwidth]{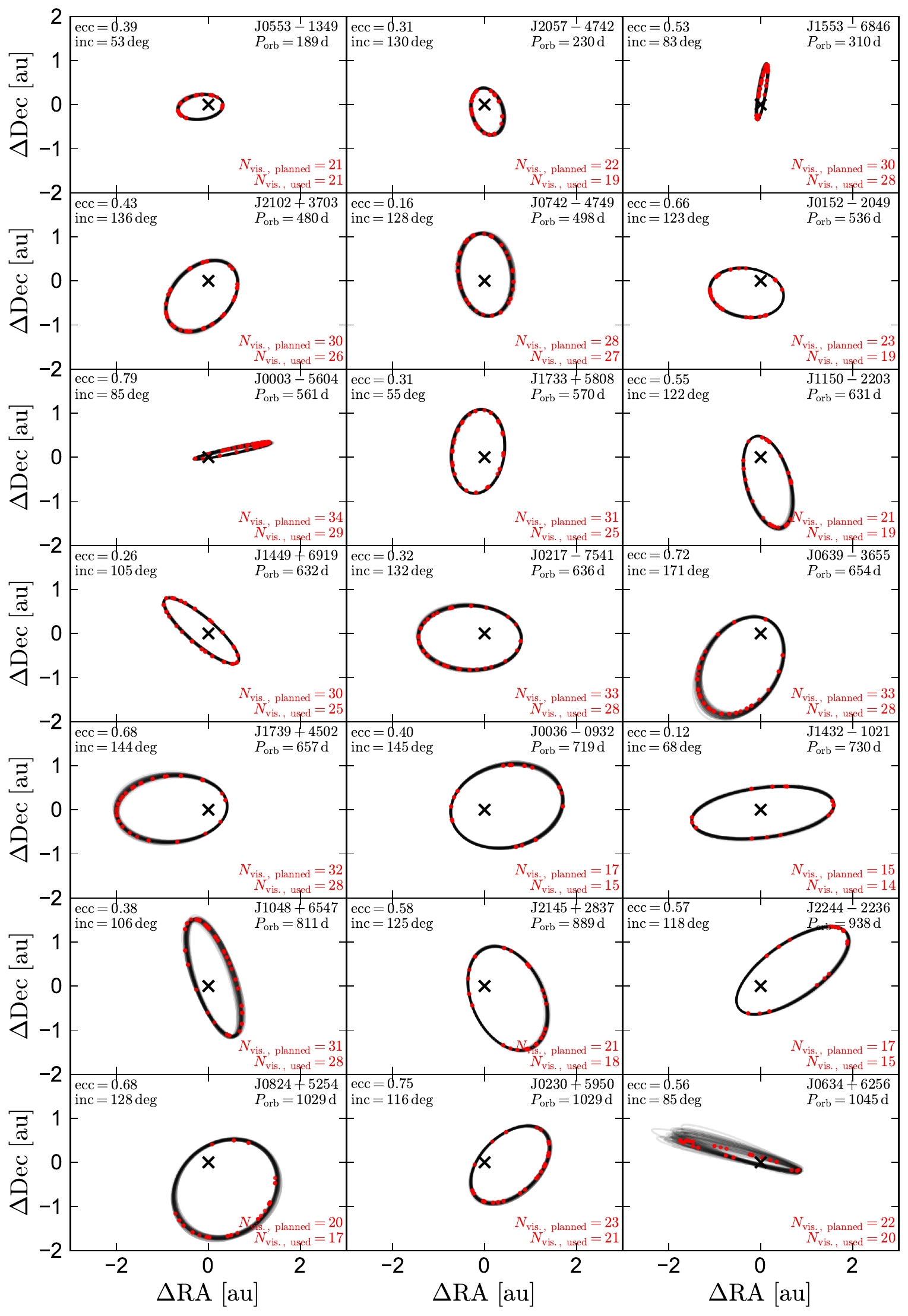}
    \caption{Sky-projected orbits of the luminous stars. The angular orbits have been multiplied by distance, so that the displayed orbital sizes correspond to relative physical sizes. Red points mark {\it Gaia} observations predicted by \texttt{GOST}, showing that all of the targets in our sample are expected to have good astrometric phase coverage. Note that we do not have access to the actual measured $\Delta$RA and $\Delta$Dec values; only to the predicted scan times. Red text in each panel indicates the number of visibility periods that are predicted by \texttt{GOST}, as well as the number actually used in calculating the astrometric solution. }
    \label{fig:gost}
\end{figure*}

\subsection{Astrometric phase coverage}
One possible concern is that the {\it Gaia} observations could have resulted in poor orbital phase coverage. This can occur, for example, for orbital periods that are close to a year, half a year, or several periods related to the {\it Gaia} scanning law \citep{El-Badry2023_bh1, Holl2023}. While  actual epoch-level astrometric data are not published in DR3, it is possible to predict when a source should have been observed using the {\it Gaia} observation scheduling tool (GOST)\footnote{\href{https://gaia.esac.esa.int/gost/}{https://gaia.esac.esa.int/gost/}}.  Given a source's coordinates, GOST returns a list of observation times when the scanning law predicts that the source will transit across the {\it Gaia} focal plane. 

Figure~\ref{fig:gost} shows sky-projected photocenter orbits for all candidates. Individual gray lines show predictions for 50 random draws from the posterior of the joint fit. Red points show interpolations of the scan times predicted by GOST onto the orbit corresponding to the marginalized posterior median.  Overall, most of the sources have good astrometric phase coverage, with {\it Gaia} observations sampling most of the predicted orbital ellipse. 

It is not guaranteed that {\it Gaia} will actually obtain data on each source at the predicted times, as GOST does not account for gaps between CCDs and issues such as micrometeoroids impacts that cause temporary gaps in the datastream. In each panel of Figure~\ref{fig:gost}, we list both the number of visibility periods predicted by GOST and the number actually used in calculating the astrometric solution (\texttt{visibility\_periods\_used}; a visibility period is a group of observations separated from other observations by at least 4 days). The number of visibility periods used ranges from $80-100\%$ of the number predicted by GOST, suggesting that the scan times shown in Figure~\ref{fig:gost} are a reasonable but not perfect approximation of those used in producing the astrometric solutions. The median number of visibility periods used is 21, significantly higher than the minimum of 12 that is required to constrain a 12-parameter astrometric binary solution.

\subsection{Astrometric solution quality diagnostics}
\label{sec:gof}

\begin{figure*}
    \centering
    \includegraphics[width=\textwidth]{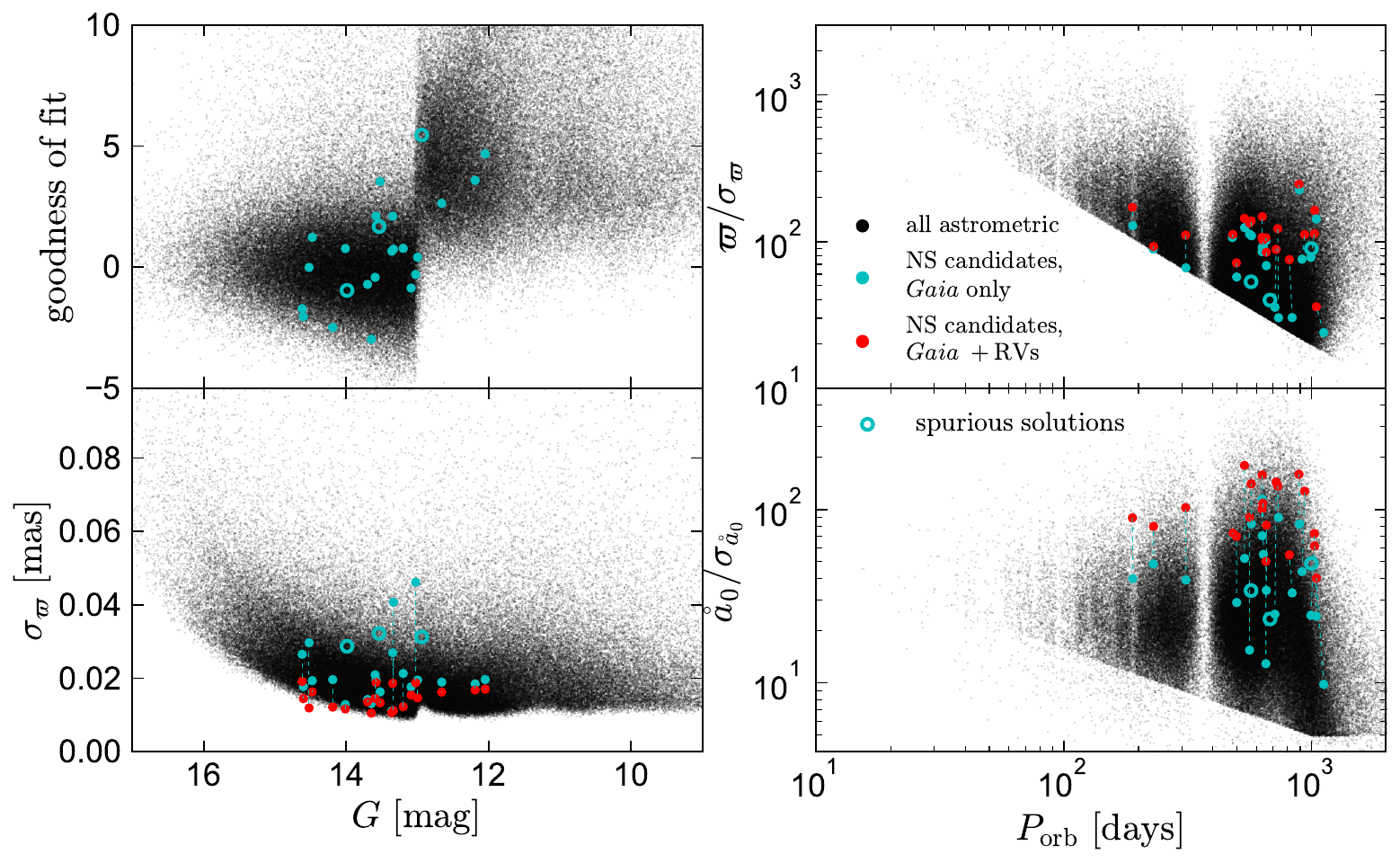} 
    \caption{Comparison of our NS candidates (cyan/red) to all sources with astrometric orbital solutions published in DR3 (black). Hollow symbols show three initial candidates we found to have spurious solutions (Appendix~\ref{sec:spurious}).  Upper left: astrometric \texttt{goodness\_of\_fit}. Large values indicate a poor formal fit. There is a discontinuity at $G=13$ owing to a quirk of the {\it Gaia} data processing: brighter sources typically have larger \texttt{goodness\_of\_fit} values because the uncertainties in their individual-epoch astrometric data are underestimated. Lower left: parallax uncertainty. Cyan points show the uncertainty reported in {\it Gaia} DR3, while red points show the uncertainty from our joint RV + astrometry fit. Right panels: parallax over error (top) and photocenter semimajor axis over error (bottom) as a function of orbital period. The diagonal ``cliff'' is a result of quality cuts imposed on the published solutions.  The NS candidates have typical \texttt{goodness\_of\_fit} values and astrometric uncertainties for their apparent magnitudes and periods, indicative of unproblematic astrometric solutions. The sources with spurious solutions have slightly larger parallax uncertainties than average but otherwise do not stand out from the sources with reliable solutions.  }
    \label{fig:gof}
\end{figure*}

As an additional check on the reliability of our candidates' astrometric solutions, Figure~\ref{fig:gof} compares several quality diagnostics of these candidates to the larger sample of all astrometric binaries from {\it Gaia} DR3. Cyan points show the {\it Gaia} astrometric solutions alone, while red points show joint {\it Gaia} + RV fits. Hollow points show three sources we found to have spurious solutions after RV follow-up (Appendix~\ref{sec:spurious}).

First considering the {\it Gaia}-only solutions, the NS candidates have \texttt{goodness\_of\_fit} values and parallax uncertainties that are broadly representative of the larger astrometric binary sample. They have higher-significance photocenter ellipse measurements (i.e., larger $\mathring{a}_0/\sigma_{\mathring{a}_0}$) than most other astrometric binaries, reflecting the fact that NS companions produce large photocenter orbits at fixed period (Figure~\ref{fig:keplers_law}).  The sources with spurious solutions have similar \texttt{goodness\_of\_fit} values  to those with good solutions, though their parallax uncertainties are on average larger at fixed apparent magnitude. 

The uncertainties from the joint {\it Gaia}+RV fits are often significantly smaller than those from {\it Gaia} alone. Even the parallax, which is not directly constrained by RVs, has smaller uncertainties in the joint {\it Gaia} + RVs fit, because the {\it Gaia} solutions include significant covariances between parallax and other orbital parameters that {\it are} directly constrained by RVs. We emphasize, however, that the uncertainties shown in Figure~\ref{fig:gof} are purely statistical. The parallax zeropoint and uncertainties of the astrometric binary solutions have yet to be investigated in detail. Nevertheless, we conclude from Figure~\ref{fig:gof} and the good agreement between measured and predicted RVs that the astrometric solutions of the candidates in our sample are robust. We also conclude that spurious solutions cannot always be easily identified on the basis of their astrometric uncertainties and quality flags. 

\section{Discussion}
\label{sec:disc}

\subsection{Nature of the unseen companions}

The orbits of the binaries in our sample are well-characterized, and the companion masses are measured with typical uncertainties of a few percent. However, we have not detected light from the companions, leaving their astrophysical nature uncertain. %Here we consider several possibilities.  
Our joint fitting of astrometry and RVs places upper limits of a few percent on the $G-$band flux ratio, $\epsilon=F_{G,2}/F_{G,\star}$
(Section~\ref{sec:flux_ratio}). As we show below, this rules out all plausible non-degenerate companions.

We begin by considering the astrometric mass function (Equation~\ref{eq:fm_ast_meaning}). For a luminous companion with flux ratio $\epsilon$ and mass ratio $q=M_2/M_\star$, this quantity can be expressed as \citep[e.g.][]{Shahaf2019}:

\begin{equation}
    \label{eq:fm_ast_def}
    f\left(M_{2}\right)_{{\rm ast}}=M_{\star}\frac{\left|q-\epsilon\right|^{3}}{\left(1+\epsilon \right)^{3}\left(1+q\right)^{2}}.
\end{equation}

\begin{figure}
    \centering
    \includegraphics[width=\columnwidth]{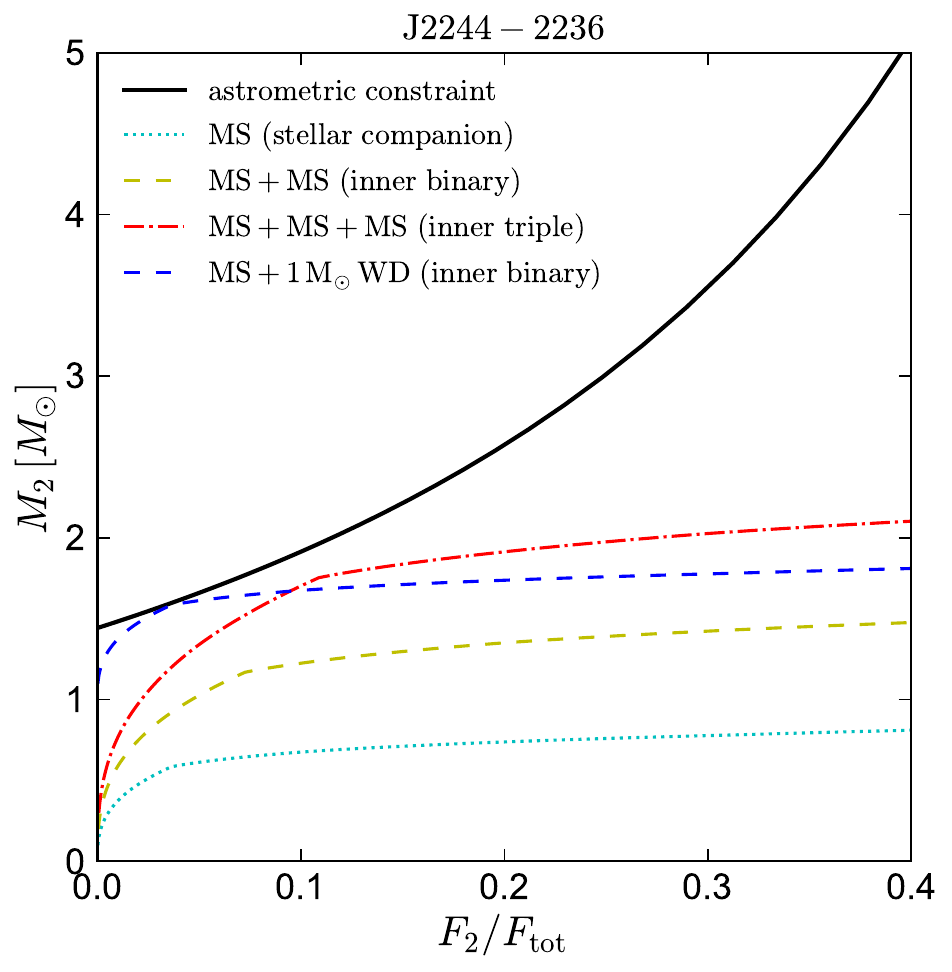}
    \caption{Constraints on the mass of the unseen companion in J2244-2236, a typical system, as a function of the $G-$band flux ratio. Solid black line shows the constraint from the astrometric mass function (Equations~\ref{eq:fm_ast_def}), assuming $M_\star = 1.00\,M_{\odot}$; a completely dark companion would imply $M_{2}\approx 1.44\,M_{\odot}$. If the companion contributes some light, its astrometrically-implied mass increases. Dotted cyan line shows the expected flux ratio and mass for a single MS companion. Because this is always below the black line, no single MS companion can explain the orbit. The same is true for an equal-mass inner binary (yellow dashed) or an equal-mass inner triple (red dot-dashed). An inner binary containing a $1\,M_{\odot}$ WD and a $\approx 0.6\,M_{\odot}$ MS star (blue dashed) could explain the astrometric mass function with the WD contributing 4\% of the light in the $G-$band, but this violates the flux ratio constraint from RVs (see text). A tight WD+WD binary (not shown, because it could have $F_2/F_{\rm tot}\approx 0$), could explain the orbit, but it is unclear how such a system could form.}
    \label{fig:scenarios}
\end{figure}

Given an observationally constrained $M_\star$ and $f\left(M_{2}\right)_{{\rm ast}}$, we can solve Equation~\ref{eq:fm_ast_def} for $q$, and thus for $M_2$, for any possible $\epsilon$. In Figure~\ref{fig:scenarios}, the black line shows this constraint for a representative source in our sample, J2244-2236, which has $M_\star = 1.00 \pm 0.03\,M_{\odot}$ and $f\left(M_{2}\right)_{{\rm ast}} =  0.503\pm0.007 \,M_{\odot}$. The dynamically-implied mass of the companion increases with its assumed luminosity. Physically, this reflects the fact that a brighter companion dilutes the light from the primary more, such that a higher companion mass is required to explain the same photocenter orbit.

We can then consider possible companion types:

\begin{enumerate}
    \item {\it A single MS star}: The dotted cyan line in Figure~\ref{fig:scenarios} shows the expected flux contribution versus mass relation for a MS companion. We take this from a 5 Gyr-old, solar-metallicity MIST isochrone. Because the mass supplied by such a companion is much less than the astrometric constraint for all flux ratios, a single MS companion is ruled out. 
    \item {\it An inner binary containing two MS stars}: The yellow dashed line in Figure~\ref{fig:scenarios} shows the mass and flux expected if the companion is a binary containing two MS stars, so the full system is a triple. We assume the two stars have equal mass, since this yields the highest mass-to-light ratio. We use the same isochrone as for a single star but assume twice the total mass and light. This still falls short of the astrometric mass constraint, so the companion cannot be a close MS+MS binary.   
    \item {\it An inner triple containing three MS stars}: The red dot-dashed line in Figure~\ref{fig:scenarios} shows the case where the companion is a triple containing three MS stars (i.e., the full system is a quadruple). We assume all three inner components have the same mass. Even this falls below the astrometric constraint for all flux ratios, so the companion cannot be an inner triple. In addition, a quadruple system with four stars in such a tight orbit would be unlikely to be stable over long timescales.
    \item {\it An inner binary containing a WD and a MS star}: The blue dashed line in Figure~\ref{fig:scenarios} corresponds to an inner binary containing a $1\,M_{\odot}$ WD and a MS star. We assume that the WD is sufficiently cold that it does not contribute any light in the optical. Such an inner binary could reproduce the observed astrometric mass function if the MS star has a mass of $\approx 0.6\,M_{\odot}$, in which case it would contribute $\approx 4\%$ of the total light. However, as discussed in Section~\ref{sec:flux_ratio}, the RVs rule out a flux ratio above 2\% for this source, because a $1.6\,M_\odot$ companion would produce larger-amplitude RV variability. Only a tight binary containing a WD with $M\gtrsim  1.1\,M_{\odot}$ and a low-mass MS star could match the data.
    %Any WD mass below $1.0\,M_{\odot}$ would fall below the astrometric constraint. %A higher WD mass could match it for two separate MS star masses.  
    \item {\it An inner binary containing two WDs}: All of the objects in our sample could in principle be explained by a tight inner binary containing two cold and relatively massive WDs. The total mass of the inner binaries would need to be near the Chandrasekhar mass. Chandrasekhar-mass close WD+WD binaries are relatively rare: despite decades of dedicated searches, {\it none} have been conclusively identified. Empirical limits suggest a space density comparable to the space density of neutron stars \citep[e.g.][]{Badenes2012}. As discussed by \citet{El-Badry2024}, there are significant evolutionary challenges to forming a massive WD+WD binary within the orbits of the observed luminous stars: such triples are likely to become dynamically unstable during their evolution, particularly given that the initial orbits of the luminous stars would have to have been significantly tighter than observed today when mass loss from the inner binary is accounted for. Nevertheless, tight WD+WD binaries cannot be ruled out based on the observed data alone.

    \item {\it A single ultramassive WD}: Several of our candidates have best-fit masses above the maximum WD mass mass, which is near $1.37\,M_\odot$ \citep{Althaus2022}. However, about half have masses below this limit, and given the astrometric uncertainties, a majority of our candidates are consistent at the few-$\sigma$ level with having $M_2 \lesssim 1.37\,M_\odot$. If the companions are single WDs, all of them would be among the highest-mass WDs known \citep[e.g.][]{Cognard2017, Caiazzo2021, Miller2023, Yamaguchi2023}.  
    
    Given that our candidates are selected from the upper tail of the inferred companion mass distribution, the possibility that companion masses are overestimated systematically should be taken seriously.  However, at least one object in the sample -- Gaia NS1 (J1432-1021) -- has a companion mass high enough that it cannot be a single WD: this object has $M_2 = 1.90\pm 0.03$, about 17$\sigma$ above the maximum WD mass. 
 
    As we discuss below, almost all the binaries in our sample have higher eccentricities than typical WD+MS binaries at similar periods. A simple interpretation is that the companions are NSs and the eccentricities are a result of natal kicks, but the data also suggest that systems containing massive WDs have higher typical eccentricities than those containing $\sim 0.6\,M_{\odot}$ WDs, so we cannot completely rule out a scenario in which many of the unseen companions are ultramassive WDs.
    
    \item {\it A single neutron star}: The inferred masses of the unseen companions in our sample are typical of neutron stars in pulsar binaries \citep[][]{Ozel2016}.  We consider this the simplest and most plausible scenario.
    
\end{enumerate}

\subsection{Period-eccentricity relation}
\label{sec:period_ecc}

\begin{figure*}[!ht]
    \centering
    \includegraphics[width=\textwidth]{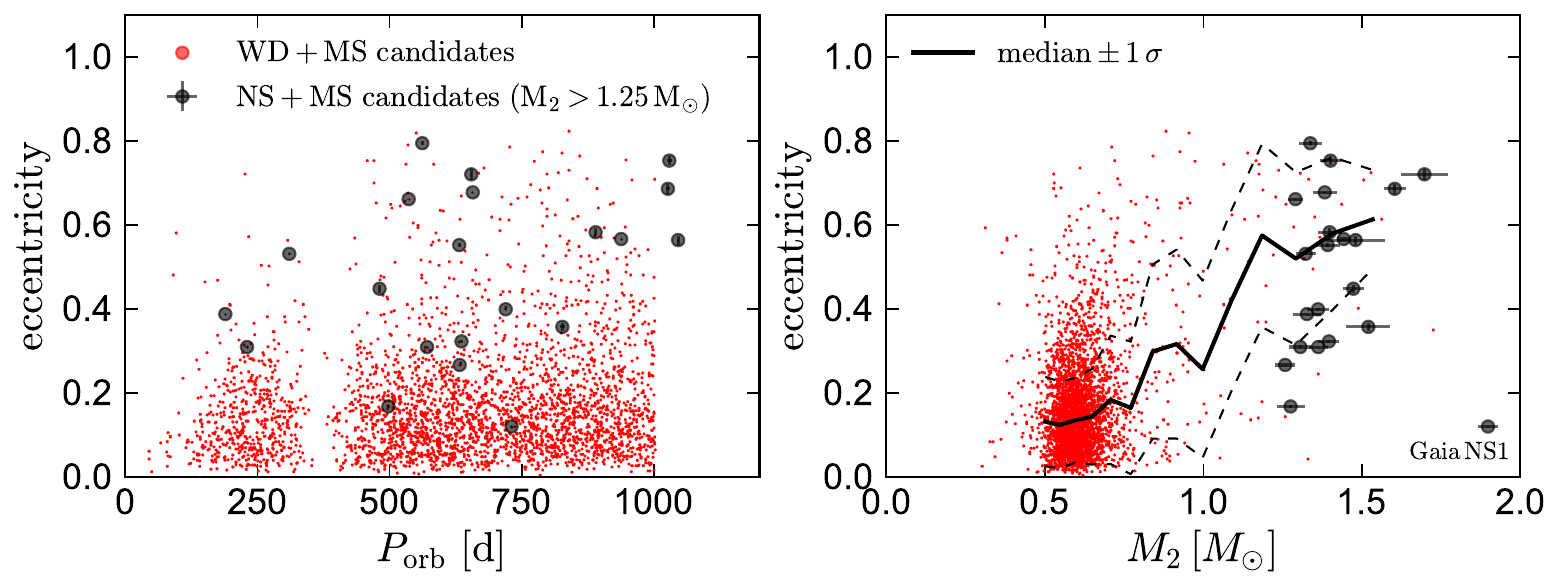}
    \caption{Left: Period-eccentricity diagram for NS+MS candidates from this work (black) compared to WD+MS binary candidates selected by \citet{Shahaf2023} from {\it Gaia} DR3 on the basis of their astrometrically-constrained mass ratios. For NS+MS candidates, uncertainties are smaller than symbols. Most of the WD candidates have low eccentricities, probably resulting from (incomplete) tidal circularization when the WD progenitor was a red giant. Our candidates have significantly higher eccentricities than the WD candidates at all periods, which may be the result of kicks during NS formation. Right: mass-eccentricity diagram. Our candidates are all selected to have $M_2 > 1.25\,M_{\odot}$. Solid and dashed lines show median and middle 68\% across both red and black points. Binaries with $M_2 = (1.0-1.2)\,M_{\odot}$, most of which likely contain high-mass WDs, also have higher eccentricities than systems containing lower-mass WDs. }
    \label{fig:period_ecc}
\end{figure*}

Figure~\ref{fig:period_ecc} (left panel) shows the period--eccentricity relation for our candidates. We compare them to the larger sample of $\sim 3000$ WD+MS binary candidates selected from {\it Gaia} astrometry by \citet{Shahaf2023b} on the basis of their AMRF. Our candidates -- which differ from the \citet{Shahaf2023b} sample only in that they have inferred companion masses above $1.25\,M_{\odot}$ -- have significantly higher eccentricities than those systems at fixed period. The low  eccentricities of the WD+MS binaries are likely a result of tidal circularization when the progenitors of the WDs were giants. If the unseen companions in our sample are NSs, their higher eccentricities could could be naturally understood as a result of natal kicks. 

Although the eccentricities of the WD+MS binary candidates in Figure~\ref{fig:period_ecc} are much lower on average than those of ordinary MS+MS binaries at the same periods, the large majority of them are {\it not} consistent with zero, and are much higher than expected for binaries that have gone through long periods of stable mass transfer \citep{Phinney1992, Lorimer2008}. This may be a consequence of these systems having first initiated mass transfer on the asymptotic giant branch rather than on the first giant branch and/or having gone through common envelope evolution rather than stable mass transfer. Similar eccentricities are observed in other populations of WD+MS binaries at these periods, such as blue stragglers and barium stars \citep{Mathieu2009, Jorissen2019, Escorza2019}.

The right panel of Figure~\ref{fig:period_ecc} shows  eccentricities and dark companion masses. A majority of the objects from the \citet{Shahaf2023b} sample with $M_2 > 1.25\,M_{\odot}$ are included in our sample: those that are not are either faint ($G>15.0$) or were excluded because our follow-up showed them to have spurious astrometric solutions or lower dark companion masses. It is evident that while most WD+MS candidates with $M_2 \lesssim 0.8\,M_{\odot}$ have low-eccentricity orbits ($e < 0.2$), systems with $M_2 \gtrsim 1.0\,M_{\odot}$ tend to be more eccentric. It is tempting to simply attribute this dichotomy to WD vs. NS companions \citep[e.g.][]{Shahaf2023}. However, the eccentric population seems to extend to lower masses than expected for NSs. We have also identified a handful of systems in the course of our follow-up that have $M_2 > 1\,M_{\odot}$, $e > 0.5$, and clear UV excess, which points to a WD companion rather unambiguously. These eccentricities could be a result of faster orbital inspiral and/or more asymmetric mass loss in the super-AGB progenitors of massive WDs \citep{Izzard2010, El-Badry2018}, or eccentricity-pumping due to massive circumbinary disks formed through the mass transfer process \citep{Dermine2013}. Kozai-Lidov oscillations in systems containing a wide tertiary are another possibility.

\subsection{Galactic orbits}

\begin{figure*}
    \centering
    \includegraphics[width=\textwidth]{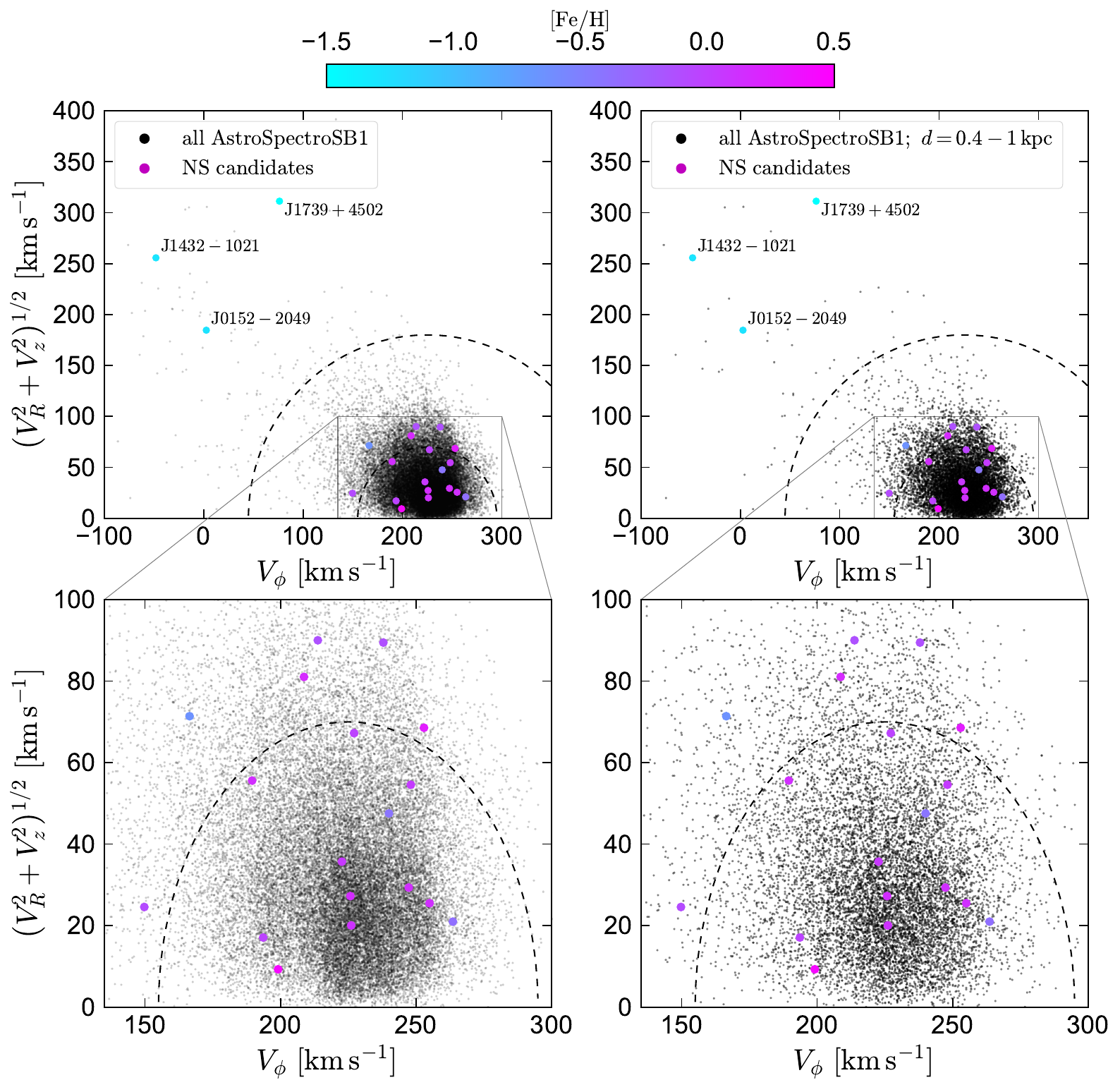}
    \caption{Toomre diagram of NS candidate binaries compared to all binaries with \texttt{AstroSpectroSB1} solutions published in {\it Gaia} DR3. In the left panels, black points show the full DR3 catalog. In the right panels, they show only binaries with distances 0.4-1 kpc, similar to the NS candidates. Points are colored by metallicity. Dashed lines show total velocities of 70 and 180 $\rm km\,s^{-1}$ with respect to the local standard of rest, approximately separating the thin disk, thick disk, and halo. The three candidates with halo orbits (labeled) all have low metallicities. Halo and thick-disk orbits are significantly over-represented in the NS candidate sample. Natal kicks may be responsible for kinematically heating the orbits of these objects somewhat, but our results imply that NS candidates are over-represented in low-metallicity populations.  }
    \label{fig:toomre}
\end{figure*}

To explore the Galactic stellar populations our NS candidates are members of and imprints of possible natal kicks, we show their locations in the Toomre diagram in Figure~\ref{fig:toomre}. For each binary, we use the center-of-mass velocity from the joint fit and the parallax and proper motion from the {\it Gaia} binary solution to calculate the current 3D motion of the binary's center of mass in Galactocentric cylindrical coordinates. We assume that the Sun is 20 pc above the Galactic midplane and 8.12 kpc from the Galactic center and has a 3D velocity vector $\left(V_{R,\odot},V_{\phi,\odot},V_{Z,\odot}\right)=\left(-12.9,245.6,7.78\right)\,{\rm km\,s^{-1}}$ \citep{Drimmel2018}. We perform the same calculation for the other 33,467 binaries with \texttt{AstroSpectroSB1} solutions. We do not consider \texttt{Orbital} solutions because their center-of-mass RVs are not known. 

The results are shown in Figure~\ref{fig:toomre}. We compare the NS candidates in our sample to all binaries with \texttt{AstroSpectroSB1} solutions (left panels) and to those with $d=0.4-1.0$ kpc (right panels; these have a similar distance distribution to the NS candidates).  Dashed lines centered on $V_\phi=225\,\rm km\,s^{-1}$ show approximate boundaries of the Galactic thin disk (total velocity $< 70\,\rm km\,s^{-1}$), thick disk ($70-180\,\rm km\,s^{-1}$), and halo ($> 180\,\rm km\,s^{-1}$) \citep[e.g.][]{Bensby2014}. Three objects in our sample -- J1739+4502, J1432-1021, and J0152-2049 -- are on halo orbits, with total velocities of 370, 350, and 290\,$\rm km\,s^{-1}$ with respect to the local standard of rest. These objects all have metallicities $\rm [Fe/H] < -1.2$, implying that their high velocities are mainly the result of membership to an old stellar population, not natal kicks. On the other hand, the 6 systems in the ``thick disk'' region of the Toomre diagram have metallicities close to solar. This suggests that natal kicks may have played a role in kinematically heating their orbits.  The median and middle 68\% range of midplane distances, $|z|$, are respectively 0.37 kpc and $(0.11-0.55)$ kpc. This makes the objects in our sample kinematically cold compared to NS LMXBs, which typically have $|z|\gtrsim 1\,{\rm kpc}$ \citep{vanParadijs1995, Jonker2004}. This is not surprising since only NSs formed with weak kicks will remain bound in the wide orbits to which astrometry is sensitive. 

Metal-poor stars on halo orbits seem to be significantly overrepresented in the NS candidate sample. In particular, 3 of the 21 objects in the sample are unambiguously on halo orbits. In comparison, 0 of the 319 WD + MS binary candidates selected by \citet{Shahaf2023} with \texttt{AstroSpectroSB1} solutions have halo-like orbits! Among 11420 \texttt{AstroSpectroSB1} orbits with $d=0.4-1.0$ kpc, 61 (i.e., 0.5\%) have halo-like orbits, 1636 (14.3\%) have thick disk-like orbits, and 9723 (85.1\%) have thin disk-like orbits. In contrast, in the NS sample, 3/21 (14\%) have halo-like orbits, 6/21 (29\%) have thick disk-like orbits, and 12/21 (57\%) have thin disk-like orbits. Since the halo orbits of the NS candidates cannot be the result of natal kicks, the high fraction of halo orbits suggest that NSs formed from low-metallicity stars are more likely to survive in {\it Gaia}-detectable binaries. As discussed by \citet{El-Badry2024_bh3}, BH companions also seem to be overrepresented in the {\it Gaia} sample at low metallicity.

\subsection{Lithium enhancement}
\label{sec:lithium}
To search for spectroscopic anomalies that could result from mass transfer from the unseen companions' progenitors, we compared their spectra to spectra of similar stars observed by the GALAH survey \citep{Buder2021}. We followed the same procedure described by \citet{El-Badry2024}: in brief, we compared the rest-frame, resolution matched, and continuum-normalized spectra of our candidates pixel-by-pixel to all high-SNR spectra in GALAH DR3 and visually inspected the closest matches. Most of our candidates have unremarkable spectra and abundances, and we defer a full abundance analysis to future work. However, we highlight one feature that is striking: all three of the metal-poor halo stars in the sample are strongly enhanced in lithium.

\begin{figure*}
    \centering
    \includegraphics[width=\textwidth]{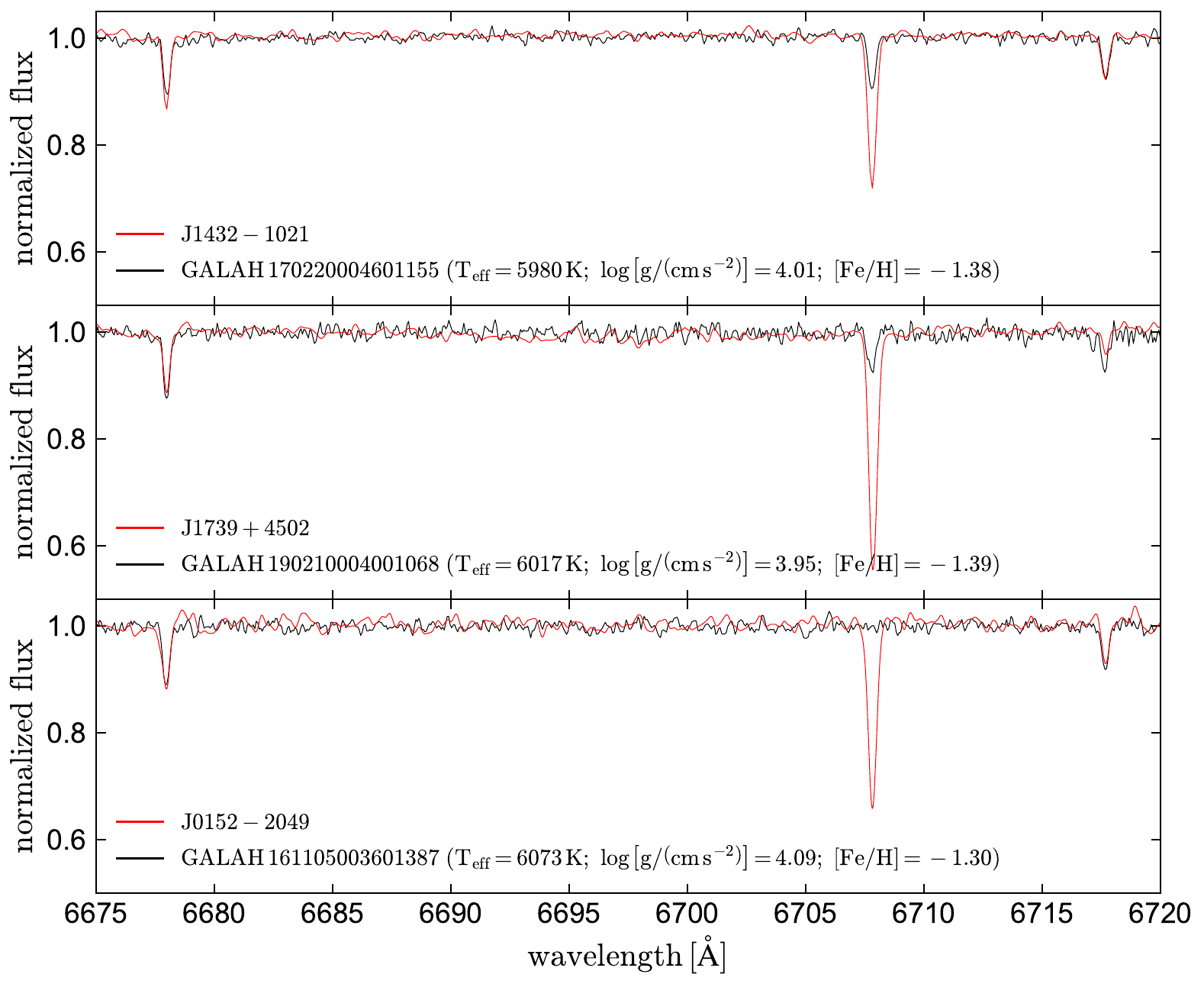}
    \caption{Evidence for lithium enhancement in the three halo stars. Red lines show the spectra of the three NS candidates. Black lines show spectra of three stars observed by the GALAH survey with similar physical parameters to each target and spectra that closely match those of the target in most lines. All three of our targets have unusually strong Li\,I\,$\lambda$6708 lines, indicative of lithium enhancement.}
    \label{fig:lithium}
\end{figure*}

Figure~\ref{fig:lithium} compares the spectra of these three stars to those of their closest spectral doppelgängers. The latter are identified using the wavelength range $5650-5850$\,\AA, which does not contain any lithium lines. The three NS candidates have Li I $\lambda$6708 lines that are clearly significantly stronger than those of the comparison stars shown in Figure~\ref{fig:lithium}, and indeed, stronger than those of {\it any} stars observed by GALAH with similar atmospheric parameters. 

\begin{figure}
    \centering
    \includegraphics[width=\columnwidth]{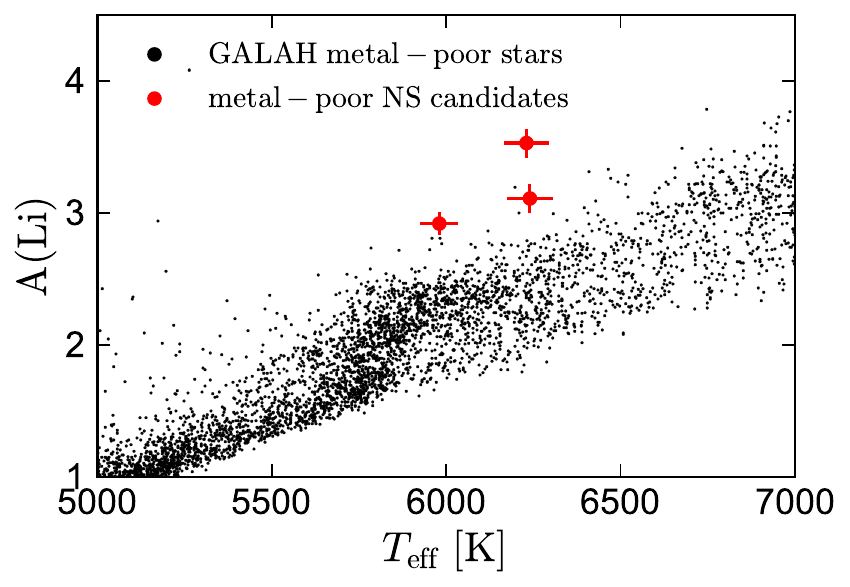}
    \caption{Lithium abundances of the three metal-poor halo stars in our sample (Figure~\ref{fig:lithium}), compared to metal-poor stars observed by the GALAH survey. All three stars are strongly enhanced in lithium, with $\rm A(Li)$ higher than any of the $\sim 10^3$ stars with similar stellar parameters observed by GALAH.}
    \label{fig:Li_abund}
\end{figure}

The sources J1432-1021, J1739+4502, and J0152-2049 have Li I $\lambda$6708 lines with equivalent widths of 114, 197, and 141 m\AA. Comparing these values to Kurucz model spectra calculated with ATLAS/SYNTHE \citep{Kurucz1979, Kurucz1993} and applying NLTE corrections from \citet{Wang2021}, we find Li abundances of $\rm A(Li) = 2.90\pm 0.08$, $\rm A(Li) = 3.53\pm 0.09$, and $\rm A(Li) = 3.11\pm 0.08$. These abundances represent more than a factor of 100 enhancement in surface lithium abundance compared to normal stars of similar temperature and metallicity (Figure~\ref{fig:Li_abund}).

The origin of the excess lithium is not yet understood. Possibilities include pollution by products of hot bottom burning in super-AGB stars, pollution by supernova ejecta, and spallation by high-energy particles \citep[see][for further discussion]{El-Badry2024}. Because metal-poor stars have thin convective envelopes, only their surface layers are expected to have been polluted. This is likely the reason we only find strong lithium enhancement in the metal-poor stars in our sample: the thickness of the convective envelope is more than $50\times$ greater at solar metallicity than at $[\rm Fe/H]=-1.5$, meaning that any accreted material will be diluted 50 times more at solar metallicity, and abundance anomalies will be much more subtle. Assuming lithium is mixed uniformly into the outer 0.2\% of the stars by mass, $(1-3)\times 10^{-11}\,M_{\odot}$ of lithium must have been accreted in order to explain the observed enhancement. 

Our inferred $\rm A(Li)$ are higher than the predicted yields of super-AGB stars with $M\lesssim 8\,M_{\odot}$, but compatible with predictions for $8\,M_{\odot}$ models (see \citealt{Ventura2010}, who note that these yields depend sensitively on the assumed winds). This could support a scenario where the companions are NSs formed through electron-capture SNe, or ultramassive WDs. However, this scenario is difficult to reconcile with the fact that one of the binaries in question has $M_2 = 1.90\pm 0.03\,M_{\odot}$, whereas electron capture SNe are expected to produce low-mass NSs. 

\begin{figure*}
    \centering   \includegraphics[width=\textwidth]{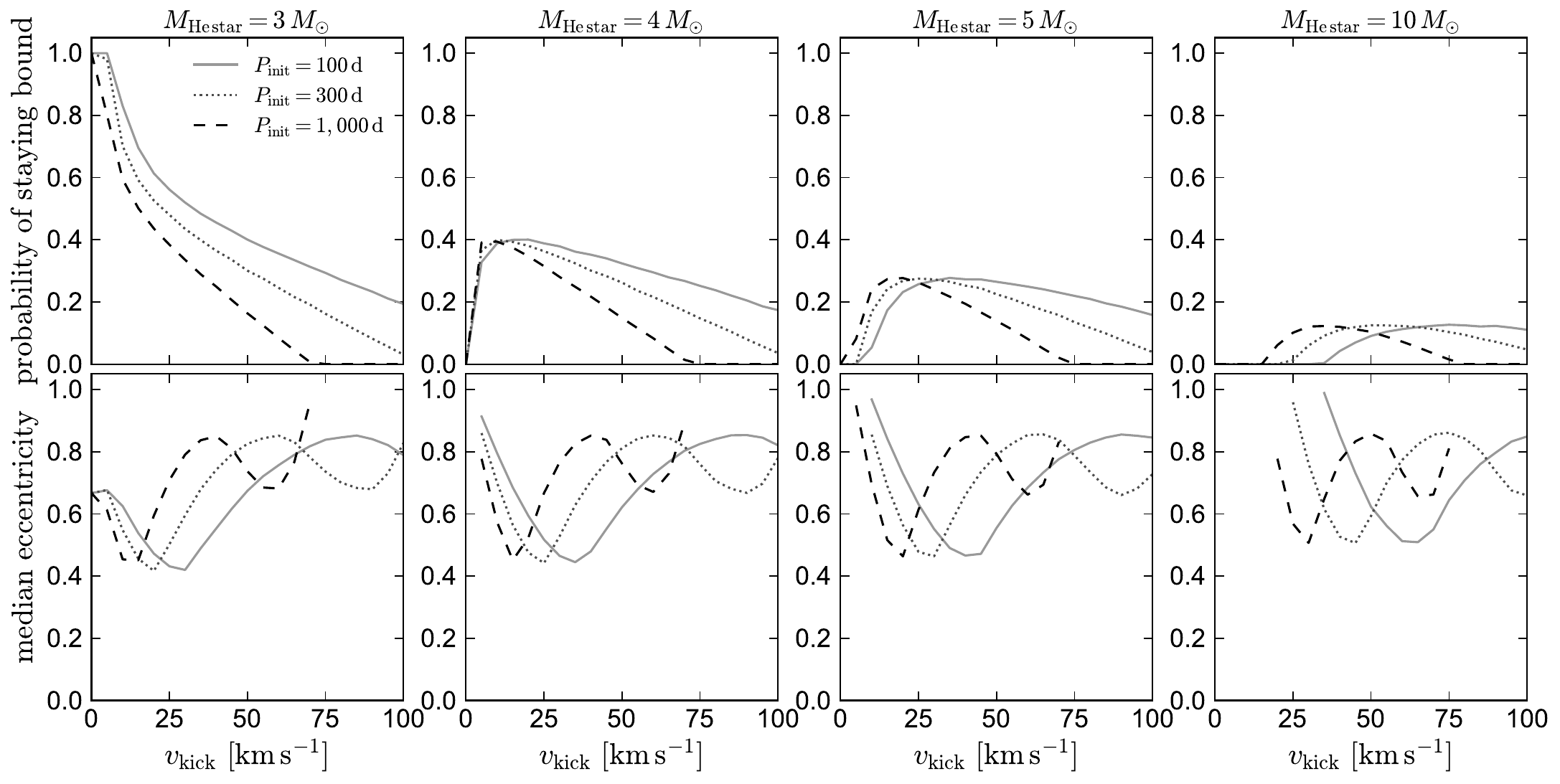}
    \caption{Survival probability (top) and median predicted eccentricity (bottom) for wide helium star + MS binaries when the He star explodes and leaves behind a NS. We assume the initial orbit is circular and consider initial helium star masses of 3, 4, and $5\,M_{\odot}$, and initial periods of 100, 300, and 1000 days. In all cases, we model the companion as a $1\,M_{\odot}$ MS star and adopt a final NS mass of $1.4\,M_{\odot}$. The NS is born with a velocity $v_{\rm kick}$ in a random direction. For a $3\,M_{\odot}$ He star, orbits are most likely to remain bound if the kick is slow. For $M_{\rm He\,star} > 3.8\,M_{\odot}$, the binary will be unbound by mass loss unless a fortuitously aligned kick allows the NS to catch the escaping companion. }
    \label{fig:kicks}
\end{figure*}

\subsection{Constraints on kicks and mass loss}
Models predict that dynamical channels are less efficient for forming wide NS binaries than for BH binaries \citep{Tanikawa2024}. Therefore, we assume the systems in our sample formed from primordial binaries and use their current orbits to place limits on their pre-SN masses and natal kicks.

We consider the evolutionary state of the binary that likely immediately preceded the current state: a circular orbit containing a stripped helium star of mass $M_{\rm He\,star}$ and a $1\,M_{\odot}$ MS companion.  Given that the orbits in our sample are too tight to accommodate red supergiants at their maximum radii of $\sim 1000\,R_{\odot}$, the binaries would likely have gone through a common envelope event prior to reaching this stage. Survival of such a common envelope event and a wide final orbit is in fact nontrivial \citep[e.g.][]{Kotko2024}, but here we consider only the effects of the SN.                                                                           

If there is no natal kick, the binary will be unbound if its total mass decreases by more than a factor of 2 during the SN \citep[e.g.][]{Blaauw1961}. The expected post-SN eccentricity in this case is 
\begin{equation}
    \label{eq:ecc}
    e= \frac{\Delta M}{M_{\rm NS}+M_{\star}},
\end{equation}
Where $M_{\rm NS}$ is the mass of the NS, $\Delta M=M_{\rm He\,star}-M_{\rm NS}$ is the mass lost during the explosion, and $M_\star$ is the mass of the luminous star. For typical targets in our sample, $e\approx 0.4$ and $M_{\star}=1\,M_{\odot}$. This would imply that only of order $1\,M_{\odot}$ was lost during the SN if natal kicks were weak.

Kicks complicate this analysis: it is possible for a well-aimed natal kick to keep a binary bound -- and even maintain a low eccentricity orbit -- with arbitrarily large mass loss during the SN. This, however, requires fine-tuning, and most orbits will simply be unbound if kicks are strong and mass loss is significant. Inferring the full distribution of kick velocities from observations of binaries like those in our sample is difficult, since NSs born with strong kicks will not make it into the sample in the first place. We can, however, still constrain the kicks and mass loss that likely occurred in the binaries that did survive.

We model the combined effects of natal kicks and instantaneous mass loss using Monte Carlo simulations. Following  \citet{Brandt1995}, we predict post-SN orbits for a range of $v_{\rm kick}$, $M_{\rm He\,star}$, and pre-SN orbital period. For each choice of these values, we generate $N=10^6$ kick directions distributed uniformly on a sphere and calculate the post-SN orbit via energy and angular momentum conservation. In all cases, we assume a $1\,M_{\odot}$ companion, a $1.4\,M_{\odot}$ NS, and an initially circular orbit.

The results are shown in Figure~\ref{fig:kicks}. The top panels show the fraction of all orbits that remain bound, and the bottom panels show the median eccentricity of those that do. The probability of remaining bound is highest for low $M_{\rm He\,star}$ and low $v_{\rm kick}$. However, a significant fraction of orbits remain bound even for $M_{\rm He\,star}$ of 5 or $10\,M_{\odot}$ when paired with a suitable kick velocity. Higher values of $M_{\rm He\,star}$ require stronger kicks to remain bound: for a given $M_{\rm He\,star}$, binaries are most likely to survive when $v_{\rm kick}$ is about half the pre-SN orbital velocity of the MS star. The median predicted post-SN eccentricity of surviving binaries is typically $0.4-0.8$. These values are slightly higher than the median eccentricity of 0.4 within our sample, but given that {\it Gaia} is less sensitive to high-eccentricity orbits, the observed eccentricity distribution appears consistent with originating mainly from kicks. 

Light curve modeling of stripped-envelope SNe suggests typical ejecta masses of $(1-3)\,M_{\odot}$ \citep[e.g.][]{Lyman2016}, which would correspond to He star masses of $\sim (3-5)\,M_{\odot}$ for the NSs in our sample. For kicks with $v_{\rm kick} \lesssim 50\,{\rm km\,s^{-1}}$, about 30\% of binaries forming from such He stars at periods typical of our sample would be expected to survive. Such kick velocities are quite low compared to typical values inferred from observations of young pulsars \citep[e.g.][]{Hobbs2005, Faucher2006}, but within the range required for NSs to be retained within globular clusters \citep{Ivanova2008}. Although only a minority of NSs formed in binaries with the periods and companions typical of our sample are likely to survive, the relatively low space density of wide NS+MS binaries suggested by our sample compared to the predicted space density of all dormant NSs\footnote{An accurate estimate of the space density of wide NS+MS binaries is presently difficult to calculate due to uncertainties in the {\it Gaia} selection function. However, the nearest NS candidate in our sample is at $d\approx 250$ pc. The can be compared to $d\approx 120$\,pc, the distance to the nearest known young NS \citep{Walter2010}, and $d\approx 20$\,pc, a predicted distance to the nearest NS based on the Galactic model of \citet{Sweeney2022}. These distances suggest that there are $\sim 10^3$ dormant NSs for every one in a binary like the objects in our sample. } allows for a scenario in which only the low-$v_{\rm kick}$ tail of the NS population survives in binaries.

Curiously, the most massive NS in the sample, Gaia NS1 (J1432-1021), has the lowest-eccentricity orbit, requiring the lowest kicks. This appears inconsistent with SNe simulations, which predict high-mass NSs to receive the strongest kicks \citep{Muller2019, Burrows2023}. It is possible that Gaia NS1 formed through a different channel than the other objects in the sample, but larger sample is required to investigate this possibility quantitatively. 

\subsection{Future evolution to symbiotic X-ray binaries}
When the MS stars in our sample leave the main sequence, they will begin transferring mass to the dark companions: first by winds, and eventually by Roche lobe overflow. \citet{Rodriguez2024} have calculated models representative of this evolution. Their calculations suggest that the binaries will be X-ray bright for at least $\sim 5$ Myr during Roche lobe overflow, and most likely for a few tens of Myr due to wind accretion at earlier times. Since the X-ray bright phase is predicted to be 100-1000 times shorter-lived than the current X-ray faint phase, one expects X-ray bright systems to be rare, but they are discoverable to large distances. 

At least a few neutron stars with solar-mass red giant companions are known to exist in symbiotic X-ray binaries \citep[e.g.][]{Hinkle2006, Hinkle2019}. One system that is likely representative of our candidates' future evolution is GX 1+4 \citep{Davidsen1977, Hinkle2006}, which contains a $\lesssim 1\,M_{\odot}$ giant in a 1161-day orbit at a distance of $\approx 4$\,kpc. Another similar system  is the symbiotic X-ray binary IGR J16194-2810 \citep{Masetti2007}, which contains a $\sim 1\,M_{\odot}$ red giant at a distance of $\approx 2.1$ kpc in a 193-day orbit with a NS companion \citep{Nagarajan2024, Hinkle2024}. Given their X-ray properties, these systems undoubtedly contain NSs. This provides some support for the interpretation that the dark objects in our binaries are also NSs, despite the uncertainties associated with forming NS+MS binaries in such wide orbits.  

\begin{figure*}
    \centering
    \includegraphics[width=\textwidth]{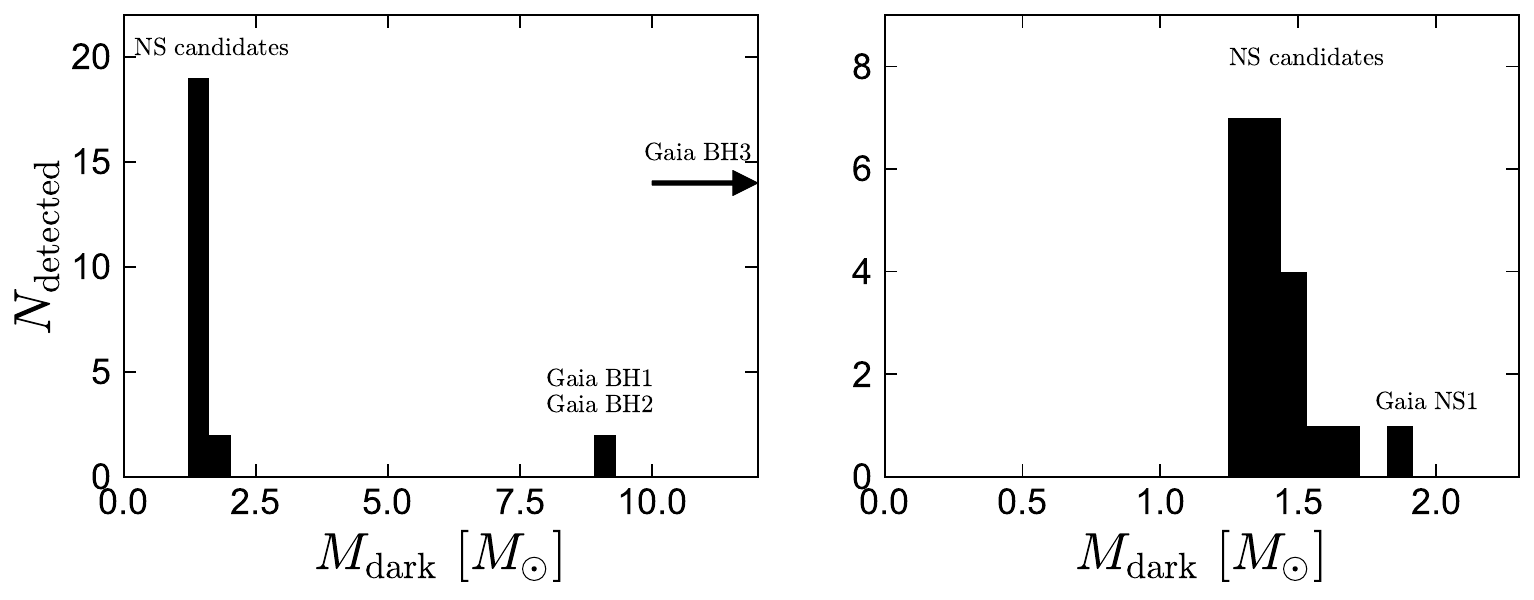}
    \caption{Mass distribution of dark companions revealed by follow-up of {\it Gaia} astrometric binaries. The 21 sources with masses $M_{\rm dark}=1.25-1.9\,M_{\odot}$ are presumed NSs introduced in this paper. The two objects with masses near $9\,M_{\odot}$ are presumed BHs \citep{El-Badry2023_bh1, El-Badry2023_bh2}. A third BH has significantly higher mass, $M_{\rm dark} \approx 33\,M_{\odot}$ \citep{Panuzzo2024}. No dark companions with mass between 2 and $8\,M_{\odot}$ have been detected, even though the effective search volume for such sources is larger than for the NSs. }
    \label{fig:mass_dist}
\end{figure*}

\subsection{Is there a BH/NS mass gap?}
\label{sec:mass_dist}

Figure~\ref{fig:mass_dist} shows the mass distribution of dark companions identified thus far from {\it Gaia} astrometry: the 21 NS candidates presented here, and three BHs studied previously \citep{El-Badry2023_bh1, El-Badry2023_bh2, Nagarajan2023, Panuzzo2024}. All the NS candidates have best-fit masses between 1.25 and $2.0\,M_{\odot}$, while the BHs have masses above $9\,M_{\odot}$. We have not detected any dark companions with intermediate masses, and our follow-up has been complete for sources published in DR3 with astrometric solutions implying $M_2 \gtrsim 2\,M_{\odot}$. The fact that we detected 21 dark companions with $M_2 < 2\,M_\odot$ -- which produce smaller astrometric wobbles than would low-mass BHs -- strongly suggests that our search would have detected $3-5\,M_{\odot}$ BHs if they existed in our search sample.

\begin{figure*}
    \centering
    \includegraphics[width=\textwidth]{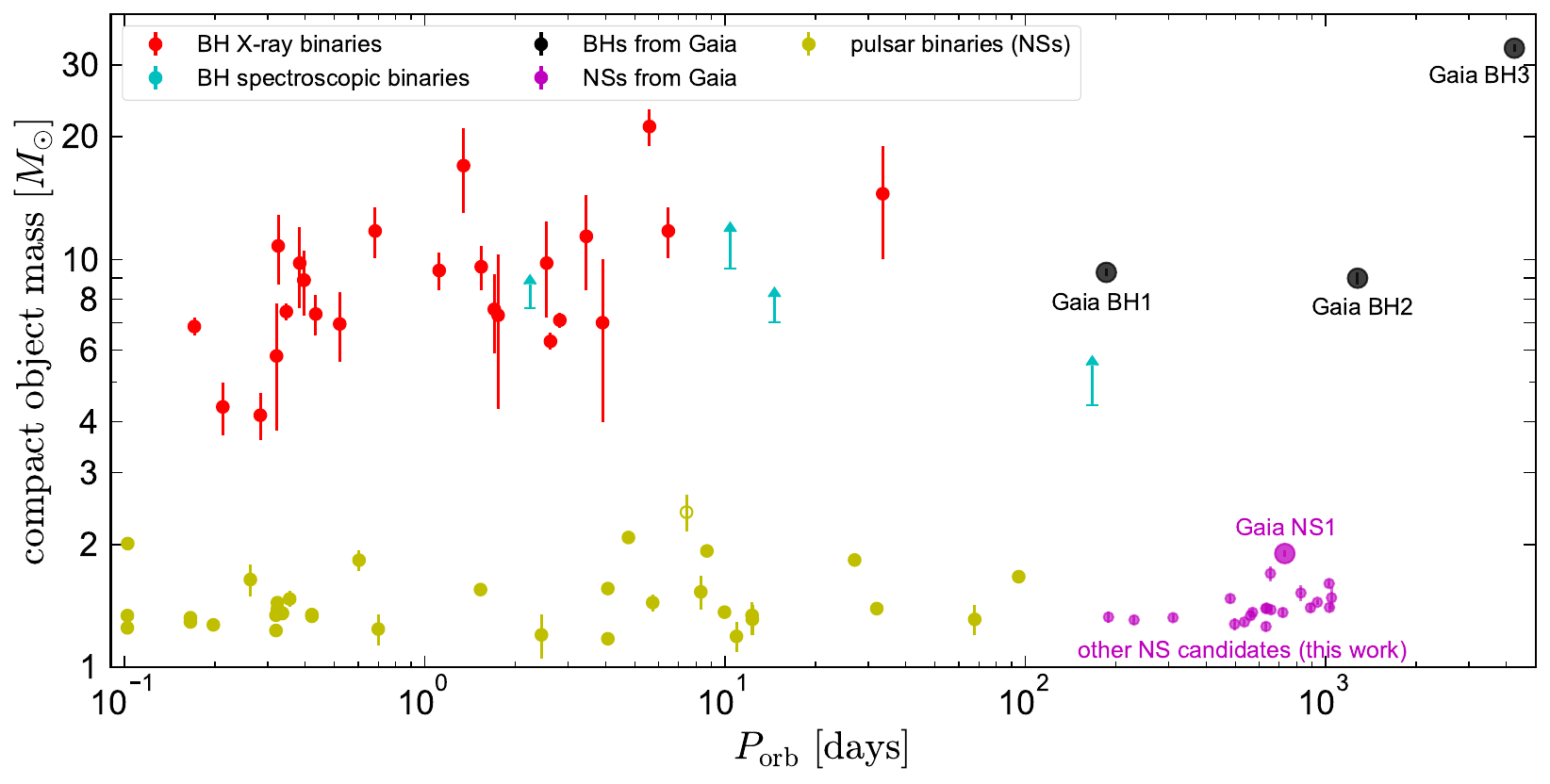}
    \caption{Masses and orbital period of Galactic BHs and NSs in binaries with well-constrained masses. The candidates presented here are shown in magenta and are compared to NSs in pulsar binaries (yellow), BHs in astrometric binaries (black), BHs in X-ray binaries (red), and BHs in spectroscopic binaries (cyan). A mass bimodality exists over at least 4 orders of magnitude in orbital period.}
    \label{fig:period_mass}
\end{figure*}

Figure~\ref{fig:period_mass} compares the masses and orbital periods of BH and NS binaries identified with {\it Gaia} astrometry to BHs and NSs found with other methods. Yellow points show NSs in radio pulsars binaries \citep{Ozel2016, Fonseca2021}. The hollow yellow symbol shows the companion to  PSR J0514-4002E, whose nature is uncertain \citep{Barr2024}. Red points show high- and low-mass BH X-ray binaries \citep{Remillard2006, Corral-Santana2016, MillerJones_2021}. Cyan lower limits show spectroscopic BH binary candidates \citep{Giesers2019, Shenar2022, Mahy2022}. The BH and NS samples are both far from complete. Nevertheless, Figure~\ref{fig:period_mass} shows rather unambiguously that the BH/NS mass distribution is bimodal over at least 4 orders of magnitude in orbital period. 

Our results thus support the presence of mass gap between NSs and BHs, similar to the gap reported for X-ray binaries \citep{Bailyn1998, Kreidberg2012} and gravitational wave sources \citep{Abbott2023}. The existence of a mass bimodality does not depend much on the nature and exact mass of objects proposed to be in the mass gap \citep[e.g.][]{Casares2022, Barr2024}, as such objects are rare, at least in binaries.  Of course, it is possible that the mass distribution of BHs in binaries is different from the mass distribution of all BHs, since (for example) low-mass BHs could experience stronger kicks and more frequently be unbound \citep{Burrows2023}.

\section{Conclusions}
\label{sec:conclusions}

The first set of orbital solutions from the {\it Gaia} mission led to the identification of more than 50 candidate neutron star (NS) + main sequence (MS) binaries in au-scale orbits. We are carrying out a spectroscopic follow-up program that yields radial velocities (RVs) and stellar parameters for these binaries. We have now covered a majority of an orbital period for most of these candidates with RVs. This allows us to tighten constraints on orbits through joint fitting of RVs and astrometry to and root out spurious astrometric orbits when the predictions of the {\it Gaia} solution do not match RVs (Appendix~\ref{sec:spurious}). Our main results are as follows: 

\begin{enumerate}
    \item {\it Summary of the sample}: We have constructed a sample of 21 candidate NS + MS binaries containing solar-type MS stars and dark companions with best-fit masses of $1.25-1.90\,M_{\odot}$ (Figure~\ref{fig:cmd}). Most systems have periods of 400 to 1000 days and distances of 0.4 to 1 kpc. We select candidates as those with photocenter wobbles too large to be explained by a normal luminous companion or an unresolved inner binary containing two luminous stars. At fixed period and luminous star mass $M_\star$, NS companions produce photocenter orbits that are larger than WDs and non-degenerate companions can produce, but smaller than those produced by BH companions (Figure~\ref{fig:keplers_law}).

    \item {\it Quality of the {\it Gaia} orbital solutions}: RVs for all the objects in our final sample are in good agreement with predictions of the {\it Gaia} orbital solutions (Figure~\ref{fig:cyan}), and joint fitting of RVs and astrometry leads to well-constrained orbits (Figure~\ref{fig:black}). For most objects, constraints from the {\it Gaia}-only and {\it Gaia}+RVs solutions are consistent within $1\sigma$ (Figure~\ref{fig:comparisons}). However, some initial candidates were removed from the sample after RV follow-up revealed tensions with the {\it Gaia} solution (Figure~\ref{fig:bad_rvs} and Table~\ref{tab:all_cands}). Most but not all of these objects have large \texttt{goodness\_of\_fit}, indicative of a problematic astrometric solution. On average, companion masses constrained by joint RV+astrometry fits are lower than the pure-astrometry estimates (Figure~\ref{fig:mass_functions}); this is an expected consequence of selecting candidates from the upper tail of the companion mass distribution. The objects all have unproblematic {\it Gaia} quality flags and astrometric uncertainties typical of their apparent magnitudes (Figure~\ref{fig:gof}). Most of the orbits are predicted to have been sampled uniformly in phase by {\it Gaia} (Figure~\ref{fig:gost}). 
    \item {\it Eccentricities}: Most of the binaries in our sample have fairly eccentric orbits, with a median eccentricity of 0.4. Their eccentricities are significantly higher than those of typical WD companions at similar periods (Figure~\ref{fig:period_ecc}), and are consistent with eccentricities expected due to kicks during NS formation (Figure~\ref{fig:kicks}). However, high-mass WDs ($M\gtrsim 1\,M_{\odot}$) {\it also} have higher eccentricities than $\sim 0.6\,M_\odot$ WDs, so the eccentricities of our candidates do not guarantee that they are NSs.
    \item {\it Metallicities and chemical abundances}: We measured metallicities for the MS stars from high-resolution spectra (Figure~\ref{fig:metallicities}). A majority of them have metallicities near solar ($-0.5<\rm [Fe/H] < 0.5$), but low metallicity halo stars are over-represented. Three out of 21 targets (14\%) have $\rm [Fe/H]\sim -1.5$ and space velocities of $\sim 300\,{\rm km\,s^{-1}}$ (Figure~\ref{fig:toomre}). Only 0.5\% of all binaries with {\it Gaia} astrometric solutions at comparable distance are on halo orbits. This suggests that low-metallicity massive stars are more likely to form NSs in wide orbits with low-mass companions. 

    All three low-metallicity stars are strongly enhanced in lithium (Figure~\ref{fig:lithium} and~\ref{fig:Li_abund}). Low-metallicity stars have much thinner convective envelopes than solar-metallicities stars of the same mass and evolutionary state, so accreted material will be diluted less and abundance anomalies will be more detectable at low metallicity. The origin of the lithium is unclear: one possibility is accretion of Li-rich winds from super-AGB stars \citep{Cameron1971, Ventura2010}. However, at least one the Li-enhanced objects has a companion with $M_2=1.90\pm0.03\,M_{\odot}$, which is considerably higher than the expected mass of high-mass WDs or NSs formed from electron capture SNe. 

    \item {\it Nature of the companions}: We have not detected radiation from the companions, but constraints on their masses and orbits can narrow down the possibilities. Joint fitting of astrometry and RVs places tight upper limits on the flux ratio, with a median 2$\sigma$ upper limit of $\epsilon < 0.03$. This rules out all plausible nondegenerate companions (Figure~\ref{fig:scenarios}). Several different possibilities remain, including single NSs, single massive WDs, and tight WD+WD, WD+NS, or WD+MS binaries. For most of the companions to be single massive WDs, their masses would have to be overestimated at the few-$\sigma$ level. Scenarios involving inner binaries are difficult to explain with evolutionary models -- triple systems tend to become unstable during their evolution when the outer orbit is as tight as our candidates -- but should not be dismissed entirely for want of imagination. It is quite possible that the sample contains more than one kind of dark companion. 

    \item {\it BH/NS mass distribution}: The {\it Gaia} mission has now enabled astrometric discovery of three BHs, 21 candidate NSs, and thousands of WDs in astrometric binaries. The mass distribution of BHs and NS candidates is shown in Figure~\ref{fig:mass_dist}. There is a conspicuous gap in the mass distribution between $2$ and $8\,M_{\odot}$. This may be in part a result of small number statistics, but the distribution does seem to disfavor a population of lower-mass BHs that significantly outnumber higher-mass BHs. Since the astrometric search volume is larger for higher-mass objects, our sample suggests that low-mass BHs are significantly less common in astrometric binaries than are NSs. When combined with samples of BHs and NSs discovered with other methods, our sample reveals a clear mass bimodality that extends over 4 orders of magnitude in orbital period (Figure~\ref{fig:period_mass}).
\end{enumerate}

It will likely remain difficult to conclusively establish the nature of the companions for some time. Radio detection of the companions as pulsars could prove that they are NSs, and efforts are underway to search for radio pulsations from most of our candidates. However, radio detection of any individual candidate seems a priori unlikely: young NSs are only detectable as radio pulsars for $\lesssim 10$ Myr -- only $\sim 0.1\%$ of the expected lifetime of these binaries -- and in the simplest evolutionary scenarios for their formation, there is no reason to expect the NSs to be recycled. Given the wide separations of the binaries and weak winds of the main-sequence companions, an X-ray detection due to accretion is also not expected \citep[e.g.][]{Rodriguez2024}. WD companions are not expected to be detectable at any wavelength unless they are fortuitously young. Models involving inner binaries can be tested with high-precision RV observations \citep{Nagarajan2023}, but only if the inner binary has a period longer than a few days.

We briefly discuss two possible avenues for determining the nature of the companions in the absence of a direct detection. First, higher-precision astrometric orbits from future {\it Gaia} data releases will tighten mass constraints. The companion mass uncertainties are in most cases limited by the astrometric inclination uncertainties, and more precise measurements could more firmly rule out the possibility that the companions are ultramassive WDs. Future {\it Gaia} releases will also include epoch astrometry for all these sources, which will allow for further checks of the consistency between the astrometry and RV data. 

Second, ongoing RV follow-up of astrometric binaries containing WDs will provide a more complete mapping of the period -- eccentricity relation for post-interaction binaries containing WDs. This relation is particularly uncertain for massive WDs in wide orbits, making it difficult to distinguish between WDs and NSs on the basis of their eccentricities. Features in the relation near the WD--NS transition mass may make it possible to statistically differentiate between WD and NS scenarios, even if the nature of companions in individual systems remain uncertain.

\section*{acknowledgments}
We thank the referee for a constructive report, and Josh Simon, Casey Lam, Kyle Kremer, Jim Fuller, and Thomas Tauris for useful discussions. We are grateful to Yuri Beletsky, Sam Kim, Angela Hempel, Régis Lachaume, Gil Esquerdo, Perry Berlind, and Mike Calkins for observing help. This research was supported by NSF grant AST-2307232. HWR acknowledges support from the European Research Council for the ERC Advanced Grant [101054731]. This research benefited from discussions in the ZTF Theory Network, funded in part by the Gordon and Betty Moore Foundation through Grant GBMF5076, and from collaboration at the ``Renaissance of Stellar Black-Hole Detections in The Local Group'' workshop hosted at the Lorentz Center in June, 2023.

This research made use of pystrometry, an open source Python package for astrometry timeseries analysis \citep[][]{Sahlmann2019}. This work made use of Astropy,\footnote{http://www.astropy.org} a community-developed core Python package and an ecosystem of tools and resources for astronomy \citep{AstropyCollaboration2022}.

This work has made use of data from the European Space Agency (ESA) mission
{\it Gaia} (\url{https://www.cosmos.esa.int/gaia}), processed by the {\it Gaia}
Data Processing and Analysis Consortium (DPAC,
\url{https://www.cosmos.esa.int/web/gaia/dpac/consortium}). Funding for the DPAC
has been provided by national institutions, in particular the institutions
participating in the {\it Gaia} Multilateral Agreement.

This paper includes data gathered with the 6.5 meter Magellan Telescopes located at Las Campanas Observatory, Chile. Some of the data presented herein were obtained at Keck Observatory, which is a private 501(c)3 non-profit organization operated as a scientific partnership among the California Institute of Technology, the University of California, and the National Aeronautics and Space Administration. The Observatory was made possible by the generous financial support of the W. M. Keck Foundation. The authors wish to recognize and acknowledge the very significant cultural role and reverence that the summit of Maunakea has always had within the Native Hawaiian community. We are most fortunate to have the opportunity to conduct observations from this mountain.

\newpage
% The text '-' below is added so that we can see
% the proper placement of the figures.
% ***********************************
% ***********************************
% ***********************************
% ------------------------------------------------
% ------------------------------------------------
% APPENDIX %
% ------------------------------------------------
% ------------------------------------------------

\newpage
\bibliography{manuscript}
%\bibliographystyle{mnras}

%\begin{multicols}{2}  % Use the multicols environment for the appendix

\clearpage

\appendix

\section{Rejection of spurious solutions and summary of all follow-up}
\label{sec:spurious}

\begin{figure}
    \centering
    \includegraphics[width=\columnwidth]{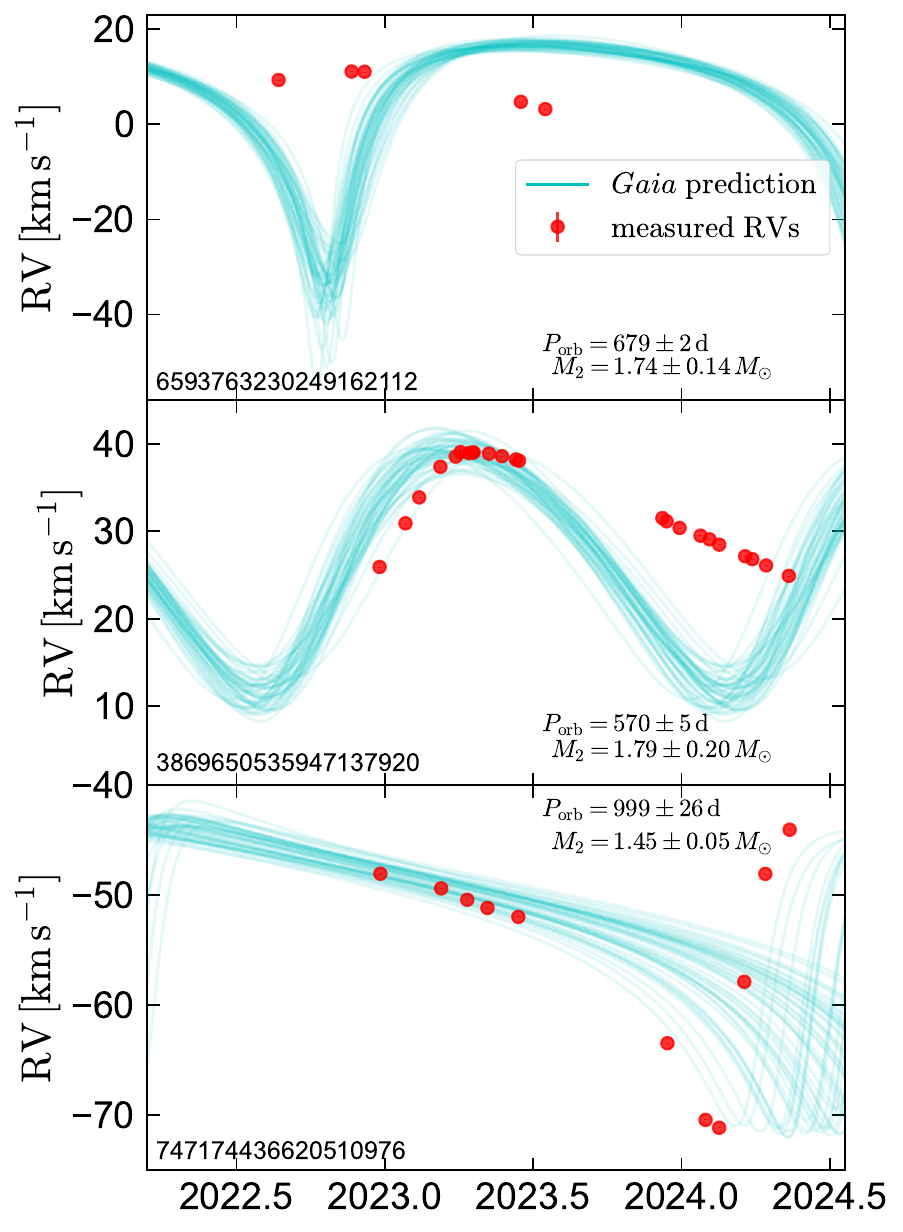}
    \caption{Examples of NS+MS candidates we rejected after RV follow-up. In the top panel, the RVs are grossly inconsistent with the predictions of the {\it Gaia} solution, suggesting that the solution is spurious, or at least has a large accumulated phase error not captured in the {\it Gaia} uncertainties. Middle panel shows an example in which the RVs evolve in a manner somewhat consistent with the {\it Gaia} solution prediction, but do not match it quantitatively. This indicates that the uncertainties of the astrometric parameters are likely underestimated, casting doubt on the companion mass constraint. Bottom panel shows a marginal case: the RVs qualitatively track the astrometric solutions predictions, but a phase offset suggests that the period uncertainty is underestimated. We exclude all these objects -- even though they may indeed host NSs or massive WDs -- retaining only sources with good agreement between astrometry and RVs. }
    \label{fig:bad_rvs}
\end{figure}

For some candidates, our RV follow-up showed the {\it Gaia} astrometric solution to be spurious or to have significantly underestimated uncertainties. Measured and predicted RVs for three such sources are shown in Figure~\ref{fig:bad_rvs}. The source shown in the top panel, with DR3 source id 6593763230249162112, was characterized as a high-quality NS candidate by both \citet{Andrews2022} and \citet{Shahaf2023}. The {\it Gaia} orbital solution has  \texttt{goodness\_of\_fit} = 1.67, indicative of an unproblematic astrometric solution. Its DR3 astrometric solution is based on 20 visibility periods, which is typical for solutions published in DR3. There does not appear to be any reason to mistrust the solution based on the quality flags published in DR3. Yet, comparison of the observed and predicted RVs leaves little doubt that the astrometric solution is seriously in error. 

The same is true for source 3869650535947137920, which is shown in the middle panel of Figure~\ref{fig:bad_rvs} and is included in the sample from \citet{Shahaf2023}. The source's \texttt{goodness\_of\_fit} is 5.46, which is quite normal for a source with $G<13$ (Figure~\ref{fig:gof}). The first several months of our follow-up showed the RVs to be evolving in a manner consistent with the astrometric solution's predictions. However, the RVs began to diverge from those predictions in the second season of our follow-up. Although the qualitatively similar shape of the predicted and observed RV curves suggests the astrometric solution is not completely spurious, the observed degree of disagreement suggests that the astrometric uncertainties are significantly underestimated. 

Finally, source 747174436620510976, shown in the bottom panel of Figure~\ref{fig:bad_rvs}, is an example of a case where the shape of the observed RV curve is consistent with the astrometric predictions, but there is a few-$\sigma$ offset in phase. Our analysis of the source's SED also suggested that the parallax is overestimated, leading us to exclude the source (Section~\ref{sec:seds}). The {\it Gaia} solution has  \texttt{goodness\_of\_fit} $= -0.96$, indicative of a good solution, and was included in both the \citet{Shahaf2023} and \citet{Andrews2022} candidate lists. 

These examples highlight the importance of RV follow-up: while the sources shown in Figure~\ref{fig:bad_rvs} clearly are binaries, and least two of the three likely have orbital parameters not too far from those inferred from astrometry alone, their astrometric uncertainties are unlikely to be reliable. Since most NSs have masses near the Chandrasekhar limit, small problems with the astrometric solution can seriously change our conclusions about the nature of a candidate. Epoch astrometry from {\it Gaia} DR4 may allow us to obtain more useful astrometric constraints from sources like those shown in Figure~\ref{fig:bad_rvs}.

In Table~\ref{tab:all_cands}, we provide a summary of our follow-up of all NS candidates from \citet{Shahaf2023} and \citet{Andrews2022}. For the \citet{Shahaf2023} sample, we list candidates for which they inferred $M_2 > 1.25\,M_{\odot}$. The \citet{Andrews2022} sample only includes targets for which they infer $M_2 > 1.4\,M_{\odot}$. A majority of the candidates ruled out by RV follow-up have significantly higher \texttt{goodness\_of\_fit} than typical sources of the same apparent magnitude (Figure~\ref{fig:gof}). Several candidates with incomplete RV follow-up, particularly those with $G<15$ that are inaccessible with FEROS and TRES, remain to be studied.

\begin{table*}
\begin{tabular}{lllllllll}
{\it Gaia} DR3 Source ID & Reference & $P_{\rm orb}$  & $G$ & GoF & status \\
 &  & [d]  & [mag]  &    \\
\hline
1522897482203494784 & S23 &  $45.516 \pm 0.005 $ & 11.05 &  1.85 & \blue{likely ultramassive WD \citep{Yamaguchi2023} }\\
3509370326763016704 & S23 &  $109.4 \pm 0.1 $ & 12.47 &  5.22 & \red{ruled out by RV follow-up} \\
6281177228434199296 & S23 &  $153.9 \pm 0.4 $ & 11.26 &  8.05 & \red{ruled out by RV follow-up} \\
4482912934572480384 & S23 &  $182.4 \pm 0.4 $ & 12.35 &  20.26 &  \red{ruled out by RV follow-up} \\
2995961897685517312 & S23 &  $189.8 \pm 0.3 $ & 13.00 &  0.41 & \green{one of our candidates} \\
2080945469200565248 & S23 &  $196.6 \pm 0.5 $ & 13.29 &  -1.25 & \red{RV follow-up implies $e < 0.001$. Likely ultramassive WD.} \\
2032579979951732736 & S23 &  $215.4 \pm 0.4 $ & 14.21 &  -1.04 & high $v\sin i $ prevented robust RV measurements \\
6481502062263141504 & S23 &  $229.7 \pm 0.5 $ & 13.58 &  2.11 & \green{one of our candidates}  \\
5820382041374661888 & S23 &  $311.1 \pm 1.4 $ & 14.19 &  -2.49 & \green{one of our candidates} \\
1525829295599805184 & A22 &  $328.2 \pm 0.8 $ & 16.18 &  2.25 & fainter than $G=15$ \\
1871419337958702720 & S23&  $479.3 \pm 1.0 $ & 13.70 &  -0.71 & \green{one of our candidates} \\
5530442371304582912 & S23&  $498.8 \pm 2.8 $ & 14.60 &  -2.04 & \green{one of our candidates} \\
3263804373319076480 & S23&  $510.7 \pm 4.7 $ & 12.67 &  5.56 & \red{ruled out by RV follow-up} \\
6601396177408279040 & S23&  $533.5 \pm 2.0 $ & 14.07 &  13.80 & \red{not followed-up due to poor GoF}\\
5136025521527939072 & S23&  $536.9 \pm 1.2 $ & 12.05 &  4.68 & \green{one of our candidates} \\
2912474227443068544 & S23&  $541.5 \pm 3.9 $ & 13.81 &  24.83 & \red{ruled out by RV follow-up} \\
4271998639836225920 & A22&  $545.3 \pm 1.1 $ & 15.62 &  0.12 & fainter than $G=15$ \\
4922744974687373440 & S23&  $562.4 \pm 2.0 $ & 14.48 &  1.23 &  \green{one of our candidates} \\
6001459821083925120 & A22&  $563.8 \pm 6.7 $ & 13.60 &  -1.54 & \red{ruled out by RV follow-up} \\
3869650535947137920 & S23&  $570.0 \pm 5.3 $ & 12.94 &  5.46 &  \red{ruled out by RV follow-up (Figure~\ref{fig:bad_rvs})} \\
1434445448240677376 & S23&  $572.4 \pm 1.8 $ & 13.65 &  -2.97 & \green{one of our candidates} \\
6802561484797464832 & S23&  $574.8 \pm 6.2 $ & 12.88 &  11.01 & \red{not followed-up due to poor GoF} \\
2196619383835483648 & S23&  $582.2 \pm 5.0 $ & 15.71 &  1.53 &  fainter than $G=15$ \\
1695294922548180224 & A22, S23&  $601.2 \pm 2.9 $ & 13.12 &  -2.74 & RV follow-up not yet completed \\ 
4744087975990080896 & A22, S23 &  $631.3 \pm 5.5 $ & 17.07 &  -0.10 &  fainter than $G=15$ \\
1694708646628402048 & S23 &  $632.0 \pm 2.8 $ & 13.20 &  0.77 & \green{one of our candidates} \\
3494029910469026432 & S23&  $632.5 \pm 1.6 $ & 12.66 &  2.64 & \green{one of our candidates} \\
4637171465304969216 & S23&  $639.2 \pm 4.4 $ & 14.01 &  0.77 & \green{one of our candidates} \\
5580526947012630912 & A22, S23&  $654.3 \pm 4.9 $ & 13.36 &  0.66 & \green{one of our candidates} \\
1350295047363872512 & A22, S23&  $657.2 \pm 4.3 $ & 13.52 &  3.53 & \green{one of our candidates} \\
6593763230249162112 & A22&  $679.9 \pm 2.8 $ & 13.54 &  1.67 & \red{ruled out by RV follow-up (Figure~\ref{fig:bad_rvs})} \\
4240540718818313984 & S23&  $691.2 \pm 2.0 $ & 14.61 &  0.41 &  RV follow-up not yet completed \\ 
2426116249713980416 & S23&  $711.7 \pm 21.5 $ & 13.02 &  -0.30 & \green{one of our candidates} \\
4578398926673187328 & S23&  $728.1 \pm 22.9 $ & 15.02 &  0.05 &  fainter than $G=15$ \\
6328149636482597888 & A22, S23&  $736.0 \pm 11.6 $ & 13.34 &  0.74 & \green{one of our candidates} \\
2885872059004028800 & S23&  $802.2 \pm 28.1 $ & 15.31 &  1.93 & fainter than $G=15$ \\
6037767138131854592 & S23&  $804.7 \pm 29.2 $ & 14.28 &  -0.91 & \red{RV follow-up not yet completed} \\ 
5590962927271507712 & A22, S23&  $817.9 \pm 5.1 $ & 15.88 &  1.86 & fainter than $G=15$ \\
1058875159778407808 & A22, S23&  $835.7 \pm 29.3 $ & 14.52 &  -0.01 & \green{one of our candidates} \\
5446310318525312768 & S23&  $866.6 \pm 10.6 $ & 10.37 &  4.27 &  \red{RV follow-up not yet completed} \\ 
6092954989675820416 & S23&  $883.1 \pm 101.5 $ & 13.67 &  0.15 & \red{ruled out by RV follow-up}  \\
3649963989549165440 & A22&  $892.5 \pm 60.2 $ & 14.30 &  0.52 & \blue{sdB + NS or WD binary \citep{Geier2023}} \\
1801110822095134848 & S23&  $893.6 \pm 2.9 $ & 12.19 &  3.58 & \green{one of our candidates} \\
4638295715945158144 & S23&  $915.4 \pm 5.8 $ & 12.93 &  7.68 & RV follow-up not yet completed \\ 
2397135910639986304 & A22, S23&  $916.0 \pm 19.1 $ & 13.35 &  2.10 & \green{one of our candidates} \\
809741149368202752 & A22, S23&  $922.4 \pm 50.5 $ & 14.91 &  1.19 & RV follow-up not yet completed \\ 
1581117310088807552 & A22, S23&  $927.3 \pm 5.7 $ & 14.51 &  2.88 & \green{one of our candidates} \\
1749013354127453696 & A22&  $932.1 \pm 77.5 $ & 14.49 &  -0.34 & \red{ruled out by RV follow-up}  \\
6588211521163024640 & S23&  $943.3 \pm 44.9 $ & 14.19 &  4.92 & \red{ruled out by RV follow-up } \\
5681911574178198400 & A22, S23&  $943.8 \pm 31.1 $ & 15.64 &  -0.41 & fainter than $G=15$ \\
5693240254808387584 & S23&  $944.4 \pm 85.1 $ & 16.40 &  -1.39 & fainter than $G=15$ \\
1028887114002082432 & S23&  $994.4 \pm 40.9 $ & 13.59 &  -0.42 & \green{one of our candidates} \\
747174436620510976 & A22, S23&  $999.4 \pm 26.5 $ & 13.99 &  -0.96 & \red{ruled out by RV follow-up (Figure~\ref{fig:bad_rvs})} \\
5593444799901901696 & A22&  $1038.8 \pm 146.0 $ & 14.42 &  0.25 & high $v\sin i $ prevented robust RV measurements \\
465093354131112960 & this work &  $1046.3 \pm 45.5 $ & 13.1 &  -0.86 & \green{one of our candidates} \\
4314242838679237120 & A22&  $1146.0 \pm 191.1 $ & 17.02 &  -0.78 & fainter than $G=15$ \\
1947292821452944896 & A22&  $1245.6 \pm 163.5 $ & 15.94 &  -0.49 & fainter than $G=15$ \\
5847919241396757888 & A22&  $1254.3 \pm 145.2 $ & 16.90 &  1.51 & fainter than $G=15$ \\
1144019690966028928 & A22&  $1401.9 \pm 61.7 $ & 13.57 &  -1.65 & RV follow-up not yet completed \\  
1854241667792418304 & A22&  $1430.3 \pm 33.2 $ & 14.87 &  1.38 & RV follow-up not yet completed \\ 
\end{tabular}
\caption{All the NS candidates from 
\citet[][A22]{Andrews2022} and \citet[][S23]{Shahaf2023}. We list the orbital period according to the {\it Gaia} astrometric solution, the apparent magnitude and \texttt{goodness\_of\_fit}, and the status of our follow-up. Candidates for which the status is colored in green are modeled in detail in this work. Those for which status is colored in red are disfavored by RV follow-up. Blue text indicates that detailed follow-up is presented elsewhere, and black text indicates that insufficient RVs have been obtained for a verdict to be reached. }
\label{tab:all_cands}
\end{table*}

\section{Blue excess due to WD companions}
\label{sec:appendix_color_excess}

As can be seen in Figure~\ref{fig:cmd}, several of our candidates are at the blue edge of the main sequence. Compared to stars near the middle of the main sequence, this amounts to a blue color excess of 0.05 to 0.1 mag. Most of our candidates have also been observed by {\it GALEX}, with NUV detections that rule out significant ($\gtrsim 0.2$ mag) UV excess. Here we investigate whether the optical blue excess could be due to a WD companion. 

\begin{figure}
    \centering
    \includegraphics[width=\columnwidth]{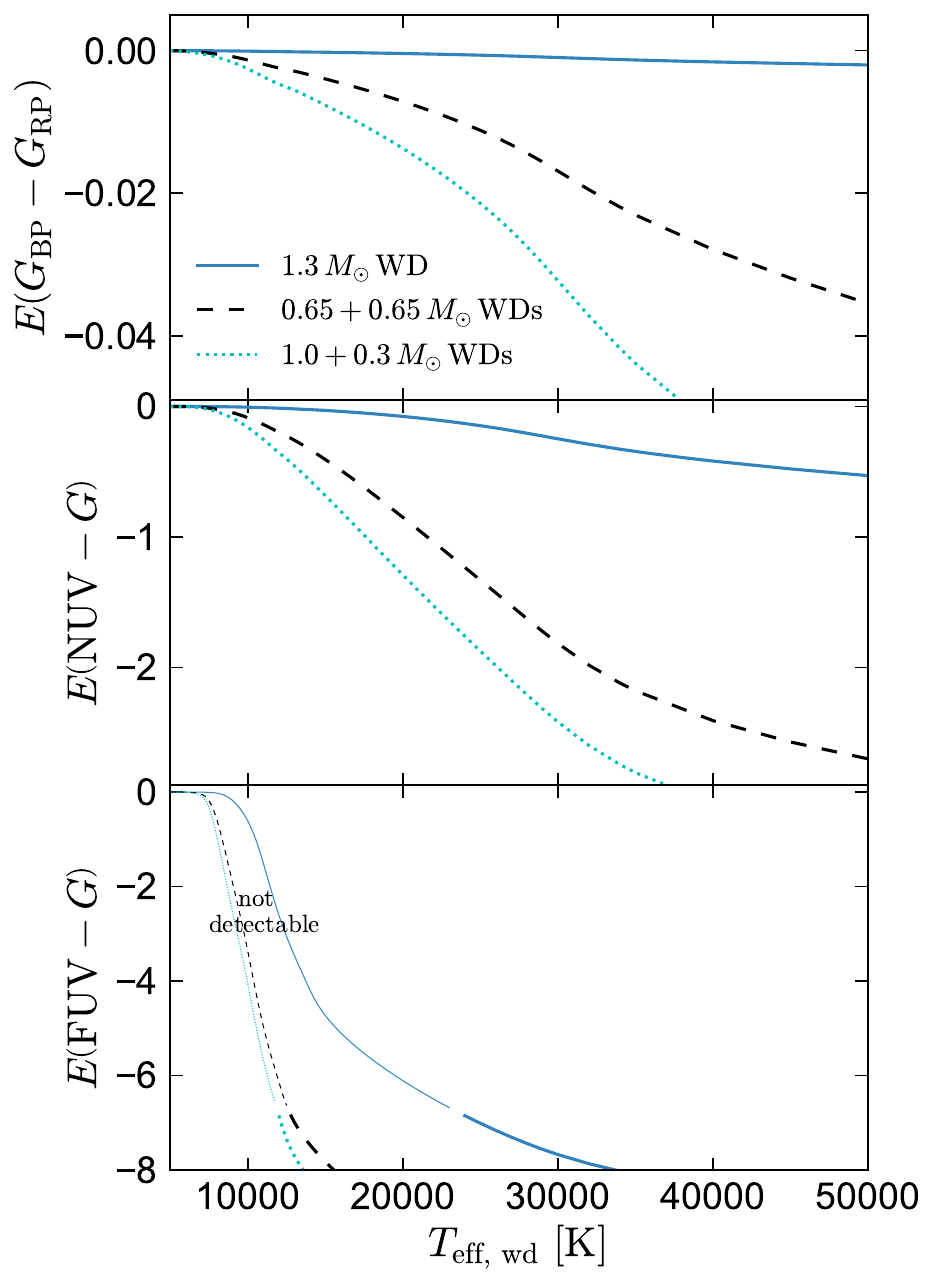}
    \caption{Predicted color excess due to a WD companion. In all cases, we assume a $1\,M_{\odot}$ solar-type primary with a $1.3\,M_{\odot}$ companion, representative of the objects in our sample. Solid blue lines show the case where the companion is a single WD, while dashed black line shows an equal-mass WD+WD binary and dotted cyan lines show a $(1.0+0.3)\,M_{\odot}$ WD binary. Top panel shows the optical blue excess in the {\it Gaia} $G_{\rm BP}-G_{\rm RP}$ bands, while the middle and bottom panels show the UV excess in the GALEX NUV and FUV bands. Lighter lines in the bottom panel show cases where the system would be fainter than 20.0 mag in the FUV band at a distance of 500 pc and would likely not be detected. WD+WD binaries lead to much more blue and UV excess than single WD companions of the same mass. Even WD companions that are quite hot produce negligible blue excess in the optical. }
    \label{fig:blue_excess}
\end{figure}

We first consider a $1\,M_{\odot}$ MS star in a binary with a single $1.3\,M_{\odot}$ WD companion. We construct the combined spectral energy distribution of the pair, modeling the WD with a Koester DA model with $\log (g/{\rm cm\,s^{-2}})=9.0$ \citep{Tremblay2009ApJ, Koester2010MmSAI} and the MS star with the same models used in Section~\ref{sec:seds}. We then calculate the combined total magnitude of both objects using \texttt{pyphot}. The solid lines in Figure~\ref{fig:blue_excess} show the resulting color excess as a function of WD effective temperature. We define the color excess in each band as the color of the combined MS + WD pair minus the color of the MS star alone. 

Because a $1.3\,M_{\odot}$ WD is very small, the predicted blue excess in the optical is negligible except at very high temperatures. On the other hand, effective temperatures above $\approx 25,000$\,K would lead to a detectable FUV source (here we assume a distance of 500 pc), which represents a strong UV excess easily distinguishable from a single MS star. A weaker, but still significant, NUV excess is also predicted. On the basis of these calculations, we conclude that -- given their high dynamical masses -- there is no plausible way for single WD companions to significantly change the optical colors of our candidates without simultaneously causing a large UV excess, which is not observed. 

Next, we consider the possibility of a close WD+WD binary of total mass $1.3\,M_{\odot}$. In this case, a broader range of combined SEDs are possible, depending on the mass ratio and cooling ages of both WDs. For simplicity, we consider the case where both WDs have the same mass and effective temperature. In this case, the blue excess is significantly stronger at fixed $T_{\rm eff}$, reflecting the fact that two $0.65\,M_{\odot}$ WDs have $\approx 18$ times the total surface area of a single $1.3\,M_{\odot}$ WD. 

With two $0.65\,M_{\odot}$ WDs, a $\approx -0.04$ mag blue excess is possible in the optical if both WDs have $T_{\rm eff}\sim 50,000$\,K. However, this would be accompanied by an excess of several magnitudes in both the NUV and FUV bands, and is thus ruled out. The same is true for a $1.0+0.3\,M_{\odot}$ WD binary. In this case, the flux is dominated by the $0.3\,M_{\odot}$ WD, which could cause a significant optical blue excess for $T_{\rm eff} \gtrsim 30,000$\,K. Here too, the excess would be accompanied by a much larger excess in the UV.  

\section{All fits}
\label{appendix:all_fits}

 Table~\ref{tab:allfits} lists constraints from joint fitting of RVs and {\it Gaia} astrometry for all sources.

\begin{longtable}{lll}
\hline\hline
\addlinespace % Adds whitespace
\multicolumn{3}{l}{\bf{J0553-1349;\,\,\,\,\,\, } Gaia\,DR3\,\,ID\,2995961897685517312 }  \\
Orbital period & $P_{\rm orb}$\,[days] & $ 189.10 \pm 0.05 $ \\
Semi-major axis & $a$\,[au] & $ 0.851 \pm 0.012$ \\
Photocenter semi-major axis & $a_0$\,[mas] & $ 1.229 \pm 0.014$ \\
Eccentricity & $e$ & $ 0.3879 \pm 0.0007$ \\
Inclination  & $i$\,[deg] & $ 53.1 \pm 0.9$ \\
Periastron time & $T_p$\,[JD-2457389] & $ -67.3 \pm 0.7$ \\
Ascending node angle & $\Omega$\,[deg] & $ 264.8 \pm 1.4$ \\
Argument of periastron & $\omega$\,[deg] & $ 162.74 \pm 0.16$ \\
Luminous star mass & $M_\star$\,[$M_{\odot}$] & $ 0.98 \pm 0.06$ \\
Neutron star mass & $M_2$\,[$M_{\odot}$] & $ 1.33 \pm 0.05$ \\
RV semi-amplitude & $K_\star$\,[$\rm km\,s^{-1}$] & $ 24.506 \pm 0.020$ \\
RV mass function & $f\left(M_{2}\right)_{{\rm RVs}}$\,[$\rm M_{\odot}$] & $ 0.2258 \pm 0.0006$ \\
Center-of-mass RV & $\gamma$\,[$\rm km\,s^{-1}$] & $ 41.033 \pm 0.015$ \\
Parallax & $\varpi$\,[mas] & $ 2.505 \pm 0.015$ \\
Goodness of fit & $\chi_{{\rm RVs}}^{2}/N_{{\rm RVs}}$ & $ 1.2$ \\
\hline
\addlinespace
\multicolumn{3}{l}{\bf{J2057-4742;\,\,\,\,\,\, } Gaia\,DR3\,\,ID\,6481502062263141504 }  \\
Orbital period & $P_{\rm orb}$\,[days] & $ 230.15 \pm 0.07 $ \\
Semi-major axis & $a$\,[au] & $ 0.978 \pm 0.008$ \\
Photocenter semi-major axis & $a_0$\,[mas] & $ 0.946 \pm 0.012$ \\
Eccentricity & $e$ & $ 0.3095 \pm 0.0026$ \\
Inclination  & $i$\,[deg] & $ 130.1 \pm 1.1$ \\
Periastron time & $T_p$\,[JD-2457389] & $ 10.7 \pm 0.9$ \\
Ascending node angle & $\Omega$\,[deg] & $ 345.1 \pm 1.3$ \\
Argument of periastron & $\omega$\,[deg] & $ 5.63 \pm 0.29$ \\
Luminous star mass & $M_\star$\,[$M_{\odot}$] & $ 1.048 \pm 0.031$ \\
Neutron star mass & $M_2$\,[$M_{\odot}$] & $ 1.31 \pm 0.04$ \\
RV semi-amplitude & $K_\star$\,[$\rm km\,s^{-1}$] & $ 20.62 \pm 0.07$ \\
RV mass function & $f\left(M_{2}\right)_{{\rm RVs}}$\,[$\rm M_{\odot}$] & $ 0.1799 \pm 0.0014$ \\
Center-of-mass RV & $\gamma$\,[$\rm km\,s^{-1}$] & $ -5.783 \pm 0.018$ \\
Parallax & $\varpi$\,[mas] & $ 1.745 \pm 0.019$ \\
Goodness of fit & $\chi_{{\rm RVs}}^{2}/N_{{\rm RVs}}$ & $ 0.3$ \\
\hline
\addlinespace
\multicolumn{3}{l}{\bf{J1553-6846;\,\,\,\,\,\, } Gaia\,DR3\,\,ID\,5820382041374661888 }  \\
Orbital period & $P_{\rm orb}$\,[days] & $ 310.17 \pm 0.11 $ \\
Semi-major axis & $a$\,[au] & $ 1.194 \pm 0.013$ \\
Photocenter semi-major axis & $a_0$\,[mas] & $ 0.899 \pm 0.009$ \\
Eccentricity & $e$ & $ 0.5314 \pm 0.0021$ \\
Inclination  & $i$\,[deg] & $ 83.2 \pm 0.9$ \\
Periastron time & $T_p$\,[JD-2457389] & $ 108.0 \pm 1.0$ \\
Ascending node angle & $\Omega$\,[deg] & $ 189.7 \pm 0.8$ \\
Argument of periastron & $\omega$\,[deg] & $ 324.52 \pm 0.21$ \\
Luminous star mass & $M_\star$\,[$M_{\odot}$] & $ 1.04 \pm 0.05$ \\
Neutron star mass & $M_2$\,[$M_{\odot}$] & $ 1.323 \pm 0.032$ \\
RV semi-amplitude & $K_\star$\,[$\rm km\,s^{-1}$] & $ 27.49 \pm 0.22$ \\
RV mass function & $f\left(M_{2}\right)_{{\rm RVs}}$\,[$\rm M_{\odot}$] & $ 0.406 \pm 0.009$ \\
Center-of-mass RV & $\gamma$\,[$\rm km\,s^{-1}$] & $ -42.95 \pm 0.11$ \\
Parallax & $\varpi$\,[mas] & $ 1.344 \pm 0.012$ \\
Goodness of fit & $\chi_{{\rm RVs}}^{2}/N_{{\rm RVs}}$ & $ 1.2$ \\
\hline
\addlinespace
\multicolumn{3}{l}{\bf{J2102+3703;\,\,\,\,\,\, } Gaia\,DR3\,\,ID\,1871419337958702720 }  \\
Orbital period & $P_{\rm orb}$\,[days] & $ 481.04 \pm 0.26 $ \\
Semi-major axis & $a$\,[au] & $ 1.632 \pm 0.010$ \\
Photocenter semi-major axis & $a_0$\,[mas] & $ 1.459 \pm 0.017$ \\
Eccentricity & $e$ & $ 0.448 \pm 0.009$ \\
Inclination  & $i$\,[deg] & $ 135.7 \pm 0.8$ \\
Periastron time & $T_p$\,[JD-2457389] & $ 138.2 \pm 1.3$ \\
Ascending node angle & $\Omega$\,[deg] & $ 230.0 \pm 1.1$ \\
Argument of periastron & $\omega$\,[deg] & $ 214.9 \pm 0.9$ \\
Luminous star mass & $M_\star$\,[$M_{\odot}$] & $ 1.033 \pm 0.03$ \\
Neutron star mass & $M_2$\,[$M_{\odot}$] & $ 1.473 \pm 0.034$ \\
RV semi-amplitude & $K_\star$\,[$\rm km\,s^{-1}$] & $ 16.9 \pm 0.5$ \\
RV mass function & $f\left(M_{2}\right)_{{\rm RVs}}$\,[$\rm M_{\odot}$] & $ 0.173 \pm 0.011$ \\
Center-of-mass RV & $\gamma$\,[$\rm km\,s^{-1}$] & $ 18.08 \pm 0.09$ \\
Parallax & $\varpi$\,[mas] & $ 1.521 \pm 0.013$ \\
Goodness of fit & $\chi_{{\rm RVs}}^{2}/N_{{\rm RVs}}$ & $ 0.5$ \\
\hline
\addlinespace
\multicolumn{3}{l}{\bf{J0742-4749;\,\,\,\,\,\, } Gaia\,DR3\,\,ID\,5530442371304582912 }  \\
Orbital period & $P_{\rm orb}$\,[days] & $ 497.6 \pm 0.4 $ \\
Semi-major axis & $a$\,[au] & $ 1.593 \pm 0.021$ \\
Photocenter semi-major axis & $a_0$\,[mas] & $ 0.965 \pm 0.014$ \\
Eccentricity & $e$ & $ 0.168 \pm 0.004$ \\
Inclination  & $i$\,[deg] & $ 129.2 \pm 1.1$ \\
Periastron time & $T_p$\,[JD-2457389] & $ 23 \pm 4$ \\
Ascending node angle & $\Omega$\,[deg] & $ 170.8 \pm 1.8$ \\
Argument of periastron & $\omega$\,[deg] & $ 331.9 \pm 1.9$ \\
Luminous star mass & $M_\star$\,[$M_{\odot}$] & $ 0.90 \pm 0.05$ \\
Neutron star mass & $M_2$\,[$M_{\odot}$] & $ 1.28 \pm 0.04$ \\
RV semi-amplitude & $K_\star$\,[$\rm km\,s^{-1}$] & $ 16.03 \pm 0.09$ \\
RV mass function & $f\left(M_{2}\right)_{{\rm RVs}}$\,[$\rm M_{\odot}$] & $ 0.204 \pm 0.004$ \\
Center-of-mass RV & $\gamma$\,[$\rm km\,s^{-1}$] & $ 41.10 \pm 0.08$ \\
Parallax & $\varpi$\,[mas] & $ 1.035 \pm 0.014$ \\
Goodness of fit & $\chi_{{\rm RVs}}^{2}/N_{{\rm RVs}}$ & $ 0.2$ \\
\hline
\addlinespace
\multicolumn{3}{l}{\bf{J0152-2049;\,\,\,\,\,\, } Gaia\,DR3\,\,ID\,5136025521527939072 }  \\
Orbital period & $P_{\rm orb}$\,[days] & $ 536.14 \pm 0.18 $ \\
Semi-major axis & $a$\,[au] & $ 1.647 \pm 0.014$ \\
Photocenter semi-major axis & $a_0$\,[mas] & $ 2.515 \pm 0.014$ \\
Eccentricity & $e$ & $ 0.6615 \pm 0.0010$ \\
Inclination  & $i$\,[deg] & $ 123.9 \pm 0.4$ \\
Periastron time & $T_p$\,[JD-2457389] & $ 166.3 \pm 1.0$ \\
Ascending node angle & $\Omega$\,[deg] & $ 291.1 \pm 0.8$ \\
Argument of periastron & $\omega$\,[deg] & $ 253.90 \pm 0.22$ \\
Luminous star mass & $M_\star$\,[$M_{\odot}$] & $ 0.782 \pm 0.030$ \\
Neutron star mass & $M_2$\,[$M_{\odot}$] & $ 1.291 \pm 0.024$ \\
RV semi-amplitude & $K_\star$\,[$\rm km\,s^{-1}$] & $ 23.041 \pm 0.033$ \\
RV mass function & $f\left(M_{2}\right)_{{\rm RVs}}$\,[$\rm M_{\odot}$] & $ 0.2866 \pm 0.0011$ \\
Center-of-mass RV & $\gamma$\,[$\rm km\,s^{-1}$] & $ 47.479 \pm 0.028$ \\
Parallax & $\varpi$\,[mas] & $ 2.453 \pm 0.017$ \\
Goodness of fit & $\chi_{{\rm RVs}}^{2}/N_{{\rm RVs}}$ & $ 0.8$ \\
\hline
\addlinespace
\multicolumn{3}{l}{\bf{J0003-5604;\,\,\,\,\,\, } Gaia\,DR3\,\,ID\,4922744974687373440 }  \\
Orbital period & $P_{\rm orb}$\,[days] & $ 561.83 \pm 0.29 $ \\
Semi-major axis & $a$\,[au] & $ 1.717 \pm 0.016$ \\
Photocenter semi-major axis & $a_0$\,[mas] & $ 2.344 \pm 0.026$ \\
Eccentricity & $e$ & $ 0.795 \pm 0.005$ \\
Inclination  & $i$\,[deg] & $ 85.9 \pm 0.8$ \\
Periastron time & $T_p$\,[JD-2457389] & $ -36.7 \pm 1.2$ \\
Ascending node angle & $\Omega$\,[deg] & $ 259.7 \pm 0.7$ \\
Argument of periastron & $\omega$\,[deg] & $ 308.4 \pm 0.5$ \\
Luminous star mass & $M_\star$\,[$M_{\odot}$] & $ 0.802 \pm 0.031$ \\
Neutron star mass & $M_2$\,[$M_{\odot}$] & $ 1.34 \pm 0.04$ \\
RV semi-amplitude & $K_\star$\,[$\rm km\,s^{-1}$] & $ 34.2 \pm 0.8$ \\
RV mass function & $f\left(M_{2}\right)_{{\rm RVs}}$\,[$\rm M_{\odot}$] & $ 0.519 \pm 0.020$ \\
Center-of-mass RV & $\gamma$\,[$\rm km\,s^{-1}$] & $ 52.58 \pm 0.07$ \\
Parallax & $\varpi$\,[mas] & $ 2.183 \pm 0.016$ \\
Goodness of fit & $\chi_{{\rm RVs}}^{2}/N_{{\rm RVs}}$ & $ 0.3$ \\
\hline
\addlinespace
\multicolumn{3}{l}{\bf{J1733+5808;\,\,\,\,\,\, } Gaia\,DR3\,\,ID\,1434445448240677376 }  \\
Orbital period & $P_{\rm orb}$\,[days] & $ 570.94 \pm 0.31 $ \\
Semi-major axis & $a$\,[au] & $ 1.835 \pm 0.018$ \\
Photocenter semi-major axis & $a_0$\,[mas] & $ 1.437 \pm 0.010$ \\
Eccentricity & $e$ & $ 0.3093 \pm 0.0010$ \\
Inclination  & $i$\,[deg] & $ 55.1 \pm 0.5$ \\
Periastron time & $T_p$\,[JD-2457389] & $ -84.6 \pm 1.5$ \\
Ascending node angle & $\Omega$\,[deg] & $ 8.1 \pm 1.1$ \\
Argument of periastron & $\omega$\,[deg] & $ 110.34 \pm 0.15$ \\
Luminous star mass & $M_\star$\,[$M_{\odot}$] & $ 1.16 \pm 0.05$ \\
Neutron star mass & $M_2$\,[$M_{\odot}$] & $ 1.362 \pm 0.030$ \\
RV semi-amplitude & $K_\star$\,[$\rm km\,s^{-1}$] & $ 16.266 \pm 0.014$ \\
RV mass function & $f\left(M_{2}\right)_{{\rm RVs}}$\,[$\rm M_{\odot}$] & $ 0.2189 \pm 0.0007$ \\
Center-of-mass RV & $\gamma$\,[$\rm km\,s^{-1}$] & $ -18.047 \pm 0.012$ \\
Parallax & $\varpi$\,[mas] & $ 1.452 \pm 0.010$ \\
Goodness of fit & $\chi_{{\rm RVs}}^{2}/N_{{\rm RVs}}$ & $ 2.3$ \\
\hline
\addlinespace
\multicolumn{3}{l}{\bf{J1150-2203;\,\,\,\,\,\, } Gaia\,DR3\,\,ID\,3494029910469026432 }  \\
Orbital period & $P_{\rm orb}$\,[days] & $ 631.81 \pm 0.22 $ \\
Semi-major axis & $a$\,[au] & $ 1.973 \pm 0.023$ \\
Photocenter semi-major axis & $a_0$\,[mas] & $ 1.859 \pm 0.018$ \\
Eccentricity & $e$ & $ 0.552 \pm 0.004$ \\
Inclination  & $i$\,[deg] & $ 122.4 \pm 0.7$ \\
Periastron time & $T_p$\,[JD-2457389] & $ 245.0 \pm 0.9$ \\
Ascending node angle & $\Omega$\,[deg] & $ 165.3 \pm 0.7$ \\
Argument of periastron & $\omega$\,[deg] & $ 180.38 \pm 0.16$ \\
Luminous star mass & $M_\star$\,[$M_{\odot}$] & $ 1.18 \pm 0.06$ \\
Neutron star mass & $M_2$\,[$M_{\odot}$] & $ 1.39 \pm 0.04$ \\
RV semi-amplitude & $K_\star$\,[$\rm km\,s^{-1}$] & $ 18.66 \pm 0.28$ \\
RV mass function & $f\left(M_{2}\right)_{{\rm RVs}}$\,[$\rm M_{\odot}$] & $ 0.246 \pm 0.009$ \\
Center-of-mass RV & $\gamma$\,[$\rm km\,s^{-1}$] & $ 29.80 \pm 0.06$ \\
Parallax & $\varpi$\,[mas] & $ 1.738 \pm 0.016$ \\
Goodness of fit & $\chi_{{\rm RVs}}^{2}/N_{{\rm RVs}}$ & $ 2.4$ \\
\hline
\addlinespace
\multicolumn{3}{l}{\bf{J1449+6919;\,\,\,\,\,\, } Gaia\,DR3\,\,ID\,1694708646628402048 }  \\
Orbital period & $P_{\rm orb}$\,[days] & $ 632.65 \pm 0.21 $ \\
Semi-major axis & $a$\,[au] & $ 1.868 \pm 0.023$ \\
Photocenter semi-major axis & $a_0$\,[mas] & $ 1.961 \pm 0.010$ \\
Eccentricity & $e$ & $ 0.2668 \pm 0.0010$ \\
Inclination  & $i$\,[deg] & $ 105.7 \pm 0.4$ \\
Periastron time & $T_p$\,[JD-2457389] & $ -257.1 \pm 1.1$ \\
Ascending node angle & $\Omega$\,[deg] & $ 132.5 \pm 0.6$ \\
Argument of periastron & $\omega$\,[deg] & $ 55.36 \pm 0.14$ \\
Luminous star mass & $M_\star$\,[$M_{\odot}$] & $ 0.91 \pm 0.05$ \\
Neutron star mass & $M_2$\,[$M_{\odot}$] & $ 1.258 \pm 0.032$ \\
RV semi-amplitude & $K_\star$\,[$\rm km\,s^{-1}$] & $ 18.584 \pm 0.018$ \\
RV mass function & $f\left(M_{2}\right)_{{\rm RVs}}$\,[$\rm M_{\odot}$] & $ 0.3767 \pm 0.0010$ \\
Center-of-mass RV & $\gamma$\,[$\rm km\,s^{-1}$] & $ -69.617 \pm 0.012$ \\
Parallax & $\varpi$\,[mas] & $ 1.812 \pm 0.010$ \\
Goodness of fit & $\chi_{{\rm RVs}}^{2}/N_{{\rm RVs}}$ & $ 2.3$ \\
\hline
\addlinespace
\multicolumn{3}{l}{\bf{J0217-7541;\,\,\,\,\,\, } Gaia\,DR3\,\,ID\,4637171465304969216 }  \\
Orbital period & $P_{\rm orb}$\,[days] & $ 636.1 \pm 0.7 $ \\
Semi-major axis & $a$\,[au] & $ 1.936 \pm 0.016$ \\
Photocenter semi-major axis & $a_0$\,[mas] & $ 1.348 \pm 0.012$ \\
Eccentricity & $e$ & $ 0.3228 \pm 0.0033$ \\
Inclination  & $i$\,[deg] & $ 132.3 \pm 0.7$ \\
Periastron time & $T_p$\,[JD-2457389] & $ 18.4 \pm 2.6$ \\
Ascending node angle & $\Omega$\,[deg] & $ 95.1 \pm 1.6$ \\
Argument of periastron & $\omega$\,[deg] & $ 29.4 \pm 0.5$ \\
Luminous star mass & $M_\star$\,[$M_{\odot}$] & $ 0.996 \pm 0.033$ \\
Neutron star mass & $M_2$\,[$M_{\odot}$] & $ 1.396 \pm 0.033$ \\
RV semi-amplitude & $K_\star$\,[$\rm km\,s^{-1}$] & $ 15.09 \pm 0.04$ \\
RV mass function & $f\left(M_{2}\right)_{{\rm RVs}}$\,[$\rm M_{\odot}$] & $ 0.1919 \pm 0.0013$ \\
Center-of-mass RV & $\gamma$\,[$\rm km\,s^{-1}$] & $ 59.641 \pm 0.027$ \\
Parallax & $\varpi$\,[mas] & $ 1.193 \pm 0.012$ \\
Goodness of fit & $\chi_{{\rm RVs}}^{2}/N_{{\rm RVs}}$ & $ 0.8$ \\
\hline
\addlinespace
\multicolumn{3}{l}{\bf{J0639-3655;\,\,\,\,\,\, } Gaia\,DR3\,\,ID\,5580526947012630912 }  \\
Orbital period & $P_{\rm orb}$\,[days] & $ 654.6 \pm 0.6 $ \\
Semi-major axis & $a$\,[au] & $ 2.132 \pm 0.028$ \\
Photocenter semi-major axis & $a_0$\,[mas] & $ 1.357 \pm 0.027$ \\
Eccentricity & $e$ & $ 0.721 \pm 0.013$ \\
Inclination  & $i$\,[deg] & $ 171.47 \pm 0.34$ \\
Periastron time & $T_p$\,[JD-2457389] & $ 52.1 \pm 1.5$ \\
Ascending node angle & $\Omega$\,[deg] & $ 362.1 \pm 1.6$ \\
Argument of periastron & $\omega$\,[deg] & $ 331.4 \pm 1.0$ \\
Luminous star mass & $M_\star$\,[$M_{\odot}$] & $ 1.32 \pm 0.06$ \\
Neutron star mass & $M_2$\,[$M_{\odot}$] & $ 1.70 \pm 0.07$ \\
RV semi-amplitude & $K_\star$\,[$\rm km\,s^{-1}$] & $ 4.25 \pm 0.24$ \\
RV mass function & $f\left(M_{2}\right)_{{\rm RVs}}$\,[$\rm M_{\odot}$] & $ 0.00175 \pm 0.00023$ \\
Center-of-mass RV & $\gamma$\,[$\rm km\,s^{-1}$] & $ -10.49 \pm 0.05$ \\
Parallax & $\varpi$\,[mas] & $ 1.130 \pm 0.011$ \\
Goodness of fit & $\chi_{{\rm RVs}}^{2}/N_{{\rm RVs}}$ & $ 0.2$ \\
\hline
\addlinespace
\multicolumn{3}{l}{\bf{J1739+4502;\,\,\,\,\,\, } Gaia\,DR3\,\,ID\,1350295047363872512 }  \\
Orbital period & $P_{\rm orb}$\,[days] & $ 657.4 \pm 0.6 $ \\
Semi-major axis & $a$\,[au] & $ 1.914 \pm 0.018$ \\
Photocenter semi-major axis & $a_0$\,[mas] & $ 1.376 \pm 0.016$ \\
Eccentricity & $e$ & $ 0.6777 \pm 0.0018$ \\
Inclination  & $i$\,[deg] & $ 144.7 \pm 0.5$ \\
Periastron time & $T_p$\,[JD-2457389] & $ 179.7 \pm 2.3$ \\
Ascending node angle & $\Omega$\,[deg] & $ 70.5 \pm 1.5$ \\
Argument of periastron & $\omega$\,[deg] & $ 334.36 \pm 0.22$ \\
Luminous star mass & $M_\star$\,[$M_{\odot}$] & $ 0.781 \pm 0.030$ \\
Neutron star mass & $M_2$\,[$M_{\odot}$] & $ 1.38 \pm 0.04$ \\
RV semi-amplitude & $K_\star$\,[$\rm km\,s^{-1}$] & $ 15.909 \pm 0.034$ \\
RV mass function & $f\left(M_{2}\right)_{{\rm RVs}}$\,[$\rm M_{\odot}$] & $ 0.1091 \pm 0.0007$ \\
Center-of-mass RV & $\gamma$\,[$\rm km\,s^{-1}$] & $ -264.202 \pm 0.026$ \\
Parallax & $\varpi$\,[mas] & $ 1.126 \pm 0.013$ \\
Goodness of fit & $\chi_{{\rm RVs}}^{2}/N_{{\rm RVs}}$ & $ 1.3$ \\
\hline
\addlinespace
\multicolumn{3}{l}{\bf{J0036-0932;\,\,\,\,\,\, } Gaia\,DR3\,\,ID\,2426116249713980416 }  \\
Orbital period & $P_{\rm orb}$\,[days] & $ 719.8 \pm 0.9 $ \\
Semi-major axis & $a$\,[au] & $ 2.075 \pm 0.021$ \\
Photocenter semi-major axis & $a_0$\,[mas] & $ 2.040 \pm 0.014$ \\
Eccentricity & $e$ & $ 0.3993 \pm 0.0021$ \\
Inclination  & $i$\,[deg] & $ 145.1 \pm 0.4$ \\
Periastron time & $T_p$\,[JD-2457389] & $ -190.8 \pm 3.4$ \\
Ascending node angle & $\Omega$\,[deg] & $ 73.8 \pm 1.6$ \\
Argument of periastron & $\omega$\,[deg] & $ 172.41 \pm 0.33$ \\
Luminous star mass & $M_\star$\,[$M_{\odot}$] & $ 0.94 \pm 0.04$ \\
Neutron star mass & $M_2$\,[$M_{\odot}$] & $ 1.362 \pm 0.034$ \\
RV semi-amplitude & $K_\star$\,[$\rm km\,s^{-1}$] & $ 11.57 \pm 0.07$ \\
RV mass function & $f\left(M_{2}\right)_{{\rm RVs}}$\,[$\rm M_{\odot}$] & $ 0.0891 \pm 0.0016$ \\
Center-of-mass RV & $\gamma$\,[$\rm km\,s^{-1}$] & $ 22.22 \pm 0.04$ \\
Parallax & $\varpi$\,[mas] & $ 1.661 \pm 0.019$ \\
Goodness of fit & $\chi_{{\rm RVs}}^{2}/N_{{\rm RVs}}$ & $ 1.4$ \\
\hline
\addlinespace
\multicolumn{3}{l}{\bf{J1432-1021;\,\,\,\,\,\, } Gaia\,DR3\,\,ID\,6328149636482597888 }  \\
Orbital period & $P_{\rm orb}$\,[days] & $ 730.9 \pm 0.5 $ \\
Semi-major axis & $a$\,[au] & $ 2.208 \pm 0.016$ \\
Photocenter semi-major axis & $a_0$\,[mas] & $ 2.132 \pm 0.016$ \\
Eccentricity & $e$ & $ 0.1203 \pm 0.0022$ \\
Inclination  & $i$\,[deg] & $ 68.8 \pm 0.5$ \\
Periastron time & $T_p$\,[JD-2457389] & $ 186.4 \pm 2.3$ \\
Ascending node angle & $\Omega$\,[deg] & $ 82.7 \pm 0.8$ \\
Argument of periastron & $\omega$\,[deg] & $ 259.4 \pm 0.7$ \\
Luminous star mass & $M_\star$\,[$M_{\odot}$] & $ 0.790 \pm 0.030$ \\
Neutron star mass & $M_2$\,[$M_{\odot}$] & $ 1.898 \pm 0.030$ \\
RV semi-amplitude & $K_\star$\,[$\rm km\,s^{-1}$] & $ 21.79 \pm 0.04$ \\
RV mass function & $f\left(M_{2}\right)_{{\rm RVs}}$\,[$\rm M_{\odot}$] & $ 0.766 \pm 0.004$ \\
Center-of-mass RV & $\gamma$\,[$\rm km\,s^{-1}$] & $ 133.457 \pm 0.033$ \\
Parallax & $\varpi$\,[mas] & $ 1.367 \pm 0.011$ \\
Goodness of fit & $\chi_{{\rm RVs}}^{2}/N_{{\rm RVs}}$ & $ 0.7$ \\
\hline
\addlinespace
\multicolumn{3}{l}{\bf{J1048+6547;\,\,\,\,\,\, } Gaia\,DR3\,\,ID\,1058875159778407808 }  \\
Orbital period & $P_{\rm orb}$\,[days] & $ 827 \pm 5 $ \\
Semi-major axis & $a$\,[au] & $ 2.344 \pm 0.031$ \\
Photocenter semi-major axis & $a_0$\,[mas] & $ 1.301 \pm 0.017$ \\
Eccentricity & $e$ & $ 0.357 \pm 0.009$ \\
Inclination  & $i$\,[deg] & $ 107.8 \pm 1.0$ \\
Periastron time & $T_p$\,[JD-2457389] & $ -47 \pm 6$ \\
Ascending node angle & $\Omega$\,[deg] & $ 159.9 \pm 1.4$ \\
Argument of periastron & $\omega$\,[deg] & $ 288.5 \pm 2.4$ \\
Luminous star mass & $M_\star$\,[$M_{\odot}$] & $ 0.99 \pm 0.05$ \\
Neutron star mass & $M_2$\,[$M_{\odot}$] & $ 1.52 \pm 0.07$ \\
RV semi-amplitude & $K_\star$\,[$\rm km\,s^{-1}$] & $ 19.0 \pm 0.5$ \\
RV mass function & $f\left(M_{2}\right)_{{\rm RVs}}$\,[$\rm M_{\odot}$] & $ 0.479 \pm 0.030$ \\
Center-of-mass RV & $\gamma$\,[$\rm km\,s^{-1}$] & $ 56.8 \pm 1.2$ \\
Parallax & $\varpi$\,[mas] & $ 0.916 \pm 0.016$ \\
Goodness of fit & $\chi_{{\rm RVs}}^{2}/N_{{\rm RVs}}$ & $ 1.4$ \\
\hline
\addlinespace
\multicolumn{3}{l}{\bf{J2145+2837;\,\,\,\,\,\, } Gaia\,DR3\,\,ID\,1801110822095134848 }  \\
Orbital period & $P_{\rm orb}$\,[days] & $ 889.5 \pm 0.7 $ \\
Semi-major axis & $a$\,[au] & $ 2.405 \pm 0.029$ \\
Photocenter semi-major axis & $a_0$\,[mas] & $ 5.923 \pm 0.026$ \\
Eccentricity & $e$ & $ 0.5840 \pm 0.0035$ \\
Inclination  & $i$\,[deg] & $ 125.29 \pm 0.24$ \\
Periastron time & $T_p$\,[JD-2457389] & $ -392.4 \pm 0.9$ \\
Ascending node angle & $\Omega$\,[deg] & $ 167.7 \pm 0.4$ \\
Argument of periastron & $\omega$\,[deg] & $ 243.03 \pm 0.33$ \\
Luminous star mass & $M_\star$\,[$M_{\odot}$] & $ 0.95 \pm 0.05$ \\
Neutron star mass & $M_2$\,[$M_{\odot}$] & $ 1.396 \pm 0.035$ \\
RV semi-amplitude & $K_\star$\,[$\rm km\,s^{-1}$] & $ 17.60 \pm 0.12$ \\
RV mass function & $f\left(M_{2}\right)_{{\rm RVs}}$\,[$\rm M_{\odot}$] & $ 0.269 \pm 0.005$ \\
Center-of-mass RV & $\gamma$\,[$\rm km\,s^{-1}$] & $ -43.19 \pm 0.11$ \\
Parallax & $\varpi$\,[mas] & $ 4.137 \pm 0.016$ \\
Goodness of fit & $\chi_{{\rm RVs}}^{2}/N_{{\rm RVs}}$ & $ 1.7$ \\
\hline
\addlinespace
\multicolumn{3}{l}{\bf{J2244-2236;\,\,\,\,\,\, } Gaia\,DR3\,\,ID\,2397135910639986304 }  \\
Orbital period & $P_{\rm orb}$\,[days] & $ 938.3 \pm 0.5 $ \\
Semi-major axis & $a$\,[au] & $ 2.527 \pm 0.018$ \\
Photocenter semi-major axis & $a_0$\,[mas] & $ 3.102 \pm 0.024$ \\
Eccentricity & $e$ & $ 0.5666 \pm 0.0011$ \\
Inclination  & $i$\,[deg] & $ 118.9 \pm 0.5$ \\
Periastron time & $T_p$\,[JD-2457389] & $ -109.3 \pm 1.6$ \\
Ascending node angle & $\Omega$\,[deg] & $ 228.6 \pm 0.5$ \\
Argument of periastron & $\omega$\,[deg] & $ 332.85 \pm 0.12$ \\
Luminous star mass & $M_\star$\,[$M_{\odot}$] & $ 1.002 \pm 0.030$ \\
Neutron star mass & $M_2$\,[$M_{\odot}$] & $ 1.443 \pm 0.023$ \\
RV semi-amplitude & $K_\star$\,[$\rm km\,s^{-1}$] & $ 18.366 \pm 0.015$ \\
RV mass function & $f\left(M_{2}\right)_{{\rm RVs}}$\,[$\rm M_{\odot}$] & $ 0.3370 \pm 0.0011$ \\
Center-of-mass RV & $\gamma$\,[$\rm km\,s^{-1}$] & $ 15.973 \pm 0.014$ \\
Parallax & $\varpi$\,[mas] & $ 2.079 \pm 0.019$ \\
Goodness of fit & $\chi_{{\rm RVs}}^{2}/N_{{\rm RVs}}$ & $ 0.5$ \\
\hline
\addlinespace
\multicolumn{3}{l}{\bf{J0824+5254;\,\,\,\,\,\, } Gaia\,DR3\,\,ID\,1028887114002082432 }  \\
Orbital period & $P_{\rm orb}$\,[days] & $ 1026.7 \pm 3.3 $ \\
Semi-major axis & $a$\,[au] & $ 2.776 \pm 0.021$ \\
Photocenter semi-major axis & $a_0$\,[mas] & $ 2.701 \pm 0.035$ \\
Eccentricity & $e$ & $ 0.686 \pm 0.012$ \\
Inclination  & $i$\,[deg] & $ 128.6 \pm 0.6$ \\
Periastron time & $T_p$\,[JD-2457389] & $ 325.8 \pm 3.3$ \\
Ascending node angle & $\Omega$\,[deg] & $ 233.3 \pm 1.0$ \\
Argument of periastron & $\omega$\,[deg] & $ 267.3 \pm 1.2$ \\
Luminous star mass & $M_\star$\,[$M_{\odot}$] & $ 1.102 \pm 0.031$ \\
Neutron star mass & $M_2$\,[$M_{\odot}$] & $ 1.604 \pm 0.034$ \\
RV semi-amplitude & $K_\star$\,[$\rm km\,s^{-1}$] & $ 18.7 \pm 0.5$ \\
RV mass function & $f\left(M_{2}\right)_{{\rm RVs}}$\,[$\rm M_{\odot}$] & $ 0.268 \pm 0.011$ \\
Center-of-mass RV & $\gamma$\,[$\rm km\,s^{-1}$] & $ -57.76 \pm 0.31$ \\
Parallax & $\varpi$\,[mas] & $ 1.643 \pm 0.015$ \\
Goodness of fit & $\chi_{{\rm RVs}}^{2}/N_{{\rm RVs}}$ & $ 0.8$ \\
\hline
\addlinespace
\multicolumn{3}{l}{\bf{J0230+5950;\,\,\,\,\,\, } Gaia\,DR3\,\,ID\,465093354131112960 }  \\
Orbital period & $P_{\rm orb}$\,[days] & $ 1029 \pm 5 $ \\
Semi-major axis & $a$\,[au] & $ 2.713 \pm 0.020$ \\
Photocenter semi-major axis & $a_0$\,[mas] & $ 3.82 \pm 0.06$ \\
Eccentricity & $e$ & $ 0.753 \pm 0.011$ \\
Inclination  & $i$\,[deg] & $ 116.5 \pm 0.5$ \\
Periastron time & $T_p$\,[JD-2457389] & $ 240.8 \pm 1.5$ \\
Ascending node angle & $\Omega$\,[deg] & $ 32.5 \pm 0.6$ \\
Argument of periastron & $\omega$\,[deg] & $ 103.1 \pm 0.8$ \\
Luminous star mass & $M_\star$\,[$M_{\odot}$] & $ 1.114 \pm 0.029$ \\
Neutron star mass & $M_2$\,[$M_{\odot}$] & $ 1.401 \pm 0.034$ \\
RV semi-amplitude & $K_\star$\,[$\rm km\,s^{-1}$] & $ 21.7 \pm 0.8$ \\
RV mass function & $f\left(M_{2}\right)_{{\rm RVs}}$\,[$\rm M_{\odot}$] & $ 0.312 \pm 0.015$ \\
Center-of-mass RV & $\gamma$\,[$\rm km\,s^{-1}$] & $ 9.4 \pm 0.4$ \\
Parallax & $\varpi$\,[mas] & $ 2.523 \pm 0.015$ \\
Goodness of fit & $\chi_{{\rm RVs}}^{2}/N_{{\rm RVs}}$ & $ 1.8$ \\
\hline
\addlinespace
\multicolumn{3}{l}{\bf{J0634+6256;\,\,\,\,\,\, } Gaia\,DR3\,\,ID\,1007185297091149824 }  \\
Orbital period & $P_{\rm orb}$\,[days] & $ 1046.0 \pm 2.1 $ \\
Semi-major axis & $a$\,[au] & $ 2.79 \pm 0.05$ \\
Photocenter semi-major axis & $a_0$\,[mas] & $ 1.072 \pm 0.027$ \\
Eccentricity & $e$ & $ 0.564 \pm 0.011$ \\
Inclination  & $i$\,[deg] & $ 85.3 \pm 1.9$ \\
Periastron time & $T_p$\,[JD-2457389] & $ 338 \pm 5$ \\
Ascending node angle & $\Omega$\,[deg] & $ 285.7 \pm 2.2$ \\
Argument of periastron & $\omega$\,[deg] & $ 233.4 \pm 1.8$ \\
Luminous star mass & $M_\star$\,[$M_{\odot}$] & $ 1.18 \pm 0.06$ \\
Neutron star mass & $M_2$\,[$M_{\odot}$] & $ 1.48 \pm 0.09$ \\
RV semi-amplitude & $K_\star$\,[$\rm km\,s^{-1}$] & $ 19.5 \pm 0.9$ \\
RV mass function & $f\left(M_{2}\right)_{{\rm RVs}}$\,[$\rm M_{\odot}$] & $ 0.45 \pm 0.05$ \\
Center-of-mass RV & $\gamma$\,[$\rm km\,s^{-1}$] & $ -58.77 \pm 0.24$ \\
Parallax & $\varpi$\,[mas] & $ 0.689 \pm 0.019$ \\
Goodness of fit & $\chi_{{\rm RVs}}^{2}/N_{{\rm RVs}}$ & $ 0.5$ \\
\hline
\addlinespace
\hline
\addlinespace
\caption{Orbit fitting results for all candidates.}
\label{tab:allfits}
\end{longtable}
%\begin{flushleft}

\section{Radial velocities}
\label{appendix:all_rvs}
Table \ref{tab:rvs} lists all the RVs used in our analysis.

\begin{longtable}{llll}
Name & HJD UTC & RV ($\rm km\,s^{-1}$) & Instrument  \\
\hline
J0553-1349 & 2459822.8517 & $39.13 \pm 0.11$ & FEROS \\
J0553-1349 & 2459891.9150 & $56.43 \pm 0.03$ & TRES \\
J0553-1349 & 2459913.9359 & $54.72 \pm 0.05$ & TRES \\
J0553-1349 & 2459915.6211 & $54.37 \pm 0.04$ & FEROS \\
J0553-1349 & 2459919.7038 & $53.35 \pm 0.04$ & FEROS \\
J0553-1349 & 2459928.8778 & $50.22 \pm 1.00$ & PEPSI \\
J0553-1349 & 2459939.8631 & $43.46 \pm 0.04$ & TRES \\
J0553-1349 & 2459973.6911 & $7.49 \pm 0.03$ & TRES \\
J0553-1349 & 2460013.6884 & $40.24 \pm 0.03$ & TRES \\
J0553-1349 & 2460019.5421 & $43.29 \pm 0.04$ & FEROS \\
J0553-1349 & 2460038.5227 & $50.55 \pm 0.50$ & MIKE \\
J0553-1349 & 2460053.4888 & $53.88 \pm 0.06$ & FEROS \\
J0553-1349 & 2460054.5185 & $54.08 \pm 0.04$ & FEROS \\
J0553-1349 & 2460187.8882 & $29.48 \pm 0.04$ & FEROS \\
J0553-1349 & 2460225.7851 & $49.87 \pm 0.03$ & FEROS \\
J0553-1349 & 2460236.9731 & $52.81 \pm 0.03$ & TRES \\
J0553-1349 & 2460256.9310 & $55.81 \pm 0.03$ & TRES \\
J0553-1349 & 2460283.8057 & $56.01 \pm 0.03$ & TRES \\
J0553-1349 & 2460299.8155 & $52.84 \pm 0.03$ & TRES \\
J0553-1349 & 2460326.7433 & $35.01 \pm 0.02$ & TRES \\
J2057-4742 & 2459824.7318 & $-19.59 \pm 0.09$ & FEROS \\
J2057-4742 & 2459905.5907 & $9.24 \pm 0.06$ & FEROS \\
J2057-4742 & 2460027.8818 & $-19.72 \pm 0.05$ & FEROS \\
J2057-4742 & 2460038.8892 & $-20.08 \pm 0.50$ & MIKE \\
J2057-4742 & 2460072.8557 & $-17.43 \pm 0.05$ & FEROS \\
J2057-4742 & 2460088.8618 & $-14.05 \pm 0.06$ & FEROS \\
J2057-4742 & 2460111.8331 & $-5.61 \pm 0.04$ & FEROS \\
J2057-4742 & 2460140.6079 & $12.65 \pm 0.04$ & FEROS \\
J2057-4742 & 2460185.6196 & $6.59 \pm 0.04$ & FEROS \\
J2057-4742 & 2460221.6285 & $-13.28 \pm 0.08$ & FEROS \\
J2057-4742 & 2460285.5516 & $-19.43 \pm 0.06$ & FEROS \\
J1553-6846 & 2459816.4861 & $-52.67 \pm 0.07$ & FEROS \\
J1553-6846 & 2459829.6119 & $-54.13 \pm 0.28$ & FEROS \\
J1553-6846 & 2459831.5985 & $-53.40 \pm 0.33$ & FEROS \\
J1553-6846 & 2459832.5539 & $-54.44 \pm 0.21$ & FEROS \\
J1553-6846 & 2459995.8453 & $-7.51 \pm 0.06$ & FEROS \\
J1553-6846 & 2459997.8394 & $-8.74 \pm 0.07$ & FEROS \\
J1553-6846 & 2460038.8012 & $-33.48 \pm 0.50$ & MIKE \\
J1553-6846 & 2460053.7555 & $-38.50 \pm 0.09$ & FEROS \\
J1553-6846 & 2460072.7728 & $-43.46 \pm 0.14$ & FEROS \\
J1553-6846 & 2460088.7001 & $-46.93 \pm 0.07$ & FEROS \\
J1553-6846 & 2460112.5953 & $-50.72 \pm 0.04$ & FEROS \\
J1553-6846 & 2460139.6335 & $-54.11 \pm 0.08$ & FEROS \\
J1553-6846 & 2460185.5423 & $-57.88 \pm 0.07$ & FEROS \\
J1553-6846 & 2460228.5196 & $-57.68 \pm 0.05$ & FEROS \\
J1553-6846 & 2460334.8082 & $-26.93 \pm 0.05$ & FEROS \\
J1553-6846 & 2460395.7137 & $-46.12 \pm 0.06$ & FEROS \\
J2102+3703 & 2459838.7411 & $10.52 \pm 0.50$ & PEPSI \\
J2102+3703 & 2459862.7453 & $5.94 \pm 0.05$ & TRES \\
J2102+3703 & 2460193.8744 & $24.49 \pm 0.04$ & TRES \\
J2102+3703 & 2460203.8485 & $23.89 \pm 0.04$ & TRES \\
J2102+3703 & 2460214.7020 & $23.14 \pm 0.04$ & TRES \\
J2102+3703 & 2460222.7492 & $22.46 \pm 0.03$ & TRES \\
J2102+3703 & 2460255.6527 & $19.54 \pm 0.03$ & TRES \\
J2102+3703 & 2460285.6430 & $16.12 \pm 0.04$ & TRES \\
J2102+3703 & 2460307.5776 & $12.99 \pm 0.04$ & TRES \\
J2102+3703 & 2460464.8749 & $20.87 \pm 0.04$ & TRES \\
J0742-4749 & 2459905.8042 & $58.32 \pm 0.06$ & FEROS \\
J0742-4749 & 2460038.5929 & $43.73 \pm 0.50$ & MIKE \\
J0742-4749 & 2460091.5001 & $35.10 \pm 0.17$ & FEROS \\
J0742-4749 & 2460228.8210 & $28.03 \pm 0.05$ & FEROS \\
J0742-4749 & 2460286.7148 & $33.71 \pm 0.07$ & FEROS \\
J0742-4749 & 2460299.7489 & $35.99 \pm 0.05$ & FEROS \\
J0742-4749 & 2460333.6623 & $43.42 \pm 0.09$ & FEROS \\
J0742-4749 & 2460395.5552 & $57.40 \pm 0.05$ & FEROS \\
J0152-2049 & 2459815.8462 & $60.86 \pm 0.11$ & FEROS \\
J0152-2049 & 2459817.8988 & $60.71 \pm 0.11$ & FEROS \\
J0152-2049 & 2459853.8745 & $57.73 \pm 0.08$ & TRES \\
J0152-2049 & 2459878.8281 & $55.80 \pm 0.08$ & TRES \\
J0152-2049 & 2459903.6764 & $54.02 \pm 0.08$ & FEROS \\
J0152-2049 & 2459921.6314 & $52.97 \pm 0.08$ & FEROS \\
J0152-2049 & 2460111.9030 & $37.40 \pm 0.06$ & FEROS \\
J0152-2049 & 2460186.7817 & $25.29 \pm 0.05$ & FEROS \\
J0152-2049 & 2460222.8075 & $21.91 \pm 0.07$ & FEROS \\
J0152-2049 & 2460263.7682 & $64.93 \pm 0.06$ & TRES \\
J0152-2049 & 2460284.6708 & $66.08 \pm 0.10$ & TRES \\
J0152-2049 & 2460284.6789 & $66.16 \pm 0.07$ & FEROS \\
J0152-2049 & 2460315.5915 & $64.05 \pm 0.19$ & TRES \\
J0152-2049 & 2460340.5730 & $61.76 \pm 0.05$ & FEROS \\
J0152-2049 & 2460345.6194 & $61.16 \pm 0.08$ & TRES \\
J0003-5604 & 2459813.8841 & $52.24 \pm 0.08$ & FEROS \\
J0003-5604 & 2459829.6990 & $51.25 \pm 0.06$ & FEROS \\
J0003-5604 & 2459832.7276 & $50.93 \pm 0.16$ & FEROS \\
J0003-5604 & 2459903.6453 & $47.00 \pm 0.07$ & FEROS \\
J0003-5604 & 2459923.5775 & $45.90 \pm 0.06$ & FEROS \\
J0003-5604 & 2460098.9090 & $35.87 \pm 0.12$ & FEROS \\
J0003-5604 & 2460113.8234 & $35.37 \pm 0.11$ & FEROS \\
J0003-5604 & 2460139.7994 & $38.17 \pm 0.05$ & FEROS \\
J0003-5604 & 2460185.6890 & $87.74 \pm 0.07$ & FEROS \\
J0003-5604 & 2460221.7244 & $71.15 \pm 0.06$ & FEROS \\
J0003-5604 & 2460284.6157 & $60.13 \pm 0.05$ & FEROS \\
J0003-5604 & 2460337.5476 & $55.07 \pm 0.07$ & FEROS \\
J1733+5808 & 2460031.8949 & $-3.46 \pm 0.04$ & TRES \\
J1733+5808 & 2460049.9539 & $-3.67 \pm 0.04$ & TRES \\
J1733+5808 & 2460071.8604 & $-4.51 \pm 0.04$ & TRES \\
J1733+5808 & 2460092.7797 & $-6.89 \pm 0.03$ & TRES \\
J1733+5808 & 2460109.9133 & $-10.09 \pm 0.04$ & TRES \\
J1733+5808 & 2460123.8108 & $-13.63 \pm 0.05$ & TRES \\
J1733+5808 & 2460157.7645 & $-24.95 \pm 0.04$ & TRES \\
J1733+5808 & 2460201.6402 & $-35.21 \pm 0.03$ & TRES \\
J1733+5808 & 2460217.6235 & $-36.03 \pm 0.04$ & TRES \\
J1733+5808 & 2460235.6037 & $-35.66 \pm 0.02$ & TRES \\
J1733+5808 & 2460353.0203 & $-23.53 \pm 0.03$ & TRES \\
J1733+5808 & 2460388.0004 & $-19.74 \pm 0.03$ & TRES \\
J1733+5808 & 2460418.9187 & $-16.62 \pm 0.03$ & TRES \\
J1150-2203 & 2459953.9957 & $36.57 \pm 0.04$ & TRES \\
J1150-2203 & 2459979.9657 & $35.69 \pm 0.03$ & TRES \\
J1150-2203 & 2460002.7907 & $34.62 \pm 0.03$ & FEROS \\
J1150-2203 & 2460008.8457 & $34.23 \pm 0.04$ & TRES \\
J1150-2203 & 2460038.6835 & $32.17 \pm 0.50$ & MIKE \\
J1150-2203 & 2460055.7516 & $30.12 \pm 0.02$ & TRES \\
J1150-2203 & 2460072.6913 & $27.65 \pm 0.02$ & FEROS \\
J1150-2203 & 2460089.6143 & $24.40 \pm 0.03$ & FEROS \\
J1150-2203 & 2460103.5881 & $20.80 \pm 0.03$ & TRES \\
J1150-2203 & 2460103.6773 & $20.94 \pm 0.03$ & TRES \\
J1150-2203 & 2460110.5077 & $18.65 \pm 0.04$ & FEROS \\
J1150-2203 & 2460286.8120 & $32.44 \pm 0.03$ & FEROS \\
J1150-2203 & 2460308.0275 & $33.92 \pm 0.03$ & TRES \\
J1150-2203 & 2460333.7586 & $35.47 \pm 0.03$ & FEROS \\
J1150-2203 & 2460344.9676 & $35.96 \pm 0.03$ & TRES \\
J1150-2203 & 2460364.9310 & $36.63 \pm 0.03$ & TRES \\
J1150-2203 & 2460386.8766 & $37.22 \pm 0.06$ & TRES \\
J1150-2203 & 2460397.6330 & $37.48 \pm 0.03$ & FEROS \\
J1150-2203 & 2460407.8571 & $37.59 \pm 0.03$ & TRES \\
J1150-2203 & 2460445.6831 & $38.11 \pm 0.03$ & TRES \\
J1449+6919 & 2459980.9622 & $-77.13 \pm 0.05$ & TRES \\
J1449+6919 & 2460016.9387 & $-73.74 \pm 0.05$ & TRES \\
J1449+6919 & 2460040.9170 & $-71.06 \pm 0.03$ & TRES \\
J1449+6919 & 2460058.9230 & $-68.98 \pm 0.05$ & TRES \\
J1449+6919 & 2460066.8727 & $-67.94 \pm 0.03$ & TRES \\
J1449+6919 & 2460090.8304 & $-64.91 \pm 0.04$ & TRES \\
J1449+6919 & 2460110.7739 & $-62.17 \pm 0.04$ & TRES \\
J1449+6919 & 2460123.7389 & $-60.43 \pm 0.08$ & TRES \\
J1449+6919 & 2460131.6774 & $-59.29 \pm 0.05$ & TRES \\
J1449+6919 & 2460161.6674 & $-55.10 \pm 0.04$ & TRES \\
J1449+6919 & 2460203.6256 & $-50.11 \pm 0.03$ & TRES \\
J1449+6919 & 2460306.0214 & $-59.32 \pm 0.04$ & TRES \\
J1449+6919 & 2460338.9694 & $-69.48 \pm 0.05$ & TRES \\
J1449+6919 & 2460359.9881 & $-75.04 \pm 0.03$ & TRES \\
J1449+6919 & 2460388.9268 & $-80.58 \pm 0.04$ & TRES \\
J1449+6919 & 2460407.9069 & $-82.97 \pm 0.02$ & TRES \\
J1449+6919 & 2460430.8414 & $-84.55 \pm 0.03$ & TRES \\
J1449+6919 & 2460442.8494 & $-85.05 \pm 0.03$ & TRES \\
J1449+6919 & 2460473.7228 & $-85.21 \pm 0.05$ & TRES \\
J0217-7541 & 2459815.8817 & $65.41 \pm 0.11$ & FEROS \\
J0217-7541 & 2459905.6561 & $77.98 \pm 0.04$ & FEROS \\
J0217-7541 & 2459984.5744 & $69.87 \pm 0.06$ & FEROS \\
J0217-7541 & 2460013.5032 & $62.79 \pm 0.08$ & FEROS \\
J0217-7541 & 2460112.9135 & $50.19 \pm 0.11$ & FEROS \\
J0217-7541 & 2460185.7575 & $48.89 \pm 0.10$ & FEROS \\
J0217-7541 & 2460222.8307 & $49.26 \pm 0.05$ & FEROS \\
J0217-7541 & 2460285.6091 & $51.48 \pm 0.03$ & FEROS \\
J0217-7541 & 2460299.7070 & $52.08 \pm 0.03$ & FEROS \\
J0217-7541 & 2460338.5986 & $54.36 \pm 0.04$ & FEROS \\
J0639-3655 & 2459903.7571 & $-12.07 \pm 0.05$ & FEROS \\
J0639-3655 & 2459919.7583 & $-12.04 \pm 0.04$ & FEROS \\
J0639-3655 & 2460038.5444 & $-8.82 \pm 0.50$ & MIKE \\
J0639-3655 & 2460076.5173 & $-4.55 \pm 0.13$ & FEROS \\
J0639-3655 & 2460089.4660 & $-5.97 \pm 0.08$ & FEROS \\
J0639-3655 & 2460185.8288 & $-9.71 \pm 0.32$ & FEROS \\
J0639-3655 & 2460225.8698 & $-10.42 \pm 0.04$ & FEROS \\
J0639-3655 & 2460284.7922 & $-10.96 \pm 0.05$ & FEROS \\
J0639-3655 & 2460333.5959 & $-11.23 \pm 0.06$ & FEROS \\
J0639-3655 & 2460395.5203 & $-11.58 \pm 0.04$ & FEROS \\
J1739+4502 & 2459838.6680 & $-268.16 \pm 0.15$ & FEROS \\
J1739+4502 & 2460035.9812 & $-270.48 \pm 0.08$ & TRES \\
J1739+4502 & 2460053.9460 & $-270.10 \pm 0.13$ & TRES \\
J1739+4502 & 2460081.9051 & $-270.17 \pm 0.10$ & TRES \\
J1739+4502 & 2460105.8478 & $-269.25 \pm 0.08$ & TRES \\
J1739+4502 & 2460121.8751 & $-268.16 \pm 0.09$ & TRES \\
J1739+4502 & 2460133.7822 & $-267.33 \pm 0.15$ & TRES \\
J1739+4502 & 2460160.7388 & $-262.50 \pm 0.19$ & TRES \\
J1739+4502 & 2460194.6856 & $-242.21 \pm 0.08$ & TRES \\
J1739+4502 & 2460205.6623 & $-238.60 \pm 0.09$ & TRES \\
J1739+4502 & 2460216.6222 & $-241.40 \pm 0.10$ & TRES \\
J1739+4502 & 2460222.6541 & $-243.83 \pm 0.08$ & TRES \\
J1739+4502 & 2460238.6060 & $-249.57 \pm 0.09$ & TRES \\
J1739+4502 & 2460354.0040 & $-263.81 \pm 0.09$ & TRES \\
J1739+4502 & 2460387.0001 & $-265.08 \pm 0.10$ & TRES \\
J1739+4502 & 2460414.8786 & $-266.23 \pm 0.07$ & TRES \\
J1739+4502 & 2460434.9001 & $-266.84 \pm 0.09$ & TRES \\
J1739+4502 & 2460477.8178 & $-267.65 \pm 0.10$ & TRES \\
J0036-0932 & 2459815.8221 & $28.72 \pm 0.03$ & FEROS \\
J0036-0932 & 2459838.8836 & $28.38 \pm 0.50$ & PEPSI \\
J0036-0932 & 2459850.8772 & $28.00 \pm 0.03$ & TRES \\
J0036-0932 & 2459878.7892 & $27.18 \pm 0.03$ & TRES \\
J0036-0932 & 2459905.6201 & $26.16 \pm 0.07$ & FEROS \\
J0036-0932 & 2459923.5550 & $25.18 \pm 0.17$ & FEROS \\
J0036-0932 & 2459928.7079 & $24.75 \pm 0.50$ & PEPSI \\
J0036-0932 & 2460087.9134 & $6.05 \pm 0.15$ & FEROS \\
J0036-0932 & 2460110.8711 & $7.94 \pm 0.04$ & FEROS \\
J0036-0932 & 2460139.8395 & $12.02 \pm 0.06$ & FEROS \\
J0036-0932 & 2460186.7594 & $18.28 \pm 0.03$ & FEROS \\
J0036-0932 & 2460221.7623 & $21.66 \pm 0.02$ & FEROS \\
J0036-0932 & 2460281.6525 & $25.41 \pm 0.03$ & TRES \\
J0036-0932 & 2460285.5846 & $25.70 \pm 0.04$ & FEROS \\
J0036-0932 & 2460311.6436 & $26.62 \pm 0.04$ & TRES \\
J0036-0932 & 2460333.5373 & $27.45 \pm 0.04$ & FEROS \\
J1432-1021 & 2459813.4918 & $139.74 \pm 0.10$ & FEROS \\
J1432-1021 & 2459831.5038 & $143.61 \pm 0.41$ & FEROS \\
J1432-1021 & 2459832.4866 & $143.60 \pm 0.25$ & FEROS \\
J1432-1021 & 2459929.0525 & $154.65 \pm 0.50$ & PEPSI \\
J1432-1021 & 2459981.0076 & $153.69 \pm 0.08$ & TRES \\
J1432-1021 & 2459990.8528 & $153.22 \pm 0.15$ & FEROS \\
J1432-1021 & 2459992.8178 & $153.16 \pm 0.15$ & FEROS \\
J1432-1021 & 2460012.8160 & $151.78 \pm 0.12$ & FEROS \\
J1432-1021 & 2460026.7002 & $150.29 \pm 0.15$ & FEROS \\
J1432-1021 & 2460036.8989 & $149.37 \pm 0.07$ & TRES \\
J1432-1021 & 2460038.7738 & $149.18 \pm 0.50$ & MIKE \\
J1432-1021 & 2460039.7097 & $149.14 \pm 0.09$ & FEROS \\
J1432-1021 & 2460046.9207 & $148.36 \pm 0.07$ & TRES \\
J1432-1021 & 2460050.6922 & $147.89 \pm 0.10$ & FEROS \\
J1432-1021 & 2460059.8472 & $146.89 \pm 0.10$ & TRES \\
J1432-1021 & 2460072.7468 & $145.13 \pm 0.20$ & FEROS \\
J1432-1021 & 2460078.7900 & $144.35 \pm 0.16$ & TRES \\
J1432-1021 & 2460085.6962 & $143.28 \pm 0.22$ & FEROS \\
J1432-1021 & 2460094.8017 & $142.60 \pm 0.08$ & TRES \\
J1432-1021 & 2460098.6573 & $141.94 \pm 0.25$ & FEROS \\
J1432-1021 & 2460109.7473 & $140.43 \pm 0.06$ & TRES \\
J1432-1021 & 2460110.6418 & $140.22 \pm 0.07$ & FEROS \\
J1432-1021 & 2460133.6779 & $136.95 \pm 0.07$ & TRES \\
J1432-1021 & 2460139.5731 & $136.22 \pm 0.13$ & FEROS \\
J1432-1021 & 2460185.4878 & $128.99 \pm 0.21$ & FEROS \\
J1432-1021 & 2460186.4889 & $128.99 \pm 0.15$ & FEROS \\
J1432-1021 & 2460306.0444 & $113.88 \pm 0.12$ & TRES \\
J1432-1021 & 2460338.8541 & $111.97 \pm 0.10$ & FEROS \\
J1432-1021 & 2460340.8396 & $111.67 \pm 0.08$ & FEROS \\
J1432-1021 & 2460352.9597 & $111.21 \pm 0.08$ & TRES \\
J1432-1021 & 2460374.9193 & $111.29 \pm 0.09$ & TRES \\
J1432-1021 & 2460386.9141 & $111.63 \pm 0.09$ & TRES \\
J1432-1021 & 2460425.6572 & $115.04 \pm 0.11$ & FEROS \\
J1432-1021 & 2460479.6871 & $124.69 \pm 0.16$ & FEROS \\
J1048+6547 & 2459984.9336 & $73.39 \pm 0.07$ & TRES \\
J1048+6547 & 2460032.7230 & $69.17 \pm 0.06$ & TRES \\
J1048+6547 & 2460072.7749 & $65.57 \pm 0.04$ & TRES \\
J1048+6547 & 2460110.6813 & $62.65 \pm 0.09$ & TRES \\
J1048+6547 & 2460328.9819 & $46.76 \pm 0.05$ & TRES \\
J1048+6547 & 2460353.8875 & $45.28 \pm 0.04$ & TRES \\
J1048+6547 & 2460390.7877 & $43.39 \pm 0.08$ & TRES \\
J1048+6547 & 2460407.8115 & $42.52 \pm 0.06$ & TRES \\
J1048+6547 & 2460443.7177 & $40.92 \pm 0.05$ & TRES \\
J2145+2837 & 2459891.7403 & $-32.63 \pm 0.04$ & TRES \\
J2145+2837 & 2460103.8742 & $-39.70 \pm 0.03$ & TRES \\
J2145+2837 & 2460120.9421 & $-40.25 \pm 0.03$ & TRES \\
J2145+2837 & 2460162.8141 & $-41.78 \pm 0.03$ & TRES \\
J2145+2837 & 2460196.8219 & $-43.24 \pm 0.02$ & TRES \\
J2145+2837 & 2460206.7777 & $-43.58 \pm 0.02$ & TRES \\
J2145+2837 & 2460239.6458 & $-45.07 \pm 0.03$ & TRES \\
J2145+2837 & 2460247.7816 & $-45.37 \pm 0.02$ & TRES \\
J2145+2837 & 2460287.6278 & $-47.36 \pm 0.03$ & TRES \\
J2145+2837 & 2460307.6060 & $-48.43 \pm 0.02$ & TRES \\
J2145+2837 & 2460493.8605 & $-64.19 \pm 0.02$ & TRES \\
J2244-2236 & 2459824.7906 & $6.87 \pm 0.06$ & FEROS \\
J2244-2236 & 2459864.7070 & $7.04 \pm 0.03$ & TRES \\
J2244-2236 & 2459903.6224 & $7.72 \pm 0.05$ & FEROS \\
J2244-2236 & 2459921.5779 & $8.16 \pm 0.04$ & FEROS \\
J2244-2236 & 2460080.9073 & $37.09 \pm 0.09$ & FEROS \\
J2244-2236 & 2460085.9247 & $38.94 \pm 0.07$ & FEROS \\
J2244-2236 & 2460101.8494 & $42.95 \pm 0.04$ & FEROS \\
J2244-2236 & 2460112.8120 & $43.59 \pm 0.03$ & FEROS \\
J2244-2236 & 2460139.7586 & $39.84 \pm 0.04$ & FEROS \\
J2244-2236 & 2460186.7363 & $30.45 \pm 0.02$ & FEROS \\
J2244-2236 & 2460221.6901 & $25.49 \pm 0.05$ & FEROS \\
J2244-2236 & 2460284.5677 & $19.45 \pm 0.04$ & FEROS \\
J2244-2236 & 2460288.5855 & $19.14 \pm 0.03$ & TRES \\
J0824+5254 & 2459923.9445 & $-44.76 \pm 0.04$ & TRES \\
J0824+5254 & 2459928.9130 & $-45.16 \pm 0.50$ & PEPSI \\
J0824+5254 & 2459970.8519 & $-46.99 \pm 0.06$ & TRES \\
J0824+5254 & 2460013.8030 & $-48.91 \pm 0.03$ & TRES \\
J0824+5254 & 2460037.8106 & $-49.80 \pm 0.03$ & TRES \\
J0824+5254 & 2460060.7132 & $-50.60 \pm 0.03$ & TRES \\
J0824+5254 & 2460250.9537 & $-56.58 \pm 0.03$ & TRES \\
J0824+5254 & 2460286.9620 & $-57.64 \pm 0.02$ & TRES \\
J0824+5254 & 2460328.8563 & $-58.92 \pm 0.02$ & TRES \\
J0824+5254 & 2460345.8572 & $-59.38 \pm 0.03$ & TRES \\
J0824+5254 & 2460391.7542 & $-60.76 \pm 0.03$ & TRES \\
J0824+5254 & 2460408.7250 & $-61.33 \pm 0.02$ & TRES \\
J0824+5254 & 2460458.6643 & $-62.95 \pm 0.04$ & TRES \\
J0230+5950 & 2459838.9142 & $-3.67 \pm 0.50$ & PEPSI \\
J0230+5950 & 2459928.7838 & $1.70 \pm 0.50$ & PEPSI \\
J0230+5950 & 2459939.6920 & $2.00 \pm 0.17$ & TRES \\
J0230+5950 & 2459958.6523 & $2.98 \pm 0.18$ & TRES \\
J0230+5950 & 2459970.6131 & $3.65 \pm 0.16$ & TRES \\
J0230+5950 & 2460208.9293 & $10.78 \pm 0.09$ & TRES \\
J0230+5950 & 2460222.8918 & $11.31 \pm 0.10$ & TRES \\
J0230+5950 & 2460244.8058 & $11.70 \pm 0.12$ & TRES \\
J0230+5950 & 2460255.8226 & $12.16 \pm 0.09$ & TRES \\
J0230+5950 & 2460277.8158 & $12.51 \pm 0.16$ & TRES \\
J0230+5950 & 2460284.7343 & $12.89 \pm 0.09$ & TRES \\
J0230+5950 & 2460307.6977 & $13.31 \pm 0.10$ & TRES \\
J0230+5950 & 2460328.6374 & $13.76 \pm 0.11$ & TRES \\
J0230+5950 & 2460339.6308 & $14.67 \pm 0.12$ & TRES \\
J0230+5950 & 2460359.6961 & $15.13 \pm 0.13$ & TRES \\
J0634+6256 & 2459826.0707 & $-72.36 \pm 1.00$ & ESI \\
J0634+6256 & 2459838.9491 & $-68.35 \pm 0.50$ & PEPSI \\
J0634+6256 & 2459924.8543 & $-47.66 \pm 0.08$ & TRES \\
J0634+6256 & 2459928.8945 & $-47.65 \pm 0.50$ & PEPSI \\
J0634+6256 & 2460013.7350 & $-45.74 \pm 0.12$ & TRES \\
J0634+6256 & 2460254.8754 & $-51.19 \pm 0.06$ & TRES \\
J0634+6256 & 2460286.9123 & $-52.00 \pm 0.08$ & TRES \\
J0634+6256 & 2460326.8598 & $-53.17 \pm 0.07$ & TRES \\
J0634+6256 & 2460345.7550 & $-53.80 \pm 0.06$ & TRES \\
J0634+6256 & 2460411.6931 & $-56.01 \pm 0.05$ & TRES \\
\hline 
\addlinespace
\caption{Radial velocities for all targets. A machine-readable version of the table is included as supplemental material.}
\label{tab:rvs}
\end{longtable}

%\end{multicols}  % End the multicols environment for the appendix

\end{document}